\documentclass{aa} 
\usepackage{txfonts}
\usepackage{natbib}
\usepackage{graphicx,hyperref}
\usepackage{color}
\bibpunct{(}{)}{;}{a}{}{,}

\newcommand{\ltsima}{$\; \buildrel < \over \sim \;$}
\newcommand{\simlt}{\lower.5ex\hbox{\ltsima}}
\newcommand{\gtsima}{$\; \buildrel > \over \sim \;$}
\newcommand{\simgt}{\lower.5ex\hbox{\gtsima}}
\newcommand{\be}{\begin {equation}}
\newcommand{\ee}{\end {equation}}

\newcommand{\gr}{\kern 2pt\hbox{}^\circ{\kern -2pt K}} 
\newcommand{\mytilde}{\raise.17ex\hbox{$\scriptstyle\mathtt{\sim}$}}
\newcommand\rtilde{\stackrel{\sim}{\smash{r}\rule{0pt}{0.8ex}}}

\def\msun{\rm{\,M_{\odot}}}

\def\msunh{\rm{\,M_{\odot}{\it h}^{-1}}}

\def\ergs{\rm{\,ergs^{-1}}}
\def\kev{\rm{\,keV}}
\def\numd{\rm{\,cm^{-3}}}
\def\sxu{\rm{\,cm^2}{gr^{-1}}}
\def\dms{\rm{\,gr{\,cm^{-2}}}}

\def\kms{\rm{\,km~s^{-1}}}
\def\acs{\rm{\,arcsec}}
\def\cts{\rm{\,counts~s^{-1}~cm^{-2}~arcsec^{-2} }}
\def\ctsn{\rm{\,counts~arcsec^{-2} }}
\def\kpc{\rm{\,kpc}}
\def\mpc{\rm{\,Mpc}}

\def\rcp{r_c^{\rm{prm}}}
\def\XDM{\rm{X-DM}}
\def\BDM{\rm{BCG-DM}}

\def\BX{\rm{BCG-X}}
\def\SDM{\rm{SZ-DM}}
\def\concI{\rm{c^{NW}_{200}}}
\def\concII{\rm{c^{SE}_{200}}}

\def\Log10{{\rm~Log_{10}}}

\def\lcdm{$\Lambda$CDM}
\def\om{\Omega_m}
\def\oml{\Omega_{\Lambda}}

\def\aap{A\&A}
\def\apj{ApJ}

\def\apjl{ApJ}
\def\mnras{MNRAS}

\def\aj{AJ}
\def\na{New Astronomy}

\def\physrep{Phys. Rep.}

\def\apjs{ApJS}

\def\gca{Geochim. Cosmochim. Acta}

\defcitealias{Zh15}{ZYL15}
\begin{document}

\title{
An N-body hydrodynamical simulation  study  of the merging  
cluster El Gordo: A compelling case for self-interacting dark matter? }

\subtitle{}

   \author{R. Valdarnini
          \inst{1,}
          }

   \institute{SISSA/ISAS, via Bonomea 265, I-34136 Trieste, Italy\\
              \email{valda@sissa.it}
             }

   \date{Received ; accepted }

\abstract{
We used a large set N-body/hydrodynamical simulations to  study the physical 
properties of the  merging  cluster  El Gordo. 
We find that the observed  X-ray structures, along with
other data,  can be matched  fairly well by simulations with  collision velocities 
$2,000 \kms  \simlt V \simlt 2,500 \kms$  and impact parameters 
$600 \kpc \simlt P \simlt 800 \kpc$. 
 The mass of the primary is constrained to 
be between $\sim 10^{15} \msun$ and $\sim 1.6 \cdot 10^{15} \msun$, in 
accordance with recent lensing-based mass measurements.
Moreover, a returning, post-apocenter,  scenario  is not supported by our 
head-on simulations.

We also considered merger models  that incorporate dark matter self-interactions.
The simulation results show that 
the observed spatial offsets between the different mass components are well 
reproduced in  self-interacting dark matter models with an elastic cross-section 
in the range  $\sigma_{DM}/{m_X}  \sim 4 -5 \sxu$.
In addition, the mean relative  line-of-sight  radial velocity between the
two brightest cluster galaxies  is  found to be  on the order of several
hundred $\kms$.  We argue that these findings provide an unambiguous signature 
of a dark matter behavior that exhibits  collisional properties in  a very 
energetic high-redshift cluster collision.  
 The range of  allowed  values we find for  $\sigma_{DM}/{m_X} $  is,  
however, inconsistent with present upper limits.
 To resolve this tension, we  suggest 
the possibility that the  self-interacting dark matter model used here  
be considered  as only a low-order approximation, and that  the underlying 
physical processes that describe the interaction of dark matter in major cluster 
mergers are  more complex than can be adequately represented by the commonly 
assumed approach based on the  scattering of dark matter particles.
}


\titlerunning{ Hydrodynamic  simulations of the El Gordo clusters
 }

\maketitle

%
\section{Introduction}
\label{sec:intro}

 Galaxy clusters are the largest virialized structures  known in the Universe,
and according to the  hierarchical scenario, their formation 
proceeds from the bottom up,  through the sequential accretion and merging of smaller 
structures.  They have measured virial masses in the range $M \sim 10^{14}-10^{15} \msun$.
Observations show that about $\sim 80\%$ of their mass content  consists of 
dark matter (DM), with the remaining  $\sim 20 \%$ in the form of baryons. 
The bulk of  these baryons ($\sim 90 \%$)  resides in the form of an 
 intracluster medium (ICM), with only a few per cent  in the form of galaxies
 \citep{Voit05}.

During the process of  cluster formation, high-velocity  gas 
flows dissipate the infall kinetic energy  and the gas is shock-heated to high temperatures.
Because of the masses involved,   at equilibrium these temperatures  will be 
 as high as $T\sim 10^7-10^8 $ K,  and  the ICM  will be in the form of a 
hot diffuse plasma.
Galaxy clusters are thus powerful X-ray emitters, and
their X-ray emission can be used to study the physics of the 
ICM and in turn derive cluster properties
 with which to constrain the currently accepted Lambda cold dark matter 
(CDM) cosmological model \citep{Voit05}.

Moreover,  observed X-ray maps show that a significant fraction of galaxy 
clusters exhibit a disturbed X-ray morphology  and are in a dynamically unrelaxed 
state \citep{Buote02}, with two or more subclusters still in the process of 
accretion and merging. In this context, the merging of galaxy clusters  is 
 essential in a number of ways for probing ICM physics and the underlying 
cosmology \citep{Sar02,Molnar16}.

Mergers between galaxy clusters can be broadly classified into two categories, 
according to the mass ratio
between  the two main mass subcomponents. Mergers are considered major when the mass 
ratio of the primary to the secondary is close to unity, and minor whenever the mass of 
the secondary is $ \simlt 30-40\%$ of the primary mass.
X-ray surface brightness maps of minor mergers reveal the existence of 
cold fronts and shocks  \citep{Mark07}, which can be used to derive ICM transport 
properties  \citep{Mark07,ZuH16}.

In this scenario, major mergers of massive galaxy clusters are among the most 
energetic 
events since the Big Bang, with collision energies in the 
range $ \sim 10^{63} - 10^{64} \ergs$  \citep{Sar02}. The energy involved in 
these collisions renders these objects unique laboratories for probing the 
collisional properties of their components, as well as  the likelihood of such 
events within  the \lcdm~cosmological model \citep{lee10,Lage15,As21}.

An important effect that will occur during a major merger is the 
expected offset between the dissipative ICM and the collisionless 
components:  galaxies and DM.  Additionally, self-interacting dark matter (SIDM)  
will also exhibit 
a further offset from the galaxy component, which will depend on the 
strength of the  cross-section. Observations of spatial offsets 
between different  components can then be used to constrain these models 
\citep[see][for a review ]{Tulin18}.

In recent years, the existence of major mergers between galaxy clusters
has been established by a growing number of observations:  
 for example, the Bullet Cluster  1E0657-56  at redshift $z=0.296$ 
\citep{Mark02,Clowe06}, 
the first observed to show the existence of DM, by displaying 
a clear spatial offset between the collisionless DM component and the 
X-ray emitting ICM. Other examples of such systems are  the cluster
 ACT-CL J0102-4915    \citep{Men12,Jee14,Zi13},
the Sausage Cluster CIZA J2242.8+5301 \citep{Jee15},
 ZwCl008.8+52 \citep{Gol17}, A1758N \citep{Rago12}, and A2146 \citep{Russel10}.

Because of  the  nonlinearity of the merging process and the
 variety  of the associated physical phenomena,
 N-body/hydrodynamical simulations have proved to be  a very powerful  tool 
for studying and interpreting mergers between galaxy clusters. 
 In particular,  idealized  binary merger simulations are designed to 
follow, in  a Newtonian frame, the collision evolution of two merging 
halos that are initially separated and at equilibrium.  
At variance with cosmological simulations, this approach allows  
a single merger to be studied   under carefully controlled initial conditions and
can be used to model a specific merging event \citep{Molnar16}.

Individual binary cluster merger simulations have been performed by many 
authors. These simulations have been  aimed either at studying 
the effects of mergers  in general
 \citep{Ro96,Ricker01,RT02,Poole06,MC07,Poole08,Mi09,ZuH09} or investigating some specific property of the ICM.  In particular, 
simulations of merging clusters 
have been used to study the    origin of cold fronts and gas sloshing 
\citep{He03,As06,ZuH10,Roe12,ZuH16,Sh18}, the formation of cool cores  
  \citep{ZuH11,VS21}, the offset between X-ray and  
  Sunyaev-Zel'dovich (SZ) map peaks \citep{Molnar12,Zh14}, and 
ICM transport properties \citep{Rus10,Roe13,ZuH13,ZuH15,Sch17}.

Moreover, this simulation approach allows simulations aimed at 
studying a particular merging system to be performed and therefore contrast  
the simulation results with a given  observation. 
Some examples of such specifically designed simulations include those of 
the Bullet Cluster \citep{Spr07,Mas08,La14}, the 
  cluster ACT-CL J0102-4915 \citep{Donnert14,Molnar15,Zh15,Zh18}, the Sausage Cluster \citep{Donnert17,Mol17},
Cygnus A \citep{Ha19}, A2146 \citep{Can12,Cha22}, 
 A3376 \citep{Mac13}, A1758N \citep{Mac15}, 
the cluster ZwCl008.8+52  \citep{Mol18}, and 
A2034 \citep{Mou21}.

Among these examples of extreme collisions, the  cluster ACT-CL J0102-4915 
at $z=0.870$, nicknamed  El Gordo,  
It was originally detected as the cluster with the 
strongest SZ effect in the Atacama Cosmology Telescope (ACT) survey 
\citep{Men10,Hass13}. From optical and X-ray studies, \citet{Men12} obtained a 
total mass on the 
order of $\sim 2 \cdot 10^{15} \msun$, with a galaxy 
velocity dispersion $\sigma_{gal} \sim 1,300 \kms$ and an integrated 
temperature in the range  $T_\mathrm{X}\simeq 15 $ keV. Similar mass estimates 
have  also been obtained from weak lensing \citep[WL;][]{Jee14} and strong lensing (SL) 
studies \citep{Zi13}.
This range of estimated masses  indicates that El Gordo  is a
major merger cluster as well as the most massive cluster
at $z>0.6$;   however, it was soon realized that the existence of such a massive 
cluster  at this redshift
is extremely unlikely within  the standard \lcdm~model \citep{Men12,Jee14}.

 The cluster exhibits a binary mass 
structure \citep{Jee14}, with its two subclusters denominated the northwestern
 (NW) and southeastern (SE), respectively 
\citep[see, for example, Figure 1 of ][]{Ng15}
because of their positions. 
WL analysis \citep{Jee14}  shows a 
 projected separation  between the mass centroids of the two components  
 of approximately   $ d \sim 700 \kpc$,  with a mass ratio in the range 
 $\sim 2:1$.  Dynamical estimates \citep{Men12}  suggest a   
high relative velocity for the merging infall, in the range $\sim 1,500\kms$ to $\sim 2,500 \kms$,  with the  uncertainties 
in the allowed range due to projection effects. 

A significant effect expected in high-velocity merging clusters is the presence
of large spatial offsets between the X-ray surface brightness peaks and the
 centroid of the surface mass densities \citep{Molnar12}, as well as between
the SZ and X-ray centroids.  All of these features have been observed in the 
El Gordo cluster \citep{Men12,Zi13}: the measured distance between the SZ and 
X-ray centroid  is $ d\sim 600 \kpc,$ and the SZ centroid is  offset  by 
a distance of $ \sim 150 \kpc$ from the main mass centroid.
Additionally, the peak of the X-ray emission is also offset from the position 
of the brightest cluster galaxy \citep[BCG;][]{Men12}. It is located in the SE 
cluster component  and  is offset by 
$ \sim 70 \kpc$ with respect the SE X-ray peak.

Moreover, the X-ray morphology of El Gordo presents several interesting features. 
The total X-ray luminosity is $L_X\sim 2\cdot 10^{45} \ergs$ in the 
$0.5-2$ keV band \citep{Men12}; the highest X-ray emission is in the 
SE region 
and is characterized by a single peak and two extended faint tails  elongated 
beyond the peak. The X-ray emission in the NW region is significantly weaker and 
presents a wake-like feature, indicative of  turbulent flows.
This well-defined X-ray morphology suggests that  the  merging axis is 
quite close to the plane of the sky 
\citep{Men12}; this is also consistent with the observed presence of a double 
radio relic \citep{Lind14}.

This set of  multifrequency observations  led to the conclusion that  
El Gordo is a   cluster at high redshift that is experiencing a major merger of two 
 massive clusters that collided  at  a  high velocity ($ \simgt 2,000 \kms$).
In this context, the most likely physical interpretation  of the ongoing 
merging \citep{Men12,Jee14} is  a scenario in which the dense cool
gas core of the secondary is moving from the NW to the SE, inducing 
 a wake in the ICM of the primary and  the
two-tail cometary structure  seen in the X-ray images.  The original 
 gas core of the primary is absent because it was destroyed during 
the collision with the high-velocity compact core of the secondary. The secondary 
 is  thus moving away from the primary, and the two clusters are seen after 
 the first pericenter passage; this is the so-called outgoing scenario.

These observational features indicate that the El Gordo cluster can be 
considered an extreme merging event,  like the Bullet Cluster but at a higher 
redshift. High-velocity collisions between  massive clusters  constitute a 
 severe challenge  to the commonly accepted \lcdm~ paradigm. For example,
\citet{As21}  argue that in a \lcdm~ context the detection of a cluster 
such as El Gordo   can be ruled out with a high significance.

Another noteworthy feature of El Gordo is the peak location of the 
different mass components, as probed by  various multifrequency observations 
\citep{Men12,Jee14, Zi13,Die20,Kim21}. As expected, the X-ray peak position
of the SE cluster is spatially offset from that of its mass centroid; 
however, the X-ray peak is not trailing the DM peak as in the Bullet Cluster 
but is instead leading the mass peak. This is clearly at 
variance with what is expected 
from dissipative arguments, but it can be explained in the so-called 
returning scenario, as proposed by \citet{Ng15}.  
According to those authors, the two clusters are not moving away from each other 
(unlike just after the pericenter passage); instead, the cluster is now in a 
post apocenter phase, with the two clusters  moving towards each other.

Moreover, the BCG   is not only trailing the X-ray peak but 
appears to also be spatially offset from the mass centroid.
This behavior  has been observed in other clusters \citep{Hal04,Can12}, 
and in an  SIDM scenario it is expected  to exist 
in major mergers,  between the collisional DM 
and  collisionless galaxies \citep{Kim17}.

Because of its many unusual properties, El Gordo has been the subject of 
several N-body/\linebreak hydrodynamical simulations aimed at reproducing the observed
X-ray morphology  \citep{Donnert14,Molnar15,Zh15,Zh18}.  
Simulations by \citet{Donnert14} were able to reproduce some of the 
cluster X-ray properties, such as the total X-ray luminosity and the offset
 between the X-ray peak and the SZ centroid,  but failed to replicate the observed
twin-tailed X-ray morphology.  This could be due to a number of factors, including the 
numerical scheme used   in the simulations.
 \citet{Donnert14}
carried out the simulations by employing  a smoothed particle hydrodynamics 
\citep[SPH;][]{Price2012} scheme, which, in its standard formulation,  it is 
known to suppress the 
growth of instabilities and, in turn, turbulent mixing  \citep{Ag07,Mi09}.

For this reason, in their simulations of the El Gordo cluster,
\citet{Molnar15} chose to use  the 
 Eulerian adaptative mesh refinement (AMR) code FLASH \citep{Fry20}. 
By considering a range of initial merger parameters, they were able to
reproduce the twin-tailed X-ray morphology,  but 
 with a projection angle between the plane of the sky and the
 merging axis that is higher ($\simgt 40 ^{\circ}$) 
 than that suggested by radio relic polarization arguments \citep{Ng15}.
Moreover, the total  cluster X-ray luminosity and mean temperature 
 extracted from the simulations were lower than the measured values,
while  the spatial offsets between the centroids were larger than those observed.

 The most complete simulation study of El Gordo presented so far was computed 
by \citet[][hereafter \citetalias{Zh15}]{Zh15}. By considering a wide range of 
initial merging parameter space, the authors constructed a large set of 
N-body/hydrodynamical binary merging simulations  aimed at reproducing the 
observational aspects of El Gordo. They used both SPH and AMR-based codes.

From their simulation ensemble, the model that is best able  to reproduce 
many of the observed properties of El Gordo is a merging cluster
(``model B'') with total mass $\sim 3 \cdot 10^{15} \msun$ and a high mass 
ratio ($\sim 3.6$).
The initial merging configuration is that of an off-axis merger with impact 
parameter $\sim 800 \kpc$ and a very high initial relative velocity 
($\sim 2,500 \kms$). The best match with measured offsets and
X-ray images of El Gordo,  as well as its temperature and luminosity, is obtained
$t \sim 0.14 $ Gyr after it passes the pericenter and with a modest inclination
angle ($ \sim 30 ^{\circ}$) between the merging axis and the sky plane.

Model B of \citetalias{Zh15} can match most of the observed
features of the El Gordo cluster, but it  is not entirely free
of problems. Specifically, a final single X-ray peak is found if the gas core 
of the primary
is destroyed during the collision, but this is obtained only if the initial
gas fraction of the primary is low ($\sim 5\%$). This in turn produces 
X-ray emission  in the outer region of the merging cluster  that is lower than 
observed. 
Finally, it is worth noting that the \citetalias{Zh15} simulations can reproduce
the total X-ray luminosity of El Gordo only for a primary cluster mass of 
$\sim 2.5 \cdot 10^{15} \msun$.
To date, the simulation set of \citetalias{Zh15} constitutes the most exhaustive 
simulation study of the El Gordo cluster. However, the merging scenario 
investigated by \citetalias{Zh15} might need to be  revisited 
in light of recent works that suggest 
substantially lower masses for the El Gordo cluster than those previously 
measured.

By adopting a free-form lens model,  
 \citet{Die20} estimate from their SL study a cluster total mass on the 
order of $\sim 10^{15} \msun$.  This value is about a factor of 2 lower 
then previous SL studies \citep{Zi13}.
In a recent paper, \citet{Kim21} present an improved WL analysis aimed at 
estimating the total mass of El Gordo and that of its components.
The values they report  for the NW and SE clusters are $\sim 9.9 \cdot 10^{14} 
\msun$ and $\sim 6.5 \cdot 10^{14} \msun$, respectively. These values are 
significantly lower ($\sim 60\%$) than those dynamically inferred by 
\citet{Men12}. They are also lower   ($\sim 30\%$) than the WL values 
reported by   \citet{Jee14}.

These updated mass estimates were obtained independently by both SL \citep{Die20}    and WL studies \citep{Kim21} based on  improved
lensing modeling; they both give consistently lower mass values than 
those estimated in previous works.  This clearly has the advantage of easing 
the tension with the \lcdm~model (\citealt{Kim21}; but see
\citealt{As23} for a different view).  Nonetheless, it remains to be 
tested whether this range of masses for the El Gordo cluster  is 
consistent with its observed X-ray morphology.

Therefore, the aim of this paper is to perform a series of 
N-body/hydrodynamical simulations of a merging scenario that incorporates  the
recently revised El Gordo cluster masses.
The simulations were performed for a variety of merging initial conditions: we did 
this in order to find the final merging configuration that can best reproduce 
various observations of El Gordo.
The  initial condition setup mirrors that implemented by
\citetalias{Zh15};  this was purposely  chosen 
to ease the comparison of  our results with  previous findings.

 The structure of the  paper is as follows. 
 Section \ref{sec:sims}  briefly describes  the hydrodynamic method we use,
 as well as the procedures employed to  set up the merging initial conditions, 
illustrating how to  construct particle realizations of cluster halos 
initially at equilibrium and the initial merging kinematics.
The results are presented in Section \ref{sec:results}, with   
 Section \ref{sec:opt} showing results from off-axis merger simulations  aimed at 
 reproducing  the observed twin-tailed X-ray morphology.  
In  Section \ref{sec:return}  we analyze the consistency of a returning scenario
according to the results extracted from head-on cluster collisions,  
while Section 
\ref{sec:sidm} is dedicated to merger simulations that incorporate  a BCG stellar
component and to  the impact of SIDM on some merging models.
Section \ref{sec:discuss} summarizes our main conclusions.
Throughout this work  we use a concordance \lcdm~ cosmology,  with
$\om=0.3$, $\oml=0.7,$ and Hubble constant 
$H_0=70\equiv 100h$\,km\,s$^{-1}$\,Mpc$^{-1}$.

\section{Simulation setup }
\label{sec:sims}
\subsection{Simulations}
\label{subsec:nummt}
%
 Our code consists  of an improved SPH numerical scheme for the hydro part,  
coupled with a standard treecode \citep{He89} to solve the  gravity 
problem.
  The Lagrangian SPH code employs an entropy conserving formulation, while 
in the momentum equations SPH gradients  are estimated 
 by evaluating integrals and performing a matrix inversion. 

This tensor approach \citep{ga12} has been
tested in a variety of hydrodynamical test cases \citep{V16}, showing that it 
outperforms standard SPH  \citep{Price2012} by greatly reducing gradient errors 
and in terms of code performances  
 can be considered  competitive 
with results obtained from AMR codes \citep{V16}. The code has  
previously been applied  to the study of turbulence in galaxy clusters 
\citep{V19} 
and to that of cool-cores  in merging  clusters \citep{VS21}.
In particular,  we refer to this last study for a comparison with results 
extracted
 from merger simulations based on Eulerian AMR schemes and for a full 
description of the code.
In the following we refer to the  SPH scheme  used here as 
integral SPH (ISPH).

For the sake of completeness, we introduce here the SPH gas density equation, 
to which  we will later refer for further generalization.
In SPH there are  $N$ gas particles with mass $m_i$, position $\bf r_i$, and 
velocity $\bf v_i$ that represent the fluid.

The SPH gas density $\rho({\bf r}) $ at the particle position $\bf r_i$   
 \begin{equation}
 \rho({\bf r_i}) =\sum_j m_j W(|{\bf r}_{ij}|,h_i)~
    \label{rho.eq}
 \end{equation}
gives the particle gas density $\rho(\bf r_i) \equiv \rho_i$. The summation
 should be over all the set of $N$ particles, but because the adopted kernel
 $W(| {\bf r}_{ij} |,h_i)$ has compact support it reduces to only the  
 $N_{nn}$ neighboring particles $j$  
for which   $|{\bf r}_i-{\bf r}_j| \leq  2 h_i$.  We use here the  $M_4$ kernel
\citep{Price2012}. 
  The smoothing length $h_i$  is obtained by finding the root of
   \begin{equation}
  \frac{4 \pi (2h_i)^3 \rho_i}{3}=N_{nn} m_i,
    \label{hrho.eq}
   \end{equation}for which   we set $N_{nn}= 32\pm 3$.


%
\subsection{Initial conditions }
\label{subsec:icsetp}

To study the merging cluster El Gordo, we performed a series of
 N-body/hydrodynamical ISPH  simulations. 
The merging system consists  of a primary cluster with mass $M_1$ and
 a secondary with mass $M_2$, and the mass ratio is $q=M_1/M_2$.    

The cluster mass is defined as the mass $M_{200}$  within the  cluster 
radius $r_{200}$.  This is the   radius  $r_{\Delta}$   within which 
the total mass is
 \begin{equation}
M_{\Delta}=\frac{4 \pi}{3} \Delta  \rho_c(z) r_{\Delta}^3~,
 \label{mcl.eq}
 \end{equation}and the  mean density is $\Delta$ times the 
cosmological critical density $\rho_c(z)$.
The cluster radii  are calculated by setting $z=0.87$, the redshift of the 
El Gordo cluster.

The initial condition setup that we used for the two colliding clusters 
follows the same setting adopted by \citetalias{Zh15}. This was purposely chosen 
with the intent of validating our initial condition method by comparing results
extracted from  similar merging runs, and then comparing  different 
merger simulations against a  baseline merging model.

To construct the initial conditions for our merger simulations, we
first performed a particle realization of two individual spherical halos,  
initially in hydrostatic equilibrium. Each halo consists of 
multiple  mass components: DM, gas, and eventually star particles.

\subsubsection{ Dark matter   }
\label{subsec:icdm}
For the DM density profile  we assumed an NFW profile 
\begin{equation}
\begin{array}{llll}
\rho_{DM}(r)&=&\dfrac{\rho_s}{r/r_s(1+r/r_s)^2}\, ,  &   0\leq r\leq r_{200} \, , \\
 \end{array}
 \label{rhodmins.eq}
 \end{equation}
where $\rho_s$ and $r_s$ are the scaling parameters for density and radius, 
respectively. The scale radius $r_s$ is defined by $ r_s=r_{200}/c_{200}$,
where $c_{200}$ is the concentration parameter.
  We set  the value of $c_{200}$   using the $c-M$ relation of \citet{Du08}: 
 \begin{equation}
c_{200}= 5.71 a^{0.47} \left( \frac{M_{200}} {2\cdot10^{12} \msunh}
\right)^{0.084}~,
 \label{cfit.eq}
 \end{equation}
where $a=1/(1+z)$.  
The DM cluster density profile is then specified  once the mass $M_{200}$ and 
 the concentration parameter $c_{200}$ are given.

At radii larger than $r_{200}$ the cumulative mass profile diverges 
 as $ r \rightarrow \infty$. We therefore introduce 
for $ r > r_{200}$ an exponential cutoff up to a final radius 
$r_{max}=\nu r_{200}$:

\begin{equation}
\begin{array}{lll}
\rho_{DM}(r)&=&\rho_{DM}(r_{200}) (r/r_{200})^{\delta} 
\exp{-\displaystyle{\left(\frac{r-r_{200}}{r_{decay}}\right)}} \, , \\
             & &~~   r_{200} < r < r_{max}~, 
 \end{array}
 \label{rhodmext.eq}
 \end{equation}
where $r_{decay}=\eta r_{200}$ is the  truncation radius, and the parameter 
$\delta$ is set by requiring the first derivative of the DM density profile
to be continuous at $r=r_{200}$.
For the runs presented here,  we set  the truncation parameters to
 the values $(\nu, \eta) =(2,0.2)$; this choice   was motivated due to stability 
considerations  \citep[see][for a discussion about the optimum choice of these
parameters]{VS21}.

To construct a particle  realization of the DM density profile,
we first computed the  cumulative DM mass $ M_{DM}(<r)$ within the radius $r$.
We solved for the  particle radius, $r,$ by inverting 
$ q(r)=M_{DM}(<r)/  M_{DM}(<r_{rmax})=y$, where $y$ is a 
uniform random number in the interval $[0,1]$.

Particle speeds are determined according to \citet{Kaz04}.
For spherically symmetric systems the density $\rho_m(r)$ of a given 
 mass component  is given by 

 \begin{equation}
\rho_m(r)= 4 \pi \int_0^{\Psi(r)} f_m(\mathcal{E})
\sqrt{2 [\Psi(r)-\mathcal{E}]} ~ d \mathcal{E}~,
 \label{rhof.eq}
 \end{equation}
where  $\Psi(r)=-\Phi(r)$ is the relative gravitational 
potential of the system , $\mathcal{E}= \Psi- v^2/2$ is the relative energy
and $f_m(\mathcal{E})$  the corresponding  distribution function. Here 
the subscript $m=DM,s$  denotes the mass (DM or stars) component under 
consideration.


Through the Abell integral transform  \citep{Bin87} it is possible to invert 
Equation  (\ref{rhof.eq}) to obtain
 \begin{equation}
f_m(\mathcal{E})=\frac{1}{\sqrt{8} \pi^2}
  \left[
 \int_0^{\mathcal{E}} \frac{d^2 \rho_m}{d\Psi^2}
\frac {d \Psi}{\sqrt{(\mathcal{E}-\Psi)}}
\right]~,
 \label{fdist.eq}
 \end{equation}
  where   the second-order derivative $ {d^2 \rho_m}/{d\Psi^2} $ 
can be  evaluated analytically from $\rho_m(r)$ and $M(<r)$.

We constructed tables of $ f_m(\mathcal{E})$  by evaluating numerically 
 the integral (\ref{fdist.eq})  over a range of energies.
We then used an acceptance--rejection method
 to obtain the particle speed 
$v= \sqrt{2[\Psi(r)-\mathcal{E}]}$ for a particle at position $r$.  

Finally,  the directions of the particle position and velocity 
vectors  are assigned by randomly orienting   unit vectors.


\begin{table*}
\caption{  IDs and initial collision parameters of the  off-axis merger 
simulations of Section \ref{sec:opt}. $^{a}$}
\label{clparam.tab}%
\centering
\begin{tabular}{cccccccccc}
\hline
Model & $M^{(1)}_{200}$ $[\msun]$  & $r_{200}~[\mpc] $  & $\concI$ & $q$ & 
$V~ [\kms] $ & $P~[\kpc] $
  &  $(f_{g1},f_{g2})$ & {\rm gas profile} \\
\hline
B\_1 &   $ 2.5 \times 10^{15}$ & 2.02 &  2.41 & 3.6  & 2500  & 800 & (0.05,0.1) 
& Burkert \\
A\_1 &   $ 1.3 \times 10^{15}$ & 1.63 & 2.54 &  2.0  & 3000  & 300 & (0.1,0.1) & 
Burkert \\
Bf &   $ 1.6 \times 10^{15}$ & 1.74 &  2.5 & 2.32   & 2500  & 600 & (0.1,0.1)& 
$\beta$-model \\
Bg &   $ 1.6 \times 10^{15}$ & 1.74 &  2.5 & 2.32   & 2000  & 600 & (0.1,0.1)& 
$\beta$-model \\
Bh &   $ 1.6 \times 10^{15}$ & 1.74  & 2.5 & 2.32   & 1500  & 600 & (0.1,0.1)& 
$\beta$-model \\
Bk &   $ 1 \times 10^{15}$ &  1.5 & 2.6 & 1.54   & 1500  & 600 & (0.1,0.1)& 
$\beta$-model \\
Bl &   $ 1 \times 10^{15}$ &  1.5 & 2.6 & 1.54   & 2000  & 600 & (0.1,0.1)& 
$\beta$-model 
\end{tabular}
\begin{flushleft}
{\it Notes.} {$^a$ Columns from left to right:
 ID of the merging model,
halo mass $M^{(1)}_{200}$ of the primary , cluster radius $r_{200}$ at which 
$\Delta=200$, halo concentration parameter $\concI$ of the primary as given by  
Equation (\ref{cfit.eq}), 
primary-to-secondary mass ratio $q=M_1/M_2$, 
initial collision velocity, collision impact 
parameter, 
primary and secondary cluster gas mass fractions $f_g$
at  $r_{200}$, adopted model for the gas density profile of the 
primary. }.
\end{flushleft}
\end{table*}

%

\subsubsection{ Baryonic matter }
\label{subsec:icgas}

To construct  the cluster gas initial conditions, we assumed spherical symmetry 
and hydrostatic equilibrium. 
Following \citetalias{Zh15}, for the gas density we adopted the \cite{Bu95} 
density profile:
\begin{equation}
\begin{array}{llll}
\rho_{gas}(r)&=&\dfrac{\rho_0}{(1+r/r_c)\left[1+r/r_c\right]^2}\, ,  &   0\leq r\leq r_{200} \, , \\
 \end{array}
 \label{rhogins.eq}
 \end{equation}where $r_c$ is the gas core radius and $\rho_0$ the central gas density. 
 At radii larger than $r_{200}$ 
the density profile is continued  by assuming that it follows that of the 
DM   \citep{Zh14}:

\begin{equation}
\rho_{gas}(r)= \rho_{DM}(r)  \frac{\rho_{gas}(r_{200})} {\rho_{DM}(r_{200})} ,
 ~~  r_{200} < r < r_{max}~.
 \label{rhogext.eq}
 \end{equation}
The choice of the core radii  is observationally motivated by the 
X-ray morphology of El Gordo. 
In their paper, \citetalias{Zh15}  chose to set $r_c=r_s/2$ and $r_c=r_s/3$ 
for the primary and secondary cluster, respectively. 
These choices will be discussed later,  here 
for the initial gas density profile of the primary
we additionally considered  the commonly employed non-isothermal 
$\beta$-model
 \citep{Cav78}:
\begin{equation}
\rho_{gas}(r)= \rho_{0}  \left(1+\frac{r^2}{r_c^2}\right)^{-\frac{3}{2} \beta}~.
 \label{rhogbeta.eq}
 \end{equation}

For a given   gas core radius, $r_c,$ and slope parameter, $\beta$, 
to completely specify the gas density profile, it is necessary to
determine the central  density $\rho_0$.
This is done by assigning the initial gas mass fraction $f_g$ of the cluster 
at $r=r_{200}$ and then finding the corresponding  root value $\rho_0$.
The choice of the initial gas density profiles is a crucial issue  in shaping 
the final X-ray morphology  of the merging system.\ We provide a thorough discussion of
 the adopted profiles in Section \ref{subsec:mrgmodels}.

Finally, under the assumption of hydrostatic equilibrium, the gas temperature 
at radius $r$ can now be derived  \citep{Mas08,Donnert14}: 

 \begin{equation}
 T(r)  =   \frac{ \mu m_p }{ k_B  } \frac{ G  }{\rho_{gas}(r)}
\int_r^{\infty}  \frac{ \rho_{gas}(u) } { u^2} M_{tot}(<u) du ~, 
\label{thydrosa.eq}
 \end{equation}where $M_{tot}(<r)$ is the total  mass within the radius $r$ and
we assume $\mu=0.59~$.  The gas thermal energy in the SPH equations is 
obtained by setting $\gamma=5/3$ for the ideal equation of state.

The initial gas particle positions were generated starting 
from a  uniform glass distribution of points. The desired 
density profile  was then obtained by applying a  radial transformation to the particle coordinates
  such that  the corresponding mass profile 
is now consistent with the initial gas mass profile.  The initial gas 
velocities were set to zero.

\subsubsection{Stellar component}
\label{subsec:icstar}

For some of our merging runs, we also considered halos that contain
a  star matter component, in addition to the  DM and gas. 
This is supposed to mimic the BCG mass distribution,  which we initially 
placed at the center of the halo.

For the density profile of the stellar component,  we used  \citep{Mer06,ZuH19}

\begin{equation}
\begin{array}{lll}
\rho_{\star}(r)&=&\rho_{e}
\exp{\bigl\{ -d_{n_s} 
 \displaystyle{\bigl[(r/r_{e})^{1/n_s}-1\bigr]} \bigr\} 
} \, , \\
  d_{n_s} & \sim & 3 n_s -1/3 +0.0079/n_s, ~~ {\it for} ~n_s \simgt 0.5 , 
 \end{array}
 \label{rhostar.eq}
 \end{equation}where $r_e$ is the effective radius and  $\rho_e$ the stellar density at 
that radius, we set here $n_s=6$. 
The BCG mass is derived according to the relation of 
\citet[][ their Table 2]{Kr18}: 

 \begin{equation}
 \log_{10} M_{\star,BCG}   \sim    0.38 [ \log_{10} (M_{500}) -14.5].  ~ 
\label{mstar.eq}
 \end{equation}

 For $M_{500} \sim 10^{15} \msun$, which is appropriate for the 
range of halo masses considered here,  the relation
(\ref{mstar.eq}) gives $M_{\star,BCG}\sim 2.3 \cdot 10^{12}\msun$.
 We can now obtain the  density $\rho_e$  by solving for the ratio
$M_{\star,BCG}/M_{500}$ within $r_{500}$,  once $r_e$ is given.

If one adopts the relation of \citet[][ Equation 33]{SH03}, 
then for  $M_{\star,BCG}\sim 3 \cdot 10^{12}\msun$ one obtains 
$r_e \simeq 30 $ kpc.  The smaller is the value of the effective radius, the
higher must be the central density for a given $M_{\star,BCG}$. However,
the mass resolution of the simulations  sets a lower limit for
the minimum spatial scale  that can be resolved, 
so the value of $r_e$ should be  considered simulation dependent.
For example in their simulations \citet{ZuH19} set $r_e=175$ kpc, 
as a compromise we set here  $r_e=60$ kpc.  This value is 
 a bit higher than the adopted gravitational softening  length for the 
DM particles (see below).  For a $\sim 10^{15} \msun$ 
halo mass and  $M_{\star,BCG}\sim 3 \cdot 10^{12}\msun$, 
we obtain  
$\rho_{\star}(r\sim 10^{-2} r_{200}) 
\simeq 10^4 \rho_c\simeq 3.7 \cdot 10^6 \msun kpc^{-3}$ for the central value of the stellar mass density.
For a given initial star density distribution, we then   determined
positions and velocities of the star particles   according to the procedure 
described in Section \ref{subsec:icdm}.

Finally, we assigned the mass of DM and gas particles as in the merger 
simulations of \citet{VS21}:

 \begin{equation}
m_d\simeq8 \times 10^8 \msun (M^{(2)}_{200}/2 \times 10^{14} \msun)~
  \label{mdark.eq}
 \end{equation}and  $m_g=f_b m_d/(1-f_b))\simeq 0.16 m_d $, respectively. 
Here $M^{(2)}_{200}$ is the mass of the secondary and 
  $f_b\simeq0.16$  the cosmological  mass gas fraction.
With this mass resolution we have 
  $N_{DM} \simeq 3.4 \times10^5 $   DM particles   
for an halo mass of $M_{200}\sim  10^{15}\msun$, 
and  $N_g \simeq 1.7 \times10^5 $  gas particles  for 
an halo gas mass of $M_{g}\sim  10^{14}\msun$.    

\begin{table*}
\centering
\caption{IDs of the merger simulations  with initial merging parameters 
as given by the corresponding merger models of Table \ref{clparam.tab}.
For each merger model the additional subscripts refer to  simulations  
with  an initially different gas core radius $r_c$ for the primary. 
For each simulation we report the value of $r_c$ in kpc and that of the ratio 
$\zeta=r_s/r_c$, where $r_s$ is  the NFW scale radius of the primary.
The units of the collision parameters  $ \{M_1, r_s, ~q,~P, ~V\}$ 
for each model   are the same as for  Table \ref{clparam.tab}, 
 the NFW radius $r_s$ is given in Mpc.
}

\scalebox{0.8}{
\begin{tabular}{lrccccc}
\hline \hline
        Model: $ \{M_1, r_s, ~q,~P, ~V\}$ 
        & \multicolumn{6}{c}{merger simulation ID}  \\
        \cline{2-7}
  &  &  &  &   &     \\
\hline
Bf: $ \{ 1.6 \cdot 10^{15}, 0.696,~2.32,~600,~2,500 \}$ 
    & Bf$\_$rc29  & Bf$\_$rc26   & Bf$\_$rc22  & Bf$\_$rc20   &
  Bf$\_$rc17 & Bf$\_$rc14 \\
$r_c$ ($\zeta=r_s/r_c$)  & 290 (2.4)   &  260 (2.67)   &  217 (3.2)  &
        200 (3.44)  & 174 (4.) & 145 (4.81)  \\ \\
\hline
Bg: $ \{ 1.6 \cdot 10^{15}, 0.696,~2.32,~600,~2,000 \}$ 
    & Bg$\_$rc29  & Bg$\_$rc23   & Bg$\_$rc20  & Bg$\_$rc17  &  Bg$\_$rc14   \\
$r_c$ ($\zeta=r_s/r_c$)  &  
        290 (2.4) & 232 (3)  &  200 (3.437)  &  174 (4)  & 145 (4.81)  &    \\ \\
\hline
Bh: $ \{ 1.6 \cdot 10^{15}, 0.696,~2.32,~600,~1,500 \}$ 
        & Bh$\_$rc29  & Bh$\_$rc22   & Bh$\_$rc20  & Bh$\_$rc17  & 
 Bh$\_$rc16 & Bh$\_$rc14  \\
$r_c$ ($\zeta=r_s/r_c$)  &  
        290 (2.4)   &  217(3.2)   & 200(3.44)  & 174(4.)  
        & 160(4.37)  & 145(4.81)  \\ \\
\hline
Bk: $ \{ 1. \cdot 10^{15}, 0.574,~1.54,~600,~1,500 \}$ 
        & Bk$\_$rc19  & Bk$\_$rc18   & Bk$\_$rc17  & Bk$\_$rc15   & 
 Bk$\_$rc14 & Bk$\_$rc12  \\
$r_c$ ($\zeta=r_s/r_c$)  &  
        191 (3.)   &  179 (3.206)   & 167 (3.43)   &  155 (3.7)  &
        143 (4)  &  119 (4.81)   \\ \\
\hline
Bl: $ \{ 1. \cdot 10^{15}, 0.574,~1.54,~600,~2,000 \}$ 
        & Bl$\_$rc24  & Bl$\_$rc21   & Bl$\_$rc19  & Bl$\_$rc18   & 
 Bl$\_$rc17 & Bl$\_$rc14  \\
$r_c$ ($\zeta=r_s/r_c$)  &  
        240 (2.39)    & 215 (2.67)   &  191 (3.0)   & 179 (3.207) &
        167 (3.43)  &  143 (4)   \\ \\
\hline
\hline
\end{tabular}
}
\label{subcl.tab}%
\end{table*}

Previous tests \citep{V19,VS21} showed that with these
 mass assignments the simulation numerical resolution  is 
adequate to properly describe  the development of hydrodynamic instabilities
during the merging phase.

The gravitational softening parameters of  the particles 
are set  according to the scaling 
 \citep{VS21}

 \begin{equation}
\varepsilon_i =15.8\cdot ( m_i/6.2\times10^8 \msun)^{1/3}\kpc.
  \label{epsd.eq}
 \end{equation}

For merger simulations with a BCG, we decided to 
increase the number of star particles $N_{s}$ by a factor of $4$. 
To obtain the mass of star particles, we then applied 
Equation (\ref{mdark.eq}), but with a coefficient scaled down by a factor of $4$.
This is to prevent any discreteness  effects 
on the star particle distribution during the simulations. With this choice of parameters 
the lower limit $\varepsilon_{\star} \sim r_{200}/\sqrt{N_{s}}$ \citep{Pow03} 
to the gravitational softening length of the star particles 
is then consistent with the values given by Equation (\ref{epsd.eq}).
We note, however, that this lower limit is derived by requiring discreteness
effects to be negligible over a Hubble time, whereas here the timescale 
of the collisions is $ \simlt 1$ Gyr.

Particle positions and velocities are integrated using a standard hierarchical 
block timestep scheme \citep{He89}. 
According to several accuracy and stability 
requirements \citep{Spr05}, 
 each particle, $i,$ has its own optimal timestep,  
$\Delta t_i$, which is chosen from a power-of-two hierarchy 
$ \Delta t_m= \Delta t _0/2^m$, where 
 $m=0,1 \ldots$ and $\Delta t_0=1/100$ Gyr is the largest allowed timestep.
During the simulations, we recorded particle properties at simulation 
times $t_n=n \Delta t_0$, where $n=0,1,\ldots$ is the simulation  step.

\subsubsection{Initial merger kinematics}
\label{subsec:ickin}

To construct the orbits of our merger simulations, 
 we chose a Cartesian system of 
coordinates $ \{x,y,z\}$, with the mergers taking place in the $ \{x,y\}$ plane
and the center of mass of the two clusters being centered at the origin.

As previously indicated, the initial separation and relative velocity vectors 
$ \{{\bf X}^{in},{\bf V}^{in}\}$  of the two clusters are set as in 
  \citetalias{Zh15}. For a given mass ratio $q=M_1/M_2$   
the initial separations are then 
${\bf X}_1= \{d_{ini}/(1+q), P/(1+q),0 \}$ and
${\bf X}_2= \{-d_{ini} q /(1+q), -P q/(1+q),0 \}$ for the primary and 
secondary cluster, respectively. 
Here $P$ is the impact parameter of the collision 
and  $d_{ini}=2(r^1_{200}+r^2_{200})$. 
Similarly, for the initial velocities we set 
${\bf V}_1= \{-V/(1+q), 0, 0 \}$ for the primary and 
${\bf V}_2= \{V q/(1+q), 0, 0 \}$ for the secondary, 
where $V$ is the initial collision velocity.

We then  shifted the  particle positions and velocities  of the two 
halos according to  the vectors $ \{{\bf X}^{in},{\bf V}^{in}\}$; 
the merger dynamical evolution was fully determined by the merging 
parameters $ \{M_1, ~q,~ P, ~V\}$. In terms of these parameters the energy 
of the system can be expressed as

 \begin{equation}
 E =\frac{1}{2} M_1 V^2 \left[ \frac{1}{1+q}- \frac{G M_1 }{d_{ini}}
\frac{1}{V^2} \frac{2}{q} \right]~,
    \label{ecoll.eq}
 \end{equation}where for the merging system we have assumed the absence of bulk motion  and
that the initial distance $d_{ini}$ between the two clusters is greater than
the sum of their radii $r_{200}$.

\subsection{Numerical implementation of self-interacting dark matter}
\label{subsec:icsidm}

The possibility that DM has collisional properties has been investigated by 
many authors and is observationally motivated by the well-known challenges 
faced by the standard CDM model over a wide range of  scales.  We refer to 
\citet{Tulin18} for a recent review on the subject.
On cluster scales,  SIDM\ offers a natural explanation 
for the apparent spatial offsets seen in merging clusters between the galactic 
and the DM component, though it must be emphasized that this effect is not
present in every merging cluster.

Several approaches have been proposed to implement DM self-interactions 
in N-body simulations.  A  DM simulation particle is a ``super'' 
 particle that represents  a patch of DM phase-space that contains a 
large number of physical DM particles \citep{Hoc88}.
The DM self-interactions are then modeled according to a standard Monte Carlo
method, in which a scattering between two DM simulation particles   takes place  
whenever  a local scattering probability is smaller than a~$[0,1]$ random 
number.

In the following, we restricted  ourselves to the  simplest  case 
of isotropic and elastic scattering between DM particles; moreover,
we  further assumed a constant, velocity-independent DM cross-section
 $\sigma_{DM}$. 
Although an SIDM model with a constant DM cross-section fails to satisfy 
a number of constraints over a wide range of mass scales \citep{Kap16}, it can nonetheless be considered a useful first-order approximation 
as long as  SIDM simulations are restricted to the range of cluster scales
 \citep{ZuH19}.

To determine the local scattering probability, different methods have been
proposed. The approach that we followed here was introduced 
by \citet{Vog12},  
and its implementation requires the definition of a local DM density.

For each DM particle, $i,$ we defined a DM density $\rho_{DM}({\bf r_i}) $  
according to  the SPH formulation, and, as in Equation (\ref{rho.eq}),  we
thus defined a DM smoothing length $h_i^{DM}$  such that 

 \begin{equation}
 \rho_{DM}({\bf r_i}) =\sum_j m^{DM}_j W(|{\bf r}_{ij}|,h^{DM}_i)~,
    \label{rhodm.eq}
 \end{equation}
where $m^{DM}_i$ denotes the mass of the DM simulation particle $i$ and 
   the summation is over  $N_{nn}=32 \pm 3$ DM neighboring particles.
The  DM smoothing length $h_i^{DM}$  is evaluated whenever the particle
$i$ is active and its timestep is synchronized with the current simulation time.
For a DM particle  that is not active (passive), its smoothing length  is 
extrapolated to the current time from its last evaluation.
 
 For each DM particle $i$ the local scattering probability with a neighboring 
DM particle $j$, within the simulation timestep  $\Delta t_i$ 
is then determined as  \citep{Vog12}

\begin{equation}
P_{ij}= m^{DM}_i W(r_{ij}, h^{DM}_i) \frac{\sigma_{DM}}{m_X} v_{ij} \Delta t_i ~,
 \label{pijdm.eq}
 \end{equation}where $v_{ij}=|\bf {v_i}-\bf {v_j}| $ is the relative velocity between 
particles $i$ and $j$, ${m_X}$ is the  physical mass of the DM particle
 and $\Delta t_i$ the timestep of particle $i$. 


 The total probability of particle $i$    scattering 
  is $P_i= \sum_j P_{ij}/2$,
where the factor 2 accounts for the other member of the scattering  pair.
  A collision  of particle $i$ with one of its neighbors  will
occur if $P_i \leq x$, where x is a uniform random number in the range $[0-1]$.
 According to \citet{Vog12} the scattering neighbor is selected by sorting 
 the set of neighboring particles according to their distance $r_{ij}$ from 
particle $i$. The scattering neighbor is then  the first particle $k$ that 
satisfies $ x \leq \sum_j^k P_{ij}$. 

When  the collision condition is satisfied,  in the case  of 
isotropic scattering the post-scattering velocities of the  tagged pair 
are

\begin{equation}
\left\{
\begin{array}{lll}
	{\bf u_i} &=&{\bf V}  + (v_{ij}/2) {\bf e}   \\
	{\bf u_j} &=&{\bf V}  - (v_{ij}/2) {\bf e}  ~,
 \end{array}
\right .
 \label{vdmscatt.eq}
 \end{equation}
 where ${\bf V}= ({\bf v_i}+{\bf v_j})/2$ is the center-of-mass velocity,
and $\bf e $ is a unit vector randomly oriented. Note that in the simulations 
we set all the DM particles as having  the same mass.

The  scattering procedure described here  is generically valid for a serial 
code; however, its implementation  in a parallel code is not so 
straightforward to arrange;  it is described in Appendix ~\ref{appSIDM}.
We also tested the  scattering  implementation  by  showing 
results from an  isolated SIDM halo.
\begin{table*}
\caption{Initial merging parameters of  a set of additional  off-axis 
merger simulations with lower impact parameters.
The meaning of the parameters is the same as in Tables  \ref{clparam.tab} 
and \ref{subcl.tab}.} 
\label{clbj.tab}%
\centering
\begin{tabular}{ccccccc}
\hline
Model & $M^{(1)}_{200}$ $[\msun]$  & $q$  & $V [\kms] $ & $P [\kpc] $
        &  $r_c$[kpc] ($\zeta$)    \\
\hline
Bja$\_$rc14 &  $ 1 \times 10^{15}$ &  1.54   & 1500  & 400 & 143 (4)  \\
Bjb$\_$rc14&  $ 1 \times 10^{15}$ &  1.54   & 2000  & 400 & 143 (4)   \\
Bjc$\_$rc12 &  $ 1 \times 10^{15}$ &  1.54   & 1500  & 500 & 119 (4.81)   \\
\end{tabular}
\end{table*}

 Finally, in all of the merger simulations in which DM is self-interacting
the DM halos are  assumed to follow initially the same truncated NFW profiles of 
the $\sigma_{DM}=0$ simulations. We justified this assumption by assuming the impact of initially cored DM density
profiles to be
subdominant  in the evolution of post-pericenter gas structures \citep{ZuH19}.

\subsection{Observables and projection analyses}
\label{subsec:imag}

For each of our merger simulations, we saved particle properties at various 
output times. This was done to generate, for a given viewing direction 
and at the chosen epoch,  images of the surface mass density, 
X-ray surface brightness,  and SZ amplitude. To validate our merger models,
these maps could then be compared against analogous images presented in previous 
works. In particular,  to consistently compare the mock maps extracted from our
simulations with those shown by \citetalias{Zh15}, we adopted for the observer 
the same coordinate system $\{\hat x,\hat y,\hat z\}$
introduced by the authors to construct their projected maps.

In such a system the line of sight is along  the  $\hat z$  direction and 
the $\{\hat x,\hat y\}$ coordinates refer to the cluster merger  as seen in 
the  plane of the sky.
The observer frame is obtained from the simulation frame $\{ x, y, z\}$ 
by applying two rotation matrices. The first performs a rotation 
 around the $z$-axis by an angle  $\alpha$, the second rotates 
 the tilted frame  by an angle $i$ around the $x^{\prime}$-axis. 
The angle $i$ between the axes $z$ and $\hat z$ is thus the angle between 
 the merging and sky plane. Our viewing direction is thus defined by the angles
 $\{\alpha ,i \}$, we  refer to \citetalias{Zh15} for more details on the 
setup of the rotation angles.

For a given output time and viewing direction, we integrated along 
the line of sight 
to extract 2D maps  of observational quantities from the simulation particles.
Specifically, we defined the surface mass density  as

 \begin{equation}
 \Sigma_m(x,y)= \int_{los}  
\left[\rho_{gas}({\bf x})+\rho_{DM}({\bf x})\right] d z~,
\label{smass.eq}
 \end{equation}where $\rho_{gas}({\bf x})$ and $\rho_{DM}({\bf x})$ 
are defined at the position $\bf x$ according to Equations (\ref{rho.eq}) 
and (\ref{rhodm.eq}), respectively.

To properly compare the observed {\it Chandra} X-ray images \citep{Men12} 
with the mock X-ray maps extracted from their simulations,
 \citetalias{Zh15} applied the MARX software package
\footnote{http://space.mit.edu/ASC/MARX} to their input X-ray maps.
Here we followed a simpler approach and, as in \citet{Molnar15}, 
we evaluated  the X-ray surface brightness by integrating 
the X-ray emissivity   $\varepsilon(\rho_g,T_{g},Z, \nu )$    
along the line of sight and over the energy: 

 \begin{equation}
 \begin{split}
         \Sigma_{X}(x,y)  =  &  \frac{1}{4 \pi (1+z)^4}  \int_{los} \,dz  \\
  &        \int \, \varepsilon(\rho_g,T_{g},Z, \nu ) A_{eff}(\nu) \, d \nu  ~,
\end{split}
\label{sbrx.eq}
 \end{equation}where  $T_g$ is the gas temperature,  $\nu$ the frequency, $Z$ the 
metal abundance of the gas, and $A_{eff}(\nu)$ the effective area of the telescope. 
For this, we used the four-chip-averaged ACIS-I  effective area \citep{Zh04}. 

However, unlike in \citet{Molnar15},  here the X-ray 
emissivity is not calculated analytically but using  cooling 
tables taken from the MAPPINGS III library \citep{Allen08}.  As in 
 \citetalias{Zh15},  we set the gas metallicity to $Z=0.3 Z_{\odot}$; for reference, 
 $Z_{\odot}\sim0.02$ \citep{And89} is the solar value.

Finally, to consistently  compare  our maps with observations and previous 
findings (\citetalias{Zh15}),  we set  the exposure time to $t_{exp}=60ks$
for all the simulated maps.
  Our X-ray maps (\ref{sbrx.eq}) will  then be 
expressed in counts arc sec$^{-2}$ and evaluated in the energy range 
$[0.5-2]$ keV, without taking into account
the impact of the cosmic X-ray background.  

We defined X-ray temperatures according to the adopted weighting scheme:

\begin{equation}
T_{W}= \frac {\int T({\bf x} ) {\it W} d^3x } { \int {\it W} d^3x}~,
\label{tr.eq}
\end{equation}where $W$ is the  weight function. 
  We considered for $\it W({\bf x}) $  the commonly employed 
   gas density function  ${\it W}=\rho_\mathrm{g}$  
(mass-weighted temperatures, $T_\mathrm{mw}$) and 
the weight function ${\it W}=\rho^2_\mathrm{g} T^{-3/4}$,
 which defines  the corresponding spectroscopic-like temperature 
$T_\mathrm{X}$.  This choice of the weight function 
has been shown \citep{mazzotta04}  to provide an accurate approximation of 
spectroscopic temperatures extracted  from X-ray observations. 
As in previous integrals, projected X-ray temperature maps are obtained by 
integrating weighted temperatures along the line of sight.

 The  SZ surface brightness is calculated at frequency $\nu$ and including 
relativistic corrections \citep{Itoh98}:

 \begin{equation}
 \begin{split}
 \Sigma_{SZ}(x,y)  = &  \frac{\sigma_T k_B}{m_e c^2} \int_{los}  n_e T_{g}  \\
         &       \left[ g(\nu)+ \Sigma_{k=1}^{k=4} Y_k \Theta^k \right]  dz~,
 \end{split}
\label{sbsz.eq}
 \end{equation}
where  $m_e$, $k_B$, $\sigma_T$, $c$  and $n_e$ are the electron mass, the 
Boltzmann constant, the  Thomson cross section, the speed of light  and the 
electron number density, respectively. The function 
 $g(\nu)= \coth(x_{\nu}/2)-4$   is the nonrelativistic frequency function, 
where $x_{\nu}= h_P \nu /(k_B T_{cmb})$ and $T_{cmb}$ is the temperature of the
cosmic microwave background.  The second term in the square brackets is a 
sum of relativistic corrections, with the coefficients $Y_n$ given 
in \citet{Itoh98} and $\Theta \equiv k_B T_g /m_e c^2$.  
 As in \citetalias{Zh15} we set $\nu=150 $ GHz and smooth the SZ maps with a 
Gaussian kernel with width $\sigma_{SZ}=270$ kpc 
{($\sim0.55^{\prime}$ at $z=0.87$).}

We evaluated maps of projected quantities  on a
 2D mesh of $N_g^2=512^2$ grid points with coordinates $\{ x_g, y_g\}$.
Because of the Lagrangian nature of SPH, the value of a function at the
grid point $ {\bf x}_g$ is the sum 

 \begin{equation}
 f({\bf x}_g) =\sum_i f_i \frac{m_i}{\rho_i}  W(|{\bf x}_g- {\bf x}_i|,h_i)~
    \label{funcsph.eq}
 \end{equation}over the subset of SPH particles that satisfy 
 $|{\bf x}_g- {\bf x}_i|\leq 2h_i$.  To perform the integrals 
along the line of sight,  we exploited the property of the $M_n$ splines
  \citep{Price2012},   which  approximate 
a  Gaussian in the limit $n\rightarrow \infty$.
 For the $M_4$  kernel used here,  the integrals along the line-of-sight  
$z$ were then performed  analytically,  and we calculated projected 
 quantities at the grid point ${\bf x}_g$ as

 \begin{equation}
  f({\bf x}_g) =\sum_i f_i \frac{m_i}{\rho_i}  W_{2D}(x^2_{gi},\sigma^G_i)~,
    \label{func.eq}
 \end{equation}

where  $x^2_{gi}=(x_g-x_i)^2+(y_g-y_i)^2$, $\sigma_i^G \simeq  h_i/\sqrt{3} $
\citep{Deh12} and $W_{2D}$ is the 2D Gaussian, 

 \begin{equation}
W_{2D}=\frac{1} {(\sqrt{2 \pi} \sigma^G_i)^2} exp[-x^2_{gi}/2 \sigma_i^G]^2.
    \label{funcG.eq}
 \end{equation}

To locate the centroid positions of the various maps that we produced to
study the El Gordo cluster, we applied the 
shrinking circle method to the simulation particles. This a 2D version of the shrinking sphere method, 
 which is commonly employed in simulations to locate DM density peaks 
\citep{Pow03}  or X-ray temperature peaks \citep{V06}.

We describe the method in three dimensions, but the generalization to projected
 quantities is straightforward. We first introduce the concept of 
weighted mean position  for a subset of simulation particles.
This subset can consist of gas, DM, or star particles that were
initially members of one of the two colliding clusters. We define a
weighted mean position  by calculating for the particle subset 
 the averaged center ${\bf x_c}=\sum_i w_i {\bf x}_i/\sum_i w_i$ , where 
the weights $w_i$  are chosen according to the  centroid under consideration.  

To locate  the centroid position, we then started from an initially 
large radius $R^{(0)}(\sim 5 \mpc)$ that contains all of the particle subset,
 and we calculated the weighted mean position,  $\vec x_c^{(0)}$. 
We then proceeded to iteratively calculate at each step $k=1,2,..,M$ the 
centroid position $\vec x_c^{(k)}$, based on the  subset particles 
that are within a sphere of radius $R^{(k)}=f R^{(k-1)}<R^{(k-1)}$, 
located at $\vec x_c^{(k-1)}$.
The last iteration $M$ is such that there must be at least a number $N_M$ of 
 particles within $R^{(M)}$.  The cluster center is then defined as 
$\vec x_c^{(M)}$.
We found that the  results are robust for $N_M \sim 100$, but the choice of the 
shrinking  factor, $f$, is critical: it must be chosen sufficiently large 
($f \simgt 0.8$) to prevent  the algorithm from failing when there are multiple
peaks.

The method is easily applied in 2D to the projected particle positions along
the viewing direction.\ Therefore, for a given mock image, we calculated the
position of the mass centroids, as well as the location of the X-ray
emission and SZ peaks.


\subsection{Merger models}
\label{subsec:mrgmodels}

As previously explained in the Introduction, the aim of this paper is to determine
whether it is possible to construct merger models for the El Gordo cluster
that can consistently reproduce the observed 
X-ray morphology, as well as many of its physical properties.
However, for these models a fundamental constraint  must be that
 of having cluster masses in line with recent WL studies, and
not as high as those employed in previous simulations
   \citep{Donnert14,Molnar15,Zh15}.

Our  merger models will be validated by contrasting  
our simulation results against the corresponding ones presented 
  by \citetalias{Zh15} for their fiducial model B.
As already outlined, this is the model that is  able 
to match most of the observed  El Gordo physical properties.
For this reason we adopt here the same initial merging configuration, 
and extract X-ray maps from our  merger simulations when the  projected distance
 between the mass centroids is $d_{DM}\sim 700 \kpc$ after the 
pericenter passage. In all of the considered models, 
the orientation angles of the merging plane  are taken to be  the same as in 
   model B of \citetalias{Zh15}: $\{\alpha, i\}=\{-90^{\circ},30^{\circ}\}$.

To significantly simplify the collision parameter space of our merger models  
we assume for the secondary the same Burkert gas density profile 
 (\ref{rhogins.eq}) as in model B of \citetalias{Zh15},
with the gas core radius  set to $r_c=r_s/3$.  We justify this assumption by 
noting from Figure 14 of \citet{Men12}
that the  map of the deprojected electron density  
exhibits a peak value of $n_e \simeq 5 \cdot 10^{-2} \numd$
for the SE component.  This value is very close 
to the core value found for the electron number density  
of the secondary in model B, at the post-pericenter epoch, that fits the observed projected distances between the mass centroids.
We use then this finding to argue 
that the gas structure of the SE  component
is adequately described here by  adopting the initial gas density profile of 
the secondary.

We then constructed our ensemble of off-axis mergers
by considering different values 
of the primary mass $M^{(1)}_{200}$  and modifying  the mass ratio 
 $q=M_1/M_2$ such that the mass of the secondary stays 
close to the estimated value $M^{(2)}_{200}\sim 6.5 \cdot 10^{14} \msun $.
From Equation (\ref{cfit.eq}) one obtains $\concII=2.682$ for the halo concentration 
parameter of the secondary.
In accordance with recent mass estimates \citep{Die20,Kim21},
we examined mergers with the following masses for the primary:
 $M^{(1)}_{200}  = 1.6 \cdot 10^{15} \msun $ and
 $M^{(1)}_{200}  = 10^{15} \msun $.

For a chosen value of the primary mass we consider mergers whose 
initial conditions  differ in the choice of the 
initial collision velocity $V$ and impact parameter $P$.
Table \ref{clparam.tab} lists the different off-axis merger models 
that we analyze, together with the corresponding  merging parameters
 $\{ M^{(1)}_{200}, ~q,~P, ~V \}$;   models A\_1 and B\_1  are the fiducial 
models A and B of \citetalias{Zh15}.  We will also be considering 
head-on mergers (models HO), but  we defer until  Section
\ref{sec:return} our description of the initial conditions adopted for those
models.

The adopted strategy for setting up  the initial conditions of our merging runs 
is based on the following result,  found by \citetalias{Zh15}.   When the authors
considered the possibility of a merger  with initial conditions similar 
to those of model B, but with a smaller primary mass 
 ($M^{(1)}_{200}  = 1.6 \cdot 10^{15} \msun $, panel f in their Figure 5), 
they found a total X-ray  luminosity much smaller  than that given by model B. 
Then they  use this finding to conclude that a massive merger
 ($M^{(1)}_{200}  \simeq 2.5 \cdot 10^{15} \msun $)  is required in order 
to obtain an X-ray luminosity as high as that observed in the El Gordo cluster.

We argue that a viable solution for obtaining  consistency between the 
recently measured cluster masses and observed luminosities is to consider 
   the possibility   of an initial gas density profile for the primary 
that is different  from the Burkert one (\ref{rhogins.eq}) previously adopted.
Specifically, we employed the well-known 
  $\beta$-model (\ref{rhogbeta.eq}) to describe the 
initial radial gas density profile of the primary. 
For a given  central  density, $\rho_0$,  the radial behavior of the gas 
density can be  modified by varying the   gas scale radius, $r_c,$ and 
the slope parameter, $\beta$, with the aim of constructing 
 the initial conditions for a merging system that
can satisfy the observational constraints previously described.

We further simplified the parameter space of our merger models by setting 
 $\beta=2/3$, as  already adopted   in merger  simulations for the 
Bullet Cluster \citep{Mas08} and, in particular, by \citet{Donnert14} 
in his simulation study of the El Gordo cluster.
For a given set of  merging parameters $\{ M^{(1)}_{200},~\xi,~P,~V \}$, this 
choice leaves us with the gas core radius, $r_c,$  of the primary 
as the only parameter needing to be set up for the initial conditions of our
merger models.

We then constructed our ensemble of merger simulations by creating, for each 
of the off-axis merger models  listed in Table \ref{clparam.tab},  a subsample
of  merging clusters with different initial conditions.  For a specified set
of collision  parameters $\{ M^{(1)}_{200},~q,~V,~P \},$ 
we set up the initial conditions of our models by considering 
  a range of values for the core radius, $r_c$.
Table  \ref{subcl.tab} lists the merger models that we considered; the ID of each 
model is given by the ID of the corresponding  model in 
Table \ref{clparam.tab}, with a subscript referring to the chosen value of 
$r_c$. This is given in terms of the dimensionless parameter 
$\zeta=r_s/r_c$. Hereafter, the initial gas core radius of the primary will be 
denoted as $\rcp$.

Finally, we also considered an additional list of merger models. The 
collision parameters of these models are given in Table  \ref{clbj.tab}.
\begin{figure*}[!ht]
\centering
\includegraphics[width=0.95\textwidth]{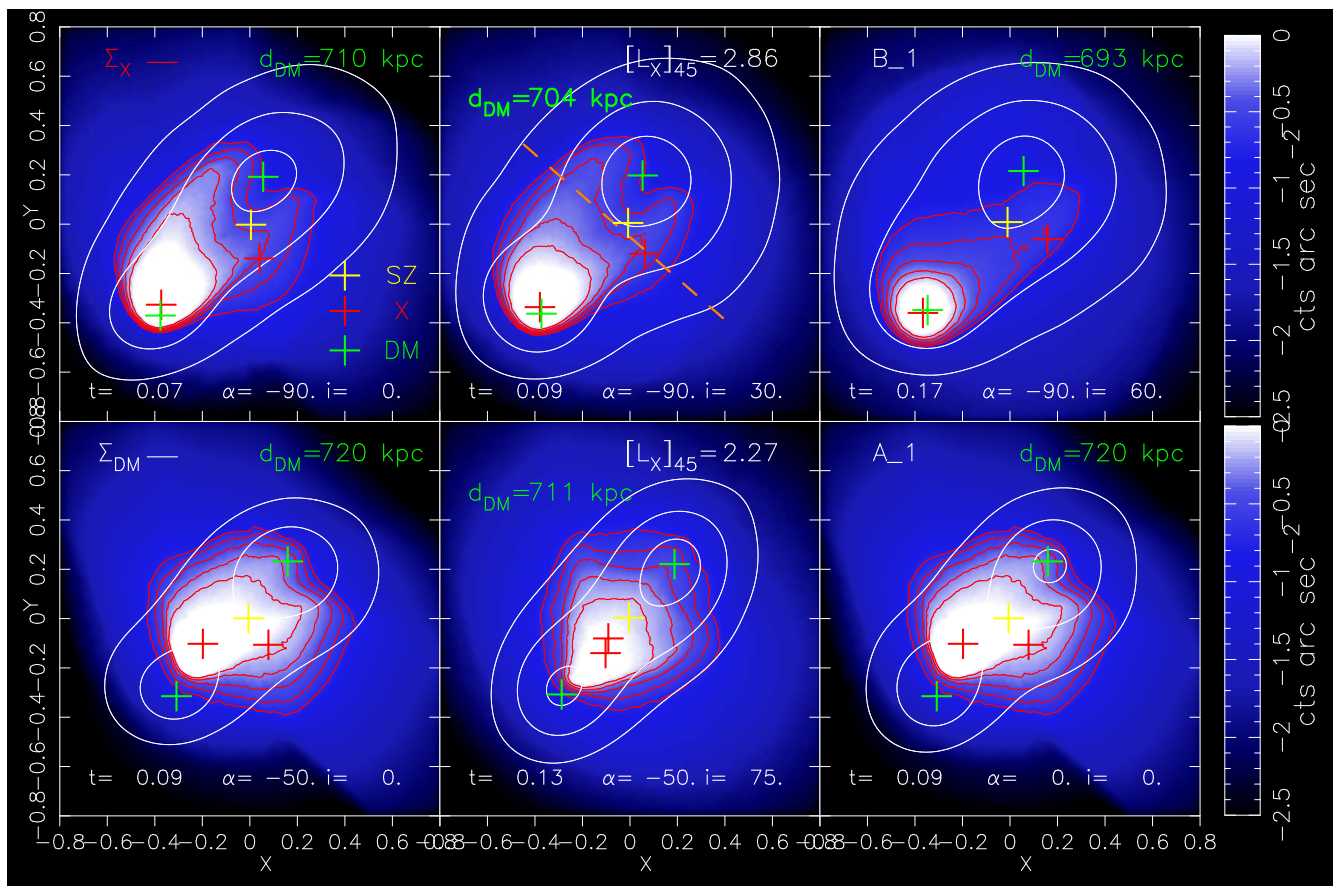}
\caption{  X-ray surface brightness images extracted from simulations 
of models B\_1 and A\_1, for different viewing directions. The box size 
is $1.6  \mpc$ and time is in Gyr, with $t=0$ at the pericenter passage.  
In each panel 
the epoch, $t,$ is when the projected distance, $d_{DM}$, between the 
mass centroids is approximately $d_{DM}\sim 700 \kpc$; the orientation
angles $\{\alpha, i\}$ of the merging plane are defined in 
Section \ref{subsec:imag}. 
In each panel the log-spaced contour levels of the 
projected X-ray surface brightness (red) and mass density (white) are overlaid.
From the inside to outside, the contour levels of the X-ray surface brightness  
and of the surface mass density are: $(6.6,4.4,2.9,1.9,1.2)\cdot 10^{-1} \ctsn $ 
and $(5.6, 3.1,1.8) \cdot 10^{-1} \dms$. 
The crosses indicate the projected spatial locations of the mass 
(green) and X-ray surface brightness (red) centroids; the yellow 
cross shows the position of the SZ centroid.
The maps can be directly compared with Figures 1 
and 3 of \citetalias{Zh15}, the orientation along the viewing direction  of 
the merging plane in the sky being the same. In particular,  
the mid panels of models A\_1 and B\_1 correspond to the fiducial models 
A and B of \citetalias{Zh15} (see their Table 2). For these models,
 the X-ray luminosity in the $0.5-2$ keV band  is given in units of 
$10^{45} \ergs$. For the  fiducial model B\_1  (top mid panel), 
the dashed orange line shows the  location and orientation of
the spatial region used to extract the X-ray surface brightness  profile.
This model is chosen as the baseline model of the simulations.
\label{fig:planeA1}
}
\end{figure*}

\section{Results}
\label{sec:results}

This section is dedicated to the presentation of our main results, which
we divide into three subsections. In Section \ref{sec:opt} we present 
results obtained by performing off-axis merging simulations of the 
El Gordo cluster, but with a mass of the primary in line with recent
 lensing  estimates and lower than that considered in previous works.
 In Section \ref{sec:return} we investigate the possibility of a returning 
scenario to explain the observed X-ray morphology and
spatial offsets between different centroids. 
Finally, in Section \ref{sec:sidm} some of the merging runs of Section
\ref{sec:opt} are revisited by performing the simulations 
with a BCG stellar component added to the initial
halo configurations, and we  also  consider the possibility 
of merging simulations within an SIDM framework.

\subsection{Search for the optimal merger models} 
\label{sec:opt}
\subsubsection{Code validation and physical processes leading to the 
formation of the observed X-ray morphology } 
\label{sec:optcode}

We first discuss simulation results obtained from the merging 
models A\_1 and B\_1   of Table \ref{clparam.tab}; the initial 
conditions of these models are the same as those for the corresponding 
models of \citetalias{Zh15}.  The purpose of these simulations is to
provide a sanity check of the initial condition setup adopted here 
for our merging simulations, as well as for the
analysis procedures presented in Section \ref{subsec:imag}.
The hydrodynamic performances of the
ISPH code have already been tested in previous papers
  \citep{V16,VS21} against results extracted from  Eulerian-based AMR codes.

For these two models in Figure \ref{fig:planeA1} 
we show, at different epochs,  
 the projected surface mass density and X-ray surface brightness maps.
The bottom panels are for model A\_1 and top panels for model B\_1, respectively. 
Each panel is for a different viewing direction
 $\{\alpha, i\}$ of the merging plane, as in 
 \citetalias{Zh15}.  The bottom panels (model A\_1) can then be compared 
directly with those in Figure 1 of \citetalias{Zh15}, while the top panels 
(model B\_1) correspond to those of their Figure 3. 
 Each panel refers to the post-pericenter epoch when the projected 
distance between the mass centroids is $ d_{DM} \sim 700 \kpc$. 

To ease comparison with the  \citetalias{Zh15} findings, in each panel 
we show contour levels of DM (white) and X-ray surface brightness (red) 
maps. Additionally, we also show the spatial location of the DM (green), 
X-ray (red) and SZ (yellow) centroids.

A visual inspection reveals that our test simulations are in substantial 
agreement 
with the corresponding ones presented in \citetalias{Zh15}, with  the mock X-ray 
images of Figures \ref{fig:planeA1}, which exhibit the same morphological 
features seen in Figures 1 and 3 of \citetalias{Zh15}. In particular, for
model A\_1 the bottom-left ($\{\alpha, i\}=\{-50^{\circ},0^{\circ}\}$)
and bottom-right ($\{\alpha, i\}=\{0^{\circ},0^{\circ}\}$) panels show the
same X-ray structures seen in Figures 1a and 1c of \citetalias{Zh15}:
a  strongly asymmetric tail (Figure 1a) and a small asymmetric 
wake-like feature (Figure 1c). The bottom-middle panel 
($\{\alpha, i\}=\{-50^{\circ},75^{\circ}\}$) of 
 Figure \ref{fig:planeA1}  corresponds to the fiducial model A of 
 \citetalias{Zh15} (Figure 1b). The X-ray morphology is more symmetric here 
and very similar to that seen in Figure 1b, with a bullet-like shape. 

Similarly, the top panels of Figure \ref{fig:planeA1}   can be 
compared with  those displayed in Figure 3 of \citetalias{Zh15}
for their model B.  The top-left ($\{\alpha, i\}=\{-90^{\circ},0^{\circ}\}$) 
and top-right ($\{\alpha, i\}=\{-90^{\circ},60^{\circ}\}$)  panels
show the existence of a wake structure trailing the secondary, 
which is asymmetric and oriented either to the left or to the right, respectively.
This is the same behavior as seen in Figure 3a and 3c of 
 \citetalias{Zh15}.   The top mid panel of Figure \ref{fig:planeA1}  
corresponds to Figure 3b of \citetalias{Zh15} and to their fiducial model B.
A comparison with this model is particularly significant, since this is the 
model that best matches the observations. In this panel 
 ($\{\alpha, i\}=\{-90^{\circ},30^{\circ}\}$)  the X-ray morphology
is clearly characterized by a well-defined twin-tailed structure, as seen in 
observations and in their Figure 3b.

The consistency between our model B\_1, with viewing angles 
 $\{\alpha, i\}=\{-90^{\circ},30^{\circ}\}$, and the fiducial model B 
of \citetalias{Zh15} can be put in a more quantitative way by comparing the 
respective X-ray luminosities $L_X$ and spectral temperatures 
$T_\mathrm{X}$.   
The  total X-ray luminosity in the $0.5-2$ keV band  
is evaluated by means of a standard SPH estimator, using the cooling tables 
implemented in our code.
For the spectroscopic-like temperature 
$T_\mathrm{X}$  we use Equation (\ref{tr.eq}). 
\begin{figure*}[!ht]
\centering
\includegraphics[width=0.95\textwidth]{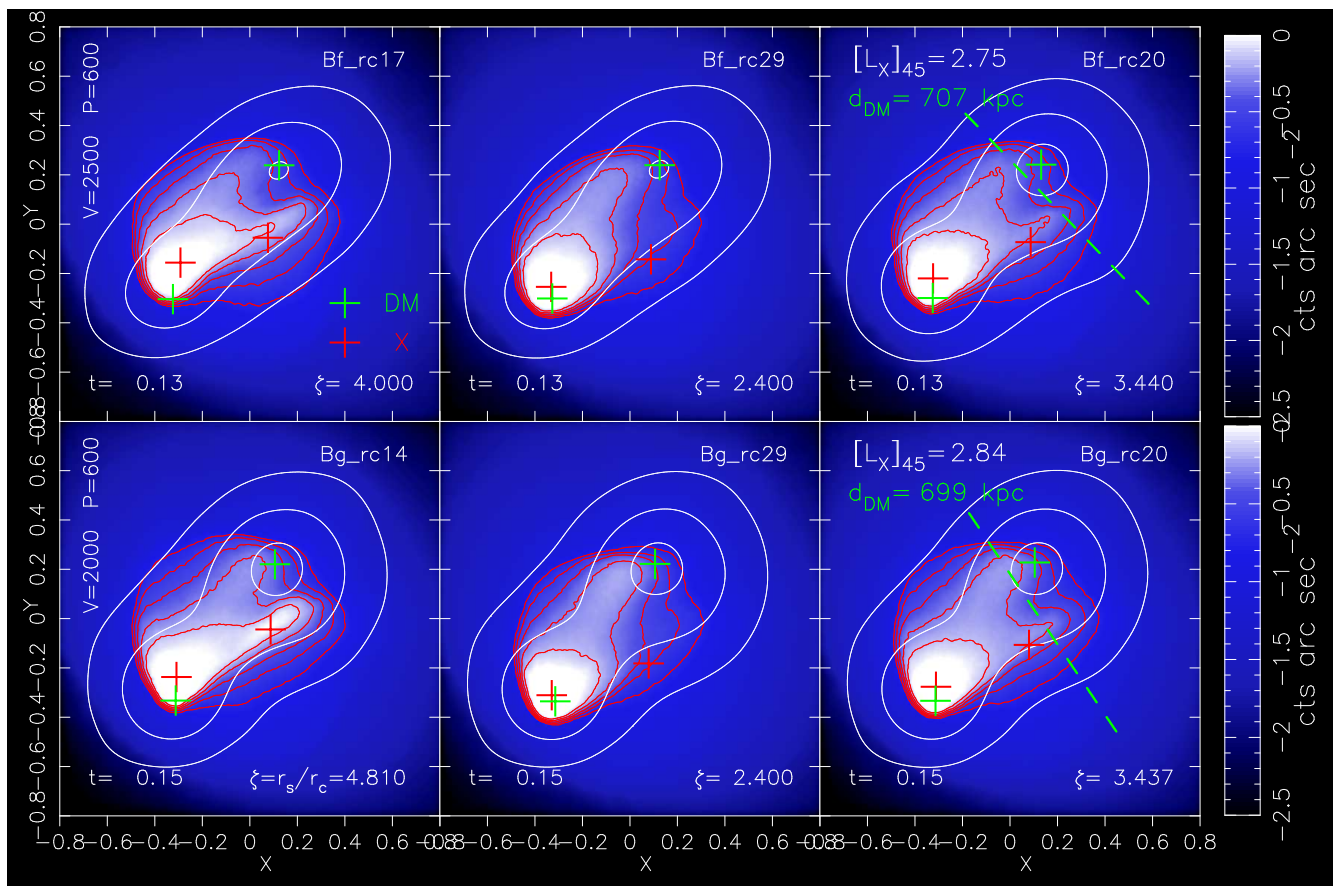}
\caption{Same as Figure \ref{fig:planeA1} but for merger models Bf and Bg. 
 Each panel refers to a simulation with a different gas core radius 
$r_c=r_s/\zeta$ of the primary (see Table \ref{subcl.tab}). 
In all of the panels maps are extracted along the same 
viewing direction ($\{\alpha, i\}=\{-90^{\circ},30^{\circ}\}$)
  of  the baseline model B\_1 in Figure  \ref{fig:planeA1}.
 The thresholds and the spacing of the contour levels are the same 
as those in Figure  \ref{fig:planeA1}.
The left panels are the chosen fiducial models 
Bf\_rc20 and Bg\_rc20 for these merging runs.
\label{fig:planeB1}
}
\end{figure*}

Additionally, a bow shock will form  in front of the   secondary as 
it reaches the pericenter and climbs its way toward the apocenter, this is 
because of the lower ICM sound speed due to the decrease in the gas  
 temperature.
The ICM gas will thus be compressed and shock-heated 
to higher temperatures, forming a leading edge oriented  
 perpendicularly to the secondary's direction of motion 
\citep{Ricker98,Mas08,Mac13,Molnar15,Sh19,Mou21,Cha22}.

The values that we obtain for these quantities are given in 
Table \ref{clres.tab} and are $L_X \simeq  2.86\cdot  10^{45} \ergs  $   
and $T_\mathrm{X}\simeq 12 \kev $, respectively. These values can be directly 
compared with 
the corresponding ones reported in Table 2 of \citetalias{Zh15}: 
$L_X \simeq  2.05\cdot  10^{45} \ergs $   and $T_\mathrm {X}  \simeq   15 \kev$.
From the same table one also has  $\delta T  \simeq  -1.1\cdot 10^{-3} \gr$
for the central SZ decrement, while from Table \ref{clres.tab} 
$\delta T  \simeq  -9\cdot 10^{-4} \gr$.
The largest relative difference is between  the X-ray luminosities, which is
naively expected given the proportionality of $L_X$ to the squared 
density and the associated uncertainties.
  
Given that these results have been obtained using two numerically different 
hydrodynamic schemes, and the uncertainties associated  with the use of 
 different estimators, we conclude that there is a fair agreement
between the two simulations.
We finally conclude that our merger model B\_1 is in accordance with
 the corresponding  fiducial model B  of \citetalias{Zh15}. 
We therefore adopted model B\_1, which  we simply refer to as 
model B from here on, as the baseline model against  which to validate 
 the simulation results obtained from our  merger models.

Before we discuss our findings from the merger models, it is useful to look first  at the processes leading to
the formation of the  twin-tailed morphology seen in the X-ray structure 
of the merger model B of Figure \ref{fig:planeA1}.
The main collision parameters that determine the X-ray structure of the
colliding clusters are the relative collision velocity $V$ and the collision 
parameter $P$, while keeping fixed the mass ratio and the viewing direction.

For a given set of collision parameters $\{ V, P \}$  the merger will be 
characterized by an infall phase in which the secondary falls in the potential 
well of the primary,   with a growing ram-pressure as the secondary 
approaches the pericenter and the relative velocity increases. This pressure 
will strip material from the secondary, leading to the formation of a 
downstream tail. Because the collision is off-center, 
 during its motion through the ICM of the primary,  the secondary will begin to 
deviate from its original trajectory and the tail will no longer  be
aligned with the  direction of motion.
\begin{figure*}[!ht]
\centering
\includegraphics[width=0.95\textwidth]{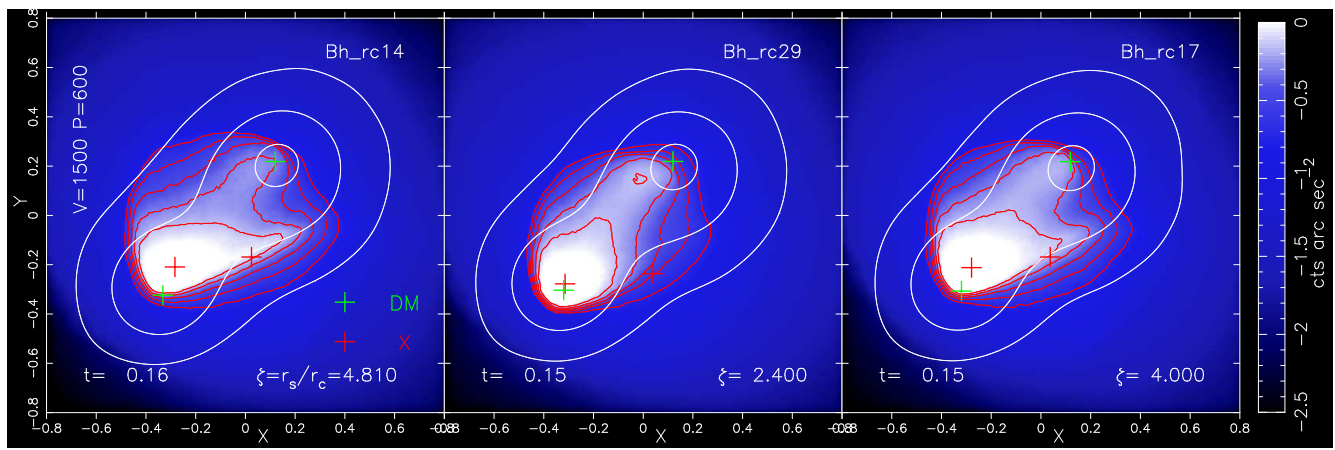}
\caption{Same as Figure \ref{fig:planeB1} but for merger models Bh. 
\label{fig:planeB1b}}
\end{figure*}

The gas morphology that is produced after the collision thus depends on the
adopted set of merging parameters $ \{M_1, ~q,~P, ~V\}$.
If $P$ gets higher while the other collision parameters are kept fixed, 
there will be a significant weakening of the ram pressure during the 
collision. This in turn will produce a strong asymmetric one-tailed 
gas structure, which is not observed; moreover, the primary would not have been 
destroyed during the collision.  Similarly,  if $P$ approaches zero, the 
collision will become axisymmetric around the collision axis, 
in contrast to the observed twin-tailed 
X-ray morphology (see Section 3.1.2). The observed asymmetric two-tail 
X-ray structure  is therefore expected to be produced  for just a certain range 
of values of the impact parameter $P$.

 Lowering the infall velocity $V$ without modifying the other collision 
parameters decreases the  amount of gas removed by ram pressure stripping, and 
the final result will be a single tail trailing the secondary. Similarly, 
if $V$ gets higher the interaction time between the two clusters will be 
strongly reduced and the resulting X-ray map will be strongly asymmetric 
(cf. Figure 5a and 5b of  \citetalias{Zh15}).

 We thus deduce that for a given set of merging parameters $\{M_1,~q,~P,~V\}$ 
  the post-collision gas structures are expected to be roughly reproduced 
 if one reduces proportionally both $P$ and $V$. The same reasoning applies 
 if the mass of the primary is decreased. Finally, projection effects can
 hide the real X-ray morphology and produce overlapping structures
along the line of sight. 

\subsubsection{Off-axis merger models with mass of the primary  
   $ M^{(1)}_{200} =1.6 \cdot 10^{15} \msun$: Models Bf, Bg, and Bh  }
\label{sec:optoffa}

We next used our findings to investigate different merger models aimed at 
 reproducing the observed El Gordo X-ray morphology  in a revisited  mass 
 estimate framework. As can be seen from Table \ref{clparam.tab}, we
 considered two families of off-axis merger models. The first family
  has the mass of the primary set to $ M^{(1)}_{200} =1.6 \cdot 10^{15} \msun $ 
  (models Bf, Bg and Bh), while $ M^{(1)}_{200} = 10^{15} \msun $ 
  for the second family (models Bk, Bl and Bj). This set of models will 
be presented in Section \ref{sec:optoffb}.
All of the models have the gas 
  fraction of the primary set to $f_{g1}=0.1$.
  In our merging runs the highest value that we use for the mass of the primary 
  is  $ M^{(1)}_{200} =1.6 \cdot 10^{15} \msun $. This is about $\sim 50\%$  
  higher than the mass estimate ($\sim 9.9 \cdot 10^{14} \msun$) reported by 
 \citet{Kim21}  for the NW cluster, but still within observational uncertainties.

  We present our simulation results following the top down ordering of Table
  \ref{clparam.tab}, with merging initial conditions that progressively deviate
  from those of the reference model B. The first model that we consider is model 
  Bf, with $P=600 \kpc$ and $V=2,500 \kms$, followed by models Bg 
($V=2,000 \kms$) and Bh ($V=1,500 \kms$).  

  Model Bf is nothing else than the merger model presented by \citetalias{Zh15}
  in their Figure 5f.  The authors rule out this model because it is unable to
 match  the required total X-ray luminosity. Our goal here is to 
 assess the extent to which a given merger model, with  a specific set of 
  merging initial conditions $ \{M_1, ~q, ~P, ~V \}$,  can reproduce the 
observational features of the El Gordo cluster.

As introduced in Section \ref{subsec:mrgmodels},  we adopt here the simulation 
strategy 
of performing a set of merging simulations with initial conditions that differ
 in the choice of the primary's   gas scale radius $\rcp$, while keeping fixed 
the other collision parameters of the chosen merger model.
Table  \ref{subcl.tab} reports the set of merging simulations  specific to a 
given model,  together with the corresponding gas core radii.

The top panels of Figure \ref{fig:planeB1}  show, from the left to the right, the 
 mock X-ray images extracted from the runs of models Bf\_rc17, Bf\_rc29, 
and Bf\_rc20. The current epoch is defined when the projected distance $d_{DM}$ 
between the mass centroids is approximately $d_{DM}\sim 700 \kpc$,  and 
as for  the fiducial model B of Figure \ref{fig:planeA1} the viewing 
angles are $\{\alpha, i\}=\{-90^{\circ},30^{\circ}\}$. Hereafter, the same
projection direction will be used to extract X-ray surface brightness  maps from 
the off-axis  merging simulations presented here. 

For the Bf models, the NFW scale radius,  $r_s$, of the primary takes the value $r_s\simeq 0.7 \mpc$, so the $\rcp$ ranges 
from $\rcp \sim 145 \kpc$ ($\zeta=4.81$: Bf\_rc14) up to 
 $\rcp \sim 290 \kpc$ ($\zeta=2.4$: Bf\_rc29).  From the maps of 
Figure \ref{fig:planeB1} it appears then that  a change of $\rcp$, within a 
certain range of values  and under certain conditions  for the primary's gas 
density profile, has the same impact on the final X-ray morphology of the merger 
as if one  were modifying the initial impact parameter $P$.
\begin{figure*}[!ht]
\centering
\includegraphics[width=0.95\textwidth]{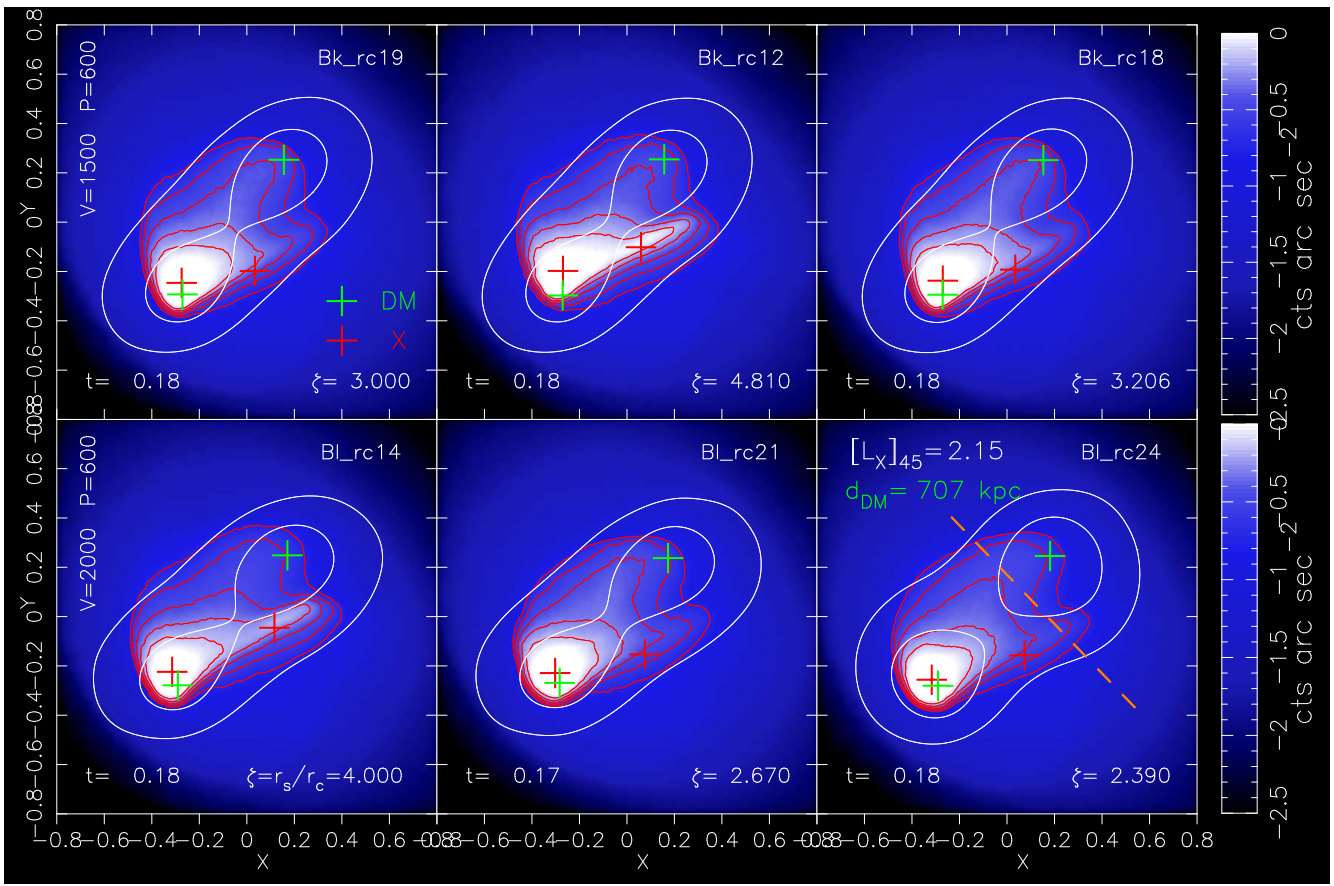}
\caption{Same as Figure \ref{fig:planeB1} but for the merger models Bk and Bl.
Model Bl\_rc24 is the chosen  fiducial model for the Bl runs.
\label{fig:planeB3}
}
\end{figure*}

 Specifically, model Bf\_rc29 clearly illustrates the effect on the post-collision 
X-ray structure of the merging when the gas core radius is increased. At the 
observer epoch, the X-ray morphology clearly exhibits only one tail trailing 
the secondary (top mid panel of  Figure \ref{fig:planeB1}).
In accordance with the previous discussion about the dependence of the final 
X-ray morphology on the collision parameters,  we argue that for model Bf\_rc29
the final one-tail X-ray feature is a consequence of the reduced ram pressure
experienced by the secondary during the collision. This is in turn due to
the primary's lower gas density because of the larger core radius.

 If instead  $\rcp$ approaches small
 values, the opposite must hold. The amount of gas stripped by the secondary 
 during its motion through the primary's ICM  will be much higher because
 of the  increased ram pressure. As can be seen from model Bf\_rc17
 (top-left panel of Figure \ref{fig:planeB1})  the X-ray morphology  at the
 observer epoch now exhibits a strongly asymmetric two-tailed structure, 
 with one tail previously absent and now with a much stronger emission than 
 the other.

It should also be stressed that such a dependence of the amount of
stripped gas on $\rcp$ takes place because the gas mass fraction is held fixed
between the models. This in turn implies that the central gas density of the
primary is anticorrelated with $\rcp$.
For a central gas density kept constant, one expects the dependence of 
the amount of stripped material on $\rcp$  to be reversed.

 These findings suggest that for the collision parameters of  the Bf 
family models, the characteristic twin-tailed X-ray structure seen in the 
El Gordo cluster can be reproduced with an appropriate choice of $\rcp$. 
Model Bf\_rc20 (top-right panel of Figure \ref{fig:planeB1}) is the 
  model whose X-ray image is best able to match the requested morphology:  
  for this model $\rcp \sim 200 \kpc$  and  its twin-tailed morphology 
    well reproduces that of fiducial model B in Figure \ref{fig:planeA1}. 
  Hereafter, we denote the fiducial model for the 
Bf family of merger models as model Bf\_rc20.

  The same reasoning can be applied to the Bg models, for which the bottom panels
  of Figure \ref{fig:planeB1}  show the mock X-ray maps extracted from three 
  runs with different values of the gas core radius of the primary,  $\rcp$.
  For these models the initial collision velocity  is lower ($V=2,000 \kms$) than
  in the Bf models,  and  model Bg\_rc20  ( $\rcp \sim 200 \kpc)$  is now the 
  fiducial model that produces  the optimal X-ray morphology. We note that 
  despite the initial velocity being lower than in model Bf\_rc20, the gas 
core radius is the same, but the X-ray morphology is a bit more asymmetric. 
\begin{figure*}[!ht]
\centering
\includegraphics[width=0.95\textwidth]{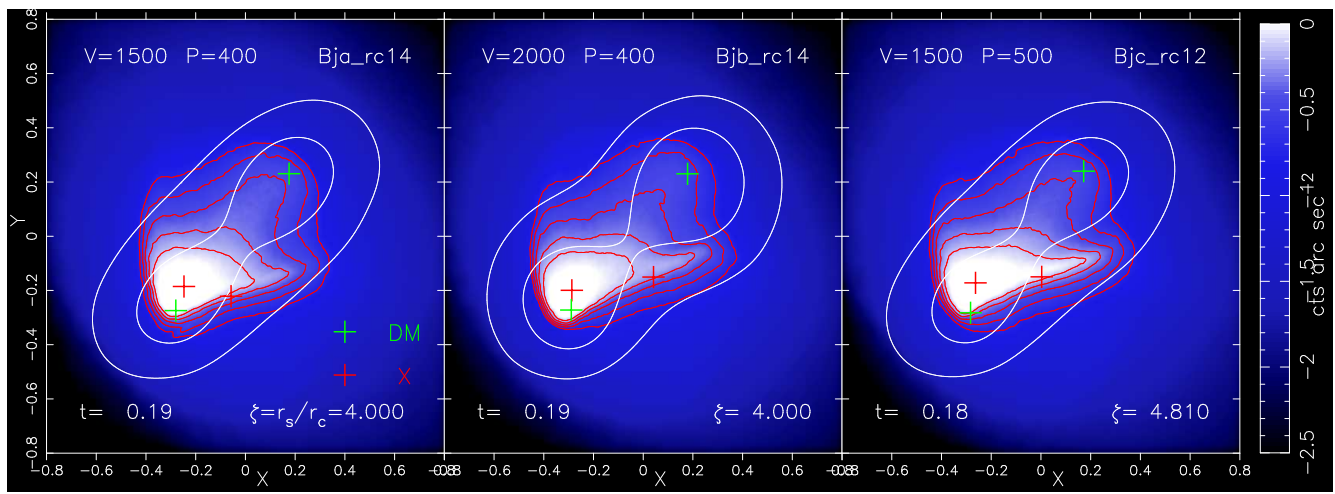}
\caption{Same as Figure \ref{fig:planeB3} but for merger models Bj 
of Table \ref{clbj.tab}. The meaning of the symbols is the same, as
is the viewing direction.
\label{fig:planeB3b}}
\end{figure*}

Figure \ref{fig:planeB1b} shows for the Bh models X-ray maps extracted from 
three merger simulations. For these  models $V=1,500 \kms$ and it is 
interesting 
to note that all of the considered models, from  $\rcp \sim 290 \kpc$ 
($\zeta=2.4$: Bh\_rc29)  down to $\rcp \sim 146 \kpc$ ($\zeta=4.81$: Bh\_rc14),   
produce strongly asymmetric X-ray structures that are in contrast with the 
observed X-ray morphology.  This is clearly a consequence of the now much weaker 
ram pressure due to the relatively low initial velocity between the two 
clusters.  

We use these findings to argue  that merger models of the El Gordo 
cluster, with a primary mass of approximately 
$M^{(1)}_{200}  \simeq 1.6 \cdot 10^{15} \msun $, require collision velocities at 
least as high as $V\simgt 2,000 \kms$
in order to be  able to  reproduce  the observed X-ray morphology.

\subsubsection{Off-axis merger models with mass of the primary  
   $ M^{(1)}_{200} =10^{15} \msun $: Models Bk, Bl, and Bj  }
\label{sec:optoffb}

For the family of merger models with mass of the 
  primary $ M^{(1)}_{200} = 10^{15} \msun $,  we have model Bk 
  ($P=600 \kpc$ and $V=1,500 \kms$) and Bl ($V=2,000 \kms$), plus the additional
  models Bj of Table \ref{clbj.tab}. 

Figures \ref{fig:planeB3}   and \ref{fig:planeB3b} show the X-ray images
constructed from simulation results of the merger models Bk, Bl and Bj. 
 The top panels of  Figure \ref{fig:planeB3}  demonstrate that the observed 
twin-tailed morphology cannot be reproduced by merger models with initial 
collision velocity  as low as  $V=1,500 \kms$ (Bk runs), regardless of the 
chosen value of the primary's gas core radius.  This behavior is similar to 
that of model Bh (Figure \ref{fig:planeB1b}), but now the mass of the primary 
is lower.

 For Bl models, the collision velocity is increased  to  $V=2,000 \kms$. The 
bottom panels of Figure \ref{fig:planeB3}  show  that  this class of merger 
model can produce an X-ray image that compares well visually  with those 
 previously used to define the list of fiducial models. 
 Model Bl\_rc24 (bottom-left panel of Figure \ref{fig:planeB3}) is the merger 
model that is best able to match the observed X-ray morphology: 
 for this model $r_s\simeq 0.58  \mpc$  and $\rcp\sim 240 \kpc $  is the 
corresponding value of its primary's gas core radius. This model is added to 
our list of fiducial models. 

 So far we have essentially considered  only merger models with initial impact parameter  
 $P=600 \kpc$; the exception  is fiducial model B 
 ($P=800 \kpc$) because of
 the high value of its primary's mass. In order to assess the viability of
 merger models with a smaller value of the impact parameter we ran an 
 additional set of models (Bj), with $P$ now ranging from $P=500 \kpc$ 
 down to $P=400 \kpc$.  For each set of collision parameters $\{V,~P \}$ 
 we performed merging simulations by considering different values of 
 $\rcp$. We show here results extracted from the
 most significant merging runs, Table  \ref{clbj.tab} lists the collision 
 parameters of these models. For all of them the mass of the primary is set to 
 $M^{(1)}_{200} = 10^{15} \msun $. 

 Figure \ref{fig:planeB3b} shows the X-ray maps extracted from the merger models
  Bj of  Table  \ref{clbj.tab}. From these maps  it is possible to deduce that 
mergers with initial impact parameter as low as $P=400 \kpc$ are unable to 
 generate the requested X-ray morphology. The X-ray structure is strongly 
 asymmetric, with only one tail present if  $V=2,000 \kms$ (Bjb\_rc14 run) 
 and approaches that of a head-on collision when $V=1,500 \kms$  (Bja\_rc14 run).
 The asymmetric one-tail structure is even more pronounced if one 
 considers model Bjc\_rc12 for which  $P=500 \kpc$. 

\subsubsection{Comparison of the results with previous findings and
observations}
\label{sec:optoffc}
In order to better constrain the fiducial merger models that have been
singled out by our set of simulations, for each of these models 
we now  compare results and observations.
Table \ref{clres.tab}   lists the measured values of several 
 quantities, together with their observational estimates. These
are taken from Table  2 of  \citetalias{Zh15}.

The spectral X-ray temperatures $T_\mathrm{X}$  given in 
Table \ref{clres.tab}  are obtained by using the estimator given by
Equation (\ref{tr.eq}).  This approach is  much simpler than that adopted
by  \citetalias{Zh15}, who estimated spectral temperatures by fitting 
mock X-ray spectra using the software package MARX; nonetheless, it has 
been employed by many authors \citep{Mas08,Molnar13,Zh18} and 
it provides a good approximation to cluster spectral fit temperatures, 
as  demonstrated by \citet{Nagai07}.

The  spectral temperature of model B 
($T_\mathrm{X}\simeq 12 \kev $) is smaller than that reported by 
 \citetalias{Zh15} for their fiducial model B ($T_\mathrm{X}\simeq 15 \kev $) 
by a factor of $ \sim 20 \%$.  
As previously discussed we do not consider this discrepancy particularly 
significant, given the use of different codes and estimators.
The temperatures of models Bf\_rc20 and Bg\_rc20 
($T_\mathrm{X}\simeq 13.5 \kev $)   
fare better than model B to approximate the
observational estimate ($T_\mathrm{X}\simeq 14.5 \kev $).   
Model Bl\_rc24 has the lowest temperature ($T_X\simeq 11 \kev$), which is 
not surprising given that this is the model with the  lowest collisional 
energy.
Finally, for  all of the models there is a remarkable agreement between the
 measured X-ray luminosities, as well as with the observational value, 
which is somewhat unexpected given the dependence of $L_X$ on the
squared density.

We now proceed to extract X-ray surface brightness  profiles
from our fiducial merger runs. We evaluate these profiles by cutting 
spatial regions across the wake behind the X-ray emission peak of the 
secondary.  As  in \citet{Molnar15},  we identified these regions  by visual
inspection,  with their spatial location and orientation changed until
 the observed spectra  best reproduce previous findings.
These regions correspond to the dashed orange lines shown in the X-ray images 
of the fiducial models.

\begin{figure}[!ht]
\centering
\includegraphics[width=8.6cm,height=8.2cm]{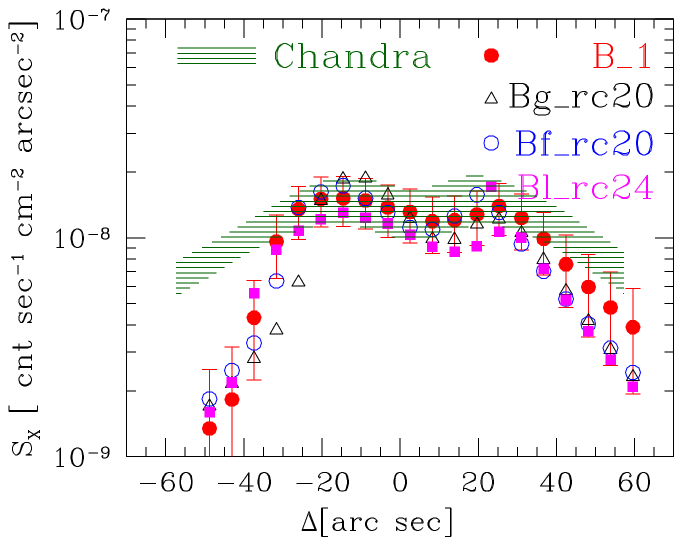}
\caption{ X-ray surface brightness  profiles across the wake 
of several cluster merger models. The extraction region of the profiles 
is shown in the 
previous images as the dashed orange line.
Distance is measured in  arc sec along the extraction path,
 model B\_1 is the corresponding B model 
 of \citetalias{Zh15}  and is used here as a reference model.
The width of the hatched region represents the uncertainties 
of the  {\it Chandra} observation, as shown by the blue error bars in Figure 8 of
 of \citetalias{Zh15} (courtesy of C. Zhang).
\label{fig:profSB}
}
\end{figure}

Figure \ref{fig:profSB}  shows the measured profiles as a function of
the distance $\Delta$ across the wake;  
these can be compared with those depicted in Figure 8 of \citetalias{Zh15} 
or  Figure 4 of  \citet{Molnar15}, after appropriate rescaling.
All of the profiles agree quite well with each other, they share a minimum
at $\Delta \sim 0 \acs$ and their two peaks  are separated by about 
 $\sim 50 \acs$. The profiles are approximately symmetric around $\Delta =0$, 
but exhibit a weaker emission in the wings, with respect 
the ones presented by \citetalias{Zh15} and  \citet{Molnar15}. 
The origin of such a discrepancy is not clear, we suggest that could be due
to possible differences in the procedures adopted to define the 
extraction region used to calculate the spectra.
\begin{table*}
\caption{Main cluster properties for the fiducial merger models.}
\label{clres.tab}%
\centering
\begin{tabular}{ccccccc}
\hline
 &   B\_1  & Bf\_rc20 & Bg\_rc20 & Bl\_rc24  &  Observations $^a$ \\
\hline
$d_{DM} [\kpc]$  &  704   &  707   & 699  & 707 & $\sim$ 700  \\
$L_X [10^{45} [\ergs] $   &  2.86   &  2.75   & 2.84  & 2.15 & 2.19 $\pm$ 0.11  \\
$T_X [\kev]$  &  11.8    &  13.2    & 13.5  &  10.9   & 14.5 $\pm$ 1\\
$d_{X-SZ} [\kpc]$  &  504   &  391   & 414  & 398 & $\sim$  600   \\
$\delta T [\mu \gr] $  &  -915   &  -959   & -997  &  -662  &  -1046 $\pm$ 116 \\
$V_r [\kms]$   &  1024   &  956   & 971  &  856  & 586 $\pm$ 96 \\
\end{tabular}
\begin{flushleft}
{\it Notes.} {Rows from top to bottom: 
projected distance between the mass centroids, 
 total X-ray luminosity in the $[0.5-2] \kev$ band,
 X-ray spectral temperature of the SE cluster,
 offset between the SE X-ray peak  and the SZ  centroid,
 central SZ decrement, 
 mean relative radial velocity between the SE and NW clusters.
$^a$: From Table 2 of \citetalias{Zh15}. 
}
\end{flushleft}
\end{table*}

 The  separation between the two maxima and their  emission level 
($ \sim 1.5 \cdot 10^{-8} \linebreak \cts$) are in accordance  with those 
exhibited by model B of \citetalias{Zh15}. However, for their model the 
X-ray emission of the minimum is almost at peak levels, whereas here it is a 
factor of $\sim 2$ lower.
This discrepancy   is less pronounced in the spectra of \citet{Molnar15}, for
which the emission level of the minimum is  about $\sim 70\%$ of the maxima.

The SZ-X-ray offset  ($d_{X-SZ} \sim 500 \kpc $ ) 
is substantial in accordance with the corresponding  value reported 
in their Table 2 by  \citetalias{Zh15} for their fiducial model B, and it is
smaller than the value estimated by observations ($\sim 600 \kpc $). 
 Following \citetalias{Zh15}  we argue that such a discrepancy is primarily due 
to the low angular resolution of the SZ measurements. For the other merger
 models the discrepancy becomes larger, as a consequence of the reduced 
collision velocity and of the primary's mass.

The SZ central temperature decrement of fiducial model B is 
$\simeq -915 \mu \gr $, in line with observational estimates 
and in accordance with the corresponding value reported by 
 \citetalias{Zh15} for their model B ($-1130 \mu \gr $). 
Models Bf\_rc20 and  Bg\_rc20 have similar decrements, while for model 
Bl\_rc24 the decrement is much smaller ($\simeq -660 \mu \gr $) and inconsistent 
with the measured value. We consider this discrepancy a clear 
 consequence of the reduced  collisional energy  of this merging model.

Another observational feature that can be compared against simulation results
is the relative radial velocity, $V_r$, along the line of sight between the 
NW and SE clusters. From the redshift distribution of member galaxies, 
 \citet{Men12}  estimate for the SE cluster a relative radial velocity 
 of $V_r^s=598 \pm 96 \kms$,  with respect to the NW component.

 The merging runs presented in this section, as well as those of
 \citetalias{Zh15}, do not incorporate a stellar component in the initial 
 condition setup. Therefore, in these simulations the observed galaxy
  velocity distribution can only be contrasted against simulation results 
  if one assumes the velocity distribution of DM particles to be a proxy
  of that observed for galaxies.  This is supposed to be a fair assumption 
  as long as the DM  is supposed to be collisionless. In 
  Section \ref{sec:sidm} we investigate the the observational consequences when this assumption
  is no longer valid.

\begin{figure}[!ht]
\includegraphics[width=8.2cm,height=8.2cm]{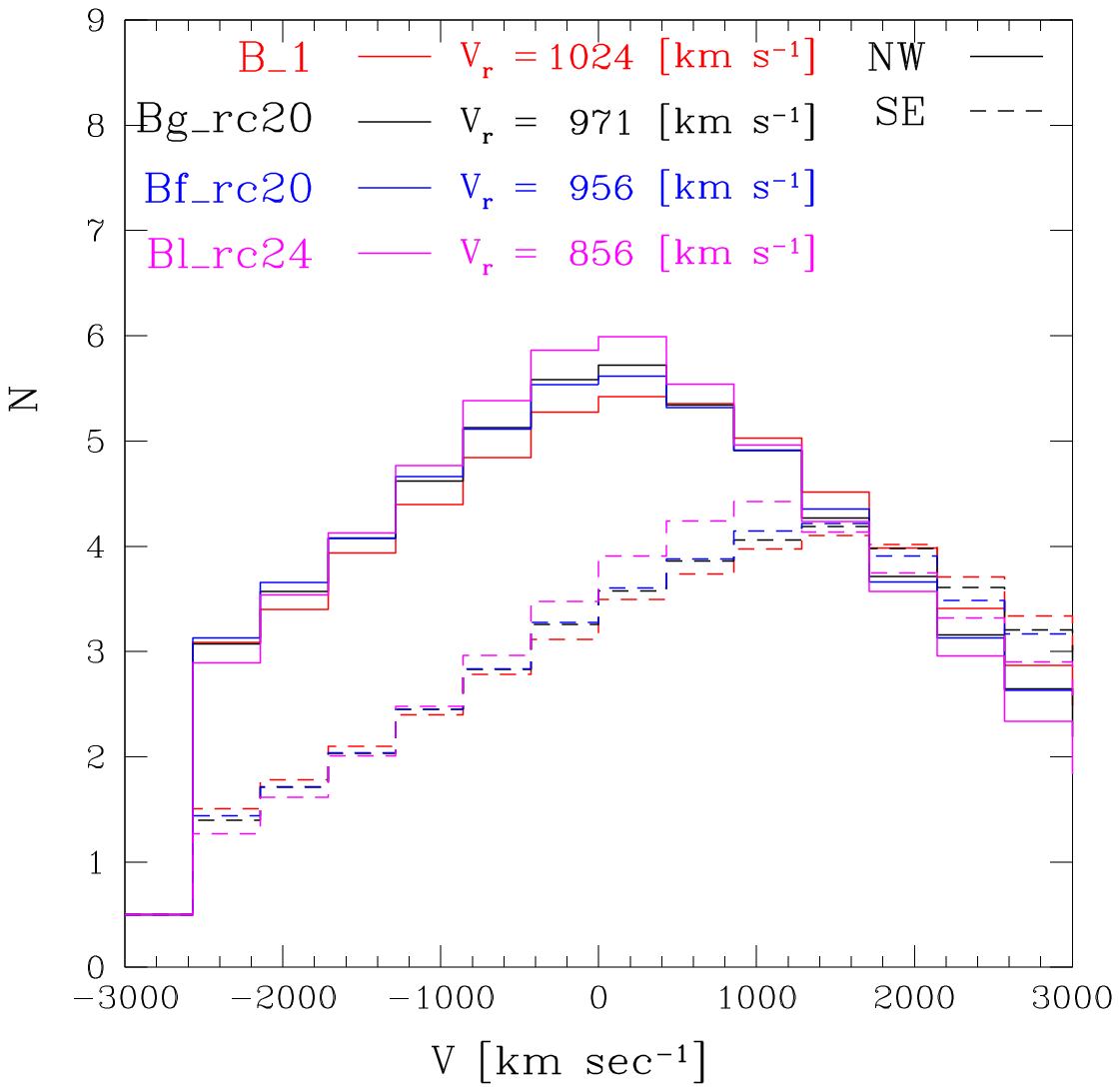}
\caption{Histograms of the DM line-of-sight velocity distribution
for the NW and SE clusters of the same merger models shown in
  Figure \ref{fig:profSB}. The NW (SE) histograms are normalized
to a total number of 51 (36) members, as in Figure 9 of  \citetalias{Zh15}. 
For each model  in the Figure is reported the 
relative mean radial velocity between the SE and NW cluster components.
\label{fig:profHV}
}
\end{figure}

  For the four fiducial merger models we show in Figure \ref{fig:profHV} 
histograms of the velocity distribution of DM particles,  for both NE and SW 
clusters.  The histograms are normalized to a total of 51 (36) members for the 
NW (SE) component, as in Figure 9 of  of \citet{Men12}. The Figure also reports 
the difference $V^{DM}_r$ along the line of sight between the mean radial 
velocities of the NW and SE clusters.  
The mean projected velocities of each component are evaluated by averaging over 
all of the DM particles that, in the observer plane, are within a distance of 
 $400 \kpc$ from its mass centroid.

 The velocities are in the range  from $V^{DM}_r \sim 1,020 \kms$ (model B) to 
 $V^{DM}_r \sim 850 \kms$ (model Bl\_rc24)  and  are significantly higher than 
the measured 
 value $V_r^s$. This  tendency has already been noted by \citetalias{Zh15}. 
 The authors argue that contamination between the different mass components, 
which occurs during the collision between the two clusters, can  
significantly bias the measurement of   radial velocities  and 
cause underestimation of  the relative real cluster velocity.

The two velocity histograms of fiducial model B can be  directly compared  
with the corresponding ones shown in Figure 9 (bottom panel) 
of \citetalias{Zh15}, the construction procedure being the same. From the two 
Figures it appears that the velocity distributions extracted from the 
two simulations are in  good agreement. In particular \citetalias{Zh15} 
report $V^{DM}_r \sim 910 \kms$ for their model B.

 To summarize, we use the results of this section  to argue that the observed 
 twin-tailed X-ray structure seen in the merging cluster El Gordo can be 
reproduced in off-axis merging simulations only for a narrow interval of  
values of the initial collision parameters.  Specifically, the observational 
morphological constraints are satisfied  for 
$2,000 \kms  \simlt V \simlt 2,500 \kms$  and 
$600 \kpc \simlt P \simlt 800 \kpc$.  The  mass of the primary is allowed 
to vary in the range 
$ 10^{15} \msun  \simlt M^{(1)}_{200} \simlt  2.5 \cdot 10^{15} \msun $, 
 with the mass ratio $q$  adjusted  to satisfy 
$  M^{(2)}_{200} \simeq  6.5 \cdot 10^{14} \msun $. 
 The lower part of the mass interval is favored by the latest mass estimates.

The fiducial merger models that have been found here to agree  with the observed
 X-ray structures also satisfy a number of observational constraints.  
 In particular models Bf\_rc20 and Bg\_rc20, which have  a  primary mass of 
 $M^{(1)}_{200}= 1.6 \cdot 10^{15} \msun$, show an  overall better agreement 
with data.

\subsection{A returning scenario? }
\label{sec:return}

One of the most challenging features to reproduce in the simulations of the 
El Gordo cluster is the observed offset between the X-ray emission and  
DM density peak of the SE component, with the latter being closer to the 
merging center of mass \citep[see, for example, Figure 1 of ][]{Ng15}.
This observational aspect is clearly at variance with what is expected 
in a outgoing, post-pericenter, scenario in which, after the collision,  
the X-ray emission peak is expected to trail the DM mass centroid.

To solve this inconsistency \citet{Ng15} suggested that the El Gordo cluster
 is presently observed at a later stage of the merger, in a returning  
 phase with the DM secondary having reached the apocenter and now falling back 
onto the primary. The authors conclude that a returning scenario is favored
by comparing available data against a large set of Monte Carlo simulations
with different input parameters. They construct different merger models 
by following the evolution of the two DM subclusters in a head-on 
scenario.

In this section we perform a suite of hydrodynamical simulations to investigate 
the viability of a returning scenario as a possible solution to the offset 
problem. To consistently compare our simulations with previous findings 
\citep{Ng15},  we considered head-on collisions and accordingly set $P=0$;
the procedures for setting up the initial conditions are those outlined in Section 
\ref{sec:sims}.

We construct our set of simulations by considering a range of collision
parameters. Specifically, for the  mass of the primary we chose
$M^{(1)}_{200}=1.3 \times  10^{15}\msun$,  as in Table 1 of 
\citet{Ng15}. The mass of the secondary is set by those authors to
$M^{(2)}_{200}=7.6 \times  10^{14}\msun$, or   $q \simeq 1.7$.
Here we instead consider the following values for the merging mass ratio: 
$q=2,~4,~10$. 

For a given value of $q,$ we considered three different values of the
collision velocity: $V=1,000,~2,000,$ and $3,000 \kms$; 
our set of simulations thus consists of nine head-on merger collisions.
The IDs of the runs are given in Table \ref{head.tab}, these
are defined according to the mass ratio $q$ and the
initial collision  velocity $V$.  
We explicitly decided to investigate a wide range of merging parameter space,
 in order to assess for which initial conditions head-on encounters
are best able to reproduce  the observed twin-tailed X-ray morphology.

As the collision parameter $P\rightarrow 0 $ the asymmetries in the X-ray
emissivity maps of the merging clusters tend to disappear and the collision
becomes axially symmetric. Previous head-on merging simulations of the El Gordo 
cluster have been conducted by \citet{Donnert14}  and \citetalias{Zh15} 
(their model A). As can be seen from Table \ref{clparam.tab}, these simulations 
were not perfectly head-on but were performed 
with an initially small impact parameter ($P=300 \kpc$).

The simulations of \citet{Donnert14}  were able to satisfy a number of 
observational constraints but failed to reproduce the twin-tailed X-ray 
morphology, the X-ray peak being followed by only one tail elongated along
the collision axis. Model A of \citetalias{Zh15}  is a highly energetic 
($V=3,000\kms$), almost head-on collision. As discussed in the previous section,
 their fiducial model A  corresponds to the  bottom-middle panel
 of Figure \ref{fig:planeA1}, and the twin-tailed morphology is absent in the 
highly symmetric X-ray structure.
A  small asymmetric wake-like feature is instead present in the bottom-right 
panel of Figure \ref{fig:planeA1}  (Figure 1c of \citetalias{Zh15}), 
which the authors interpret as  a consequence of a nonzero impact 
parameter. 

In this case it is worth noting  the different values found in the  two 
simulations for the  projected separation between the mass centroids. 
\citetalias{Zh15} report $d_{DM}\simeq 600 \kpc$, while we found here 
 $d_{DM}\simeq 720 \kpc$.
We attribute this discrepancy to the very large value of the relative 
velocity  ($V_{rel} \simeq 4,700 \kms$)  between the two 
clusters at the evolution time $t=0.09$ Gyr, when the maps are 
extracted, and the time resolution of our simulations,  
given by the timestep $\Delta t_0=1/100$ Gyr. 
The current epoch is identified at run time when 
 $d_{DM}\geq  700 \kpc$, and the snapshot of Figure \ref{fig:planeA1}
  (bottom right) corresponds to the simulation step $n=164$.
The previous output is saved at $n=162$, when $d_{DM}\simeq  630 \kpc$.

\begin{table}[t]
\centering
\caption{
 IDs and initial parameters of the head-on 
merging simulations of Section \ref{sec:return} $^{(a)}$}
\begin{tabular}{lrccc}
\hline \hline
         $V [\kms]$ & \multicolumn{3}{c}{$q=M_1/M_2$}  \\
\cline{2-4}
  & 2  & 4  & 10  &   \\
\hline
 $ 1,000$ & R01V10  & R02V10   & R03V10       \\
\hline
 $ 2,000$ & R01V20  & R02V20   & R03V20       \\
\hline
 $ 3,000$ & R01V30  & R02V30   & R03V30       \\
\hline
\hline
\end{tabular}
\label{head.tab}%
\begin{flushleft}
{\it Notes.} {$^a$: The meaning of the parameters $q$ and $V$ is the same 
as in Table \ref{clparam.tab}.
For all the simulations we set the mass of the primary to 
$M^{(1)}_{200}=1.3 \cdot10^{15} \msun$  
and the initial gas mass fraction to 
$f_g=0.1$ for both of the clusters.}
\end{flushleft}
\end{table}

\begin{figure*}[!ht]
\centering
\includegraphics[width=0.95\textwidth]{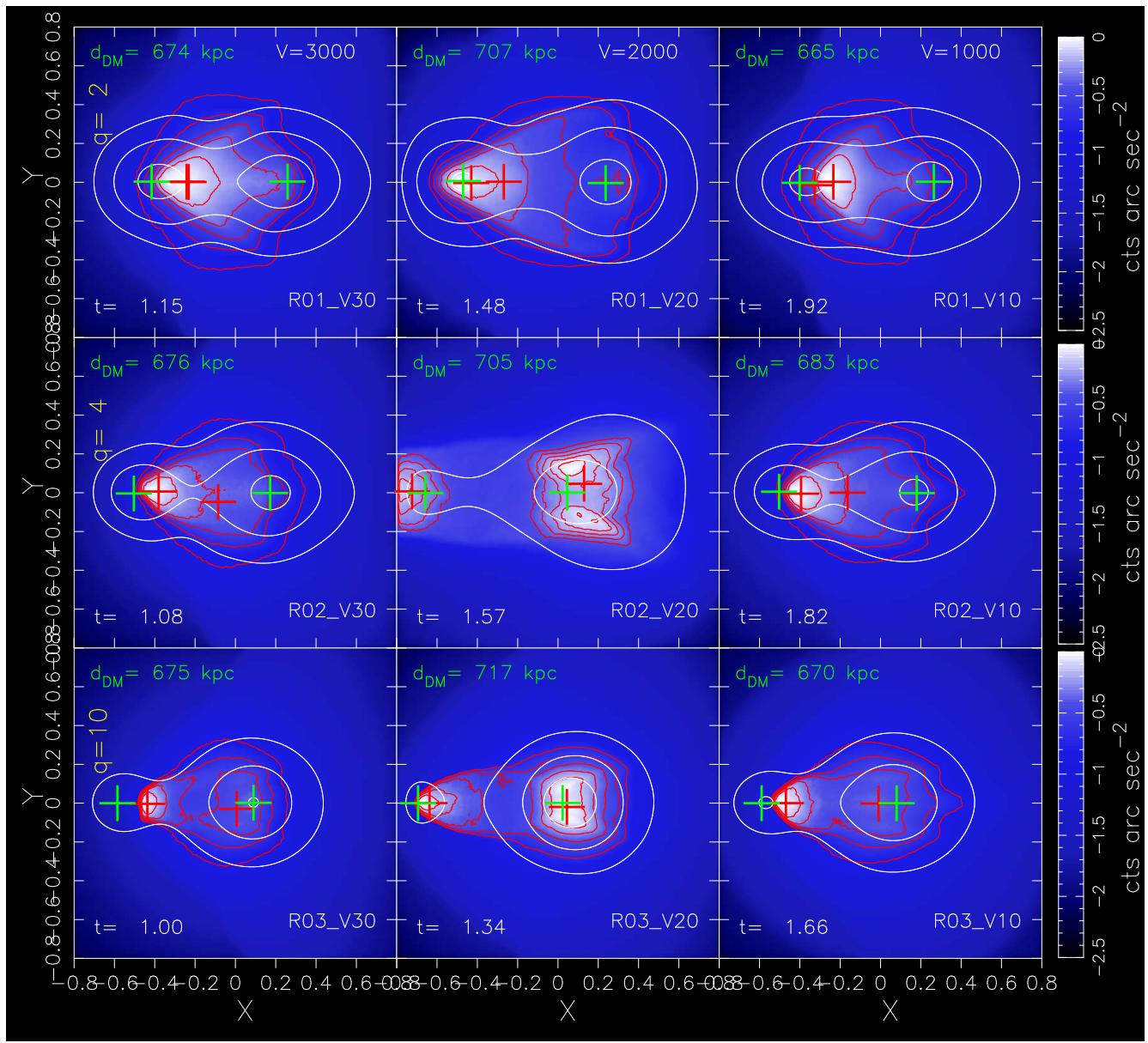}
\caption{ Surface brightness maps  for the head-on mergers
described in Section \ref{sec:return}. As in the previous maps,
time is in Gyr, with $t=0$ at the start of the simulation and the
 box size is $1.6  \mpc$. In each panel is indicated the
 collision model whose initial merging parameters are given in 
Table \ref{head.tab}. 
 Each row of panels refers to simulations with the same value of the 
mass ratio $q$, while each column is for simulations having the same value
of $V$. This is reported in $\kms$.
The maps are extracted at output times when 
the distance $d_{DM}$ between the mass centroids of the two clusters is 
approximately  $d_{DM}\simeq 700 \kpc$. 
As in Figure \ref{fig:planeA1}, 
the crosses indicate the projected spatial locations of the mass 
(green) and X-ray surface brightness (red) centroids.
\label{fig:planeHO_1}
}
\end{figure*}

To summarize, no twin-tailed X-ray structure has been reproduced so far
in head-on  merging simulations of the El Gordo cluster. We thus now discuss
 results extracted from our set of head-on merging simulations.
For each of the merging
 runs of Table \ref{head.tab}  we show in Figure \ref{fig:planeHO_1} 
the corresponding X-ray emission maps in the outgoing scenario. The 
 collisions are seen face-on and 
the secondary moves from right to left. The time $t$  displayed 
in each panel is when for the first time $d_{DM}\simeq 700 \kpc$ 
after the pericenter passage; here $t=0$ is when the simulations start.

From Figure \ref{fig:planeHO_1} it can be seen that
all of the simulations shown are axisymmetric around the collision axis. 
Moreover, not unexpectedly, the X-ray emission peaks  are always trailing 
the DM centroids. 
The only exception to this is the run R02V10,  but its initial conditions are a 
bit peculiar (see below). Finally, in all of the 
cases,  the core of the primary 
does not 
survive the encounter with the secondary, with the ICM behind the secondary 
being shaped by its motion through the main cluster.
For the R01 runs ($q=2$), the ICM morphology is arc-shaped around the 
secondary's core, in accordance with previous findings \citep{Ricker01}, 
while for higher mass ratios the post-pericenter shape of the ICM depends on the 
initial velocity $V$.
\begin{figure*}[!ht]
\centering
\includegraphics[width=0.95\textwidth]{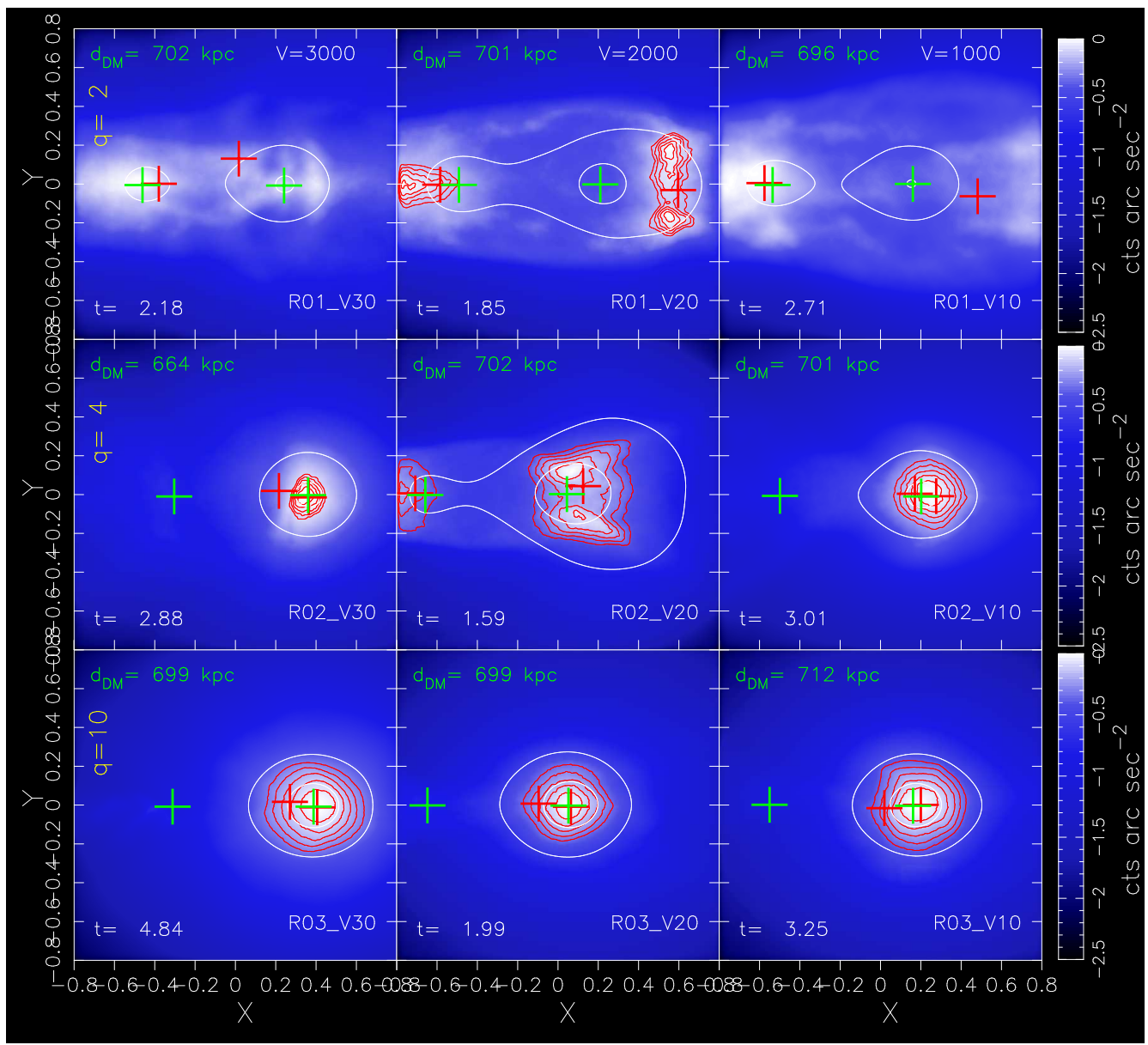}
\caption{  Same as Figure \ref{fig:planeHO_1}, 
 but here the output times are chosen so that  
 $d_{DM}\simeq 700 \kpc$ after the apocenter passage.
\label{fig:planeHO_2}
}
\end{figure*}

The only merging simulations that show  X-ray structures whose shapes
can be considered to exhibit some degree of similarity   with the 
requested twin-tailed morphology are the R01 runs ($q=2$).
In fact, the initial conditions of the R01V30 run are very close to those
of model A in Table \ref{clparam.tab}, while those of \citet{Donnert14}  
can be considered intermediate between R01V30 and R01V20.
In the other mergers with $q>2$, the smaller is the mass of the secondary, 
the less is the impact on the ICM of the primary.
 A significant X-ray structure
behind the secondary is still visible if the decrease in the mass of the 
secondary is accompanied by a corresponding reduction in the 
collision velocity (runs R02V10 and R03V10).

To assess the viability  of the returning scenario,  Figure \ref{fig:planeHO_2}
shows X-ray maps extracted from the same runs as in  Figure \ref{fig:planeHO_1}, 
but now in a post-apocenter phase.  The criterion to identify the observer 
epoch is still that of having $d_{DM}\simeq 700 \kpc$, but this when 
the secondary has reached the apocenter and is now returning toward the primary.

The time difference $\Delta t_{orb}$  between the epochs  shown in 
Figure \ref{fig:planeHO_2} and those in the corresponding panels of 
Figure \ref{fig:planeHO_1}  gives,  for a specific model,  
 the time spent by the secondary between the outgoing phase and the returning 
one.
 The time  $\Delta t_{orb}$  can assume a wide range of values: from 
 $\Delta t_{orb}\sim 0.02$  (R02V10) up to  $\Delta t_{orb}\sim 4$  Gyr 
(R03V30).
 
The time  $\Delta t_{orb}$  is a function of both $q$ and $V$; for a given 
$V$, it can be seen that it increases as $q$ gets higher. Such behavior is 
clearly a consequence of the reduced exchange of energy   
between the two clusters during the collision. 
The merging run R02V10 has a very small value of $\Delta t_{orb}$  because its 
initial velocity ($V=2,000 \kms$) is such that when the secondary reaches the
orbit's apocenter it is also when  $d_{DM}\simeq 700 \kpc$;   this is 
the reason why the  X-ray emission and DM centroids of the secondary are 
so close  in both of these two phases.  

To compare our findings with the proposed model of \citet{Ng15},   we had to choose
from Figure \ref{fig:planeHO_2} the merging runs for which the value of 
the mass ratio is closer to the value they adopted  ($q\sim1.7$).
These conditions are satisfied by the runs R01V10, R01V20, and R01V30, 
for which $\Delta t_{orb}\sim 0.8, 0.4$, and $\sim1$ Gyr, respectively.
R01V30 is, however, ruled out by the timescale 
 $\Delta t_{orb}=0.91-0.46 \sim 0.5$ Gyr favored by  \citet{Ng15}.
Their Monte Carlo simulations  include neither
dynamical friction nor tidal forces, so a comparison between 
their favored value  of $\Delta t_{orb}$ with the values we found here
is not entirely consistent.
Nonetheless, in a previous paper \citep[see Table 2 of][]{VS21} we demonstrated 
that the impact of dynamical friction and tidal forces becomes significant 
for off-axis mergers when the mass ratio gets higher ($q\simgt5$).
We then use these findings to argue that the constraint on $\Delta t_{orb}$ 
of \citet{Ng15} can be used to rule out model R01V30.
\begin{table*}
\centering
\caption{Initial merger parameters and IDs of the merger simulations  
of Figure \ref{fig:planeB4} and \ref{fig:planeSXa}. $^a$ }
\label{bsdm.tab}%
\begin{tabular}{lcccc}
\hline \hline
Model: $ \{M_1, r_s, ~q,~P, ~V,~\zeta=r_s/r_c\}$ 
        & BCGs & $ \sigma_{DM}/m_X [\sxu]$  & IDs of the merger simulations   \\
\hline
Bf$\_$rc20: $ \{ 1.6 \cdot 10^{15}, 0.696,~2.32,~600,~2,500,~3.44 \}$ 
    & $\surd$ & 0 & DBf$\_$rc20       \\
\hline
Bl$\_$rc24: $ \{ 1. \cdot 10^{15}, 0.574,~1.54,~600,~2,000,~2.39 \}$ 
        & $\surd$ & 0& DBl$\_$rc24      \\
\hline
Bf$\_$rc20  & $\surd$ & 1/2/5 & XDBf$\_$rc20     \\
Bl$\_$rc24  & $\surd$ & 1/2/5 & XDBl$\_$rc24     \\
\end{tabular}
\begin{flushleft}
{\it Notes.} {$^a$ Columns from left to right:
 collision parameters  of the mergers, 
  cluster halos  with initially a BCG  mass component,   
value of the SIDM cross-section per unit mass, 
 ID of the merger simulations. For the simulations 
with ${\sigma_{DM}}/{m_X} >0$, the  label X generically refers to
merger runs  with different values of ${\sigma_{DM}}/{m_X}$.}
\end{flushleft}
\end{table*}

This leaves us with models R01V20 and R01V10. We note  
  that for model R01V20 the  DM centroid of the secondary 
is now closer to the center of mass of the system
than its X-ray emission peak;  the distance between the two centroids  
is approximately $ \sim 100 \kpc$.
However, neither of these two merger models  is able
to reproduce the X-ray structure that we see in the El Gordo clusters. 

 In fact these models, as well as all of the others, are morphologically 
inconsistent with the observed twin-tailed X-ray morphology.  We argue that 
this behavior can be interpreted as follows. The primary's ICM undergoes a 
strong increase in temperature and luminosity as the two cluster cores approach
the pericenter, when their relative separation has  a minimum. 
Subsequently the two clusters move away from each other and the shock-heated 
gas, which is no longer in hydrostatic equilibrium, cools adiabatically and its 
temperature and density drop significantly. Because of the X-ray emissivity 
dependence on the squared density, this in turn implies  a substantial reduction 
in the X-ray brightness  of the gas structures created during the collision.

This behavior is in line with previous findings \citep{Ricker01,Poole06} and in 
particular with \citetalias{Zh15}, who argue that the two wings seen at $t=0.09$ 
after the pericenter passage in their model of Figure 1c are short-lived. 
In particular, for model R01V20 the X-ray map of Figure \ref{fig:planeHO_2}
  can be compared with the  maps presented by  
 \citet[][Figure 5 ]{Donnert14}  in his simulation study of El Gordo.
To properly compare the two merger models, it is necessary to identify among 
the various panels of Figure 5  the post-apocenter epoch when the secondary is 
falling  onto the primary, and the  relative distance  between the two clusters
is approximately of $\sim 500-700 \kpc$.
It is easily seen that this condition is  satisfied at 
$t\sim4.5$ Gyr, and a comparison between the two panels reveals the presence 
of very similar X-ray structures.

 The results presented here are valid for head-on collisions; nonetheless,
it should be emphasized that off-axis mergers are also clearly inconsistent
with a returning scenario.  This can be clearly seen by realizing that 
for these merger models the time
difference $\Delta t_{orb}$  between the pre- and post-apocenter epochs, at which 
must be satisfied the constraint $d_{DM}\simeq 700 \kpc$, are significantly 
higher  than for mergers with orbits having  zero angular momentum. 
For the fiducial models of  Section \ref{sec:opt} it is found 
$\Delta t_{orb}\sim 1-2 $ Gyr, whereas we have seen that the  post-collision 
X-ray structures are fast evolving and their lifetime lies in the 
range $\sim 0.1-0.2$ Gyr.

To summarize, the results of this section strongly support the outgoing 
scenario to explain the observed features of the merging cluster El Gordo. 
In the next section we investigate alternative possibilities 
to reproduce the observed offsets between the different mass components of
the El Gordo cluster.

\begin{figure*}[!ht]
\centering
\includegraphics[width=0.95\textwidth]{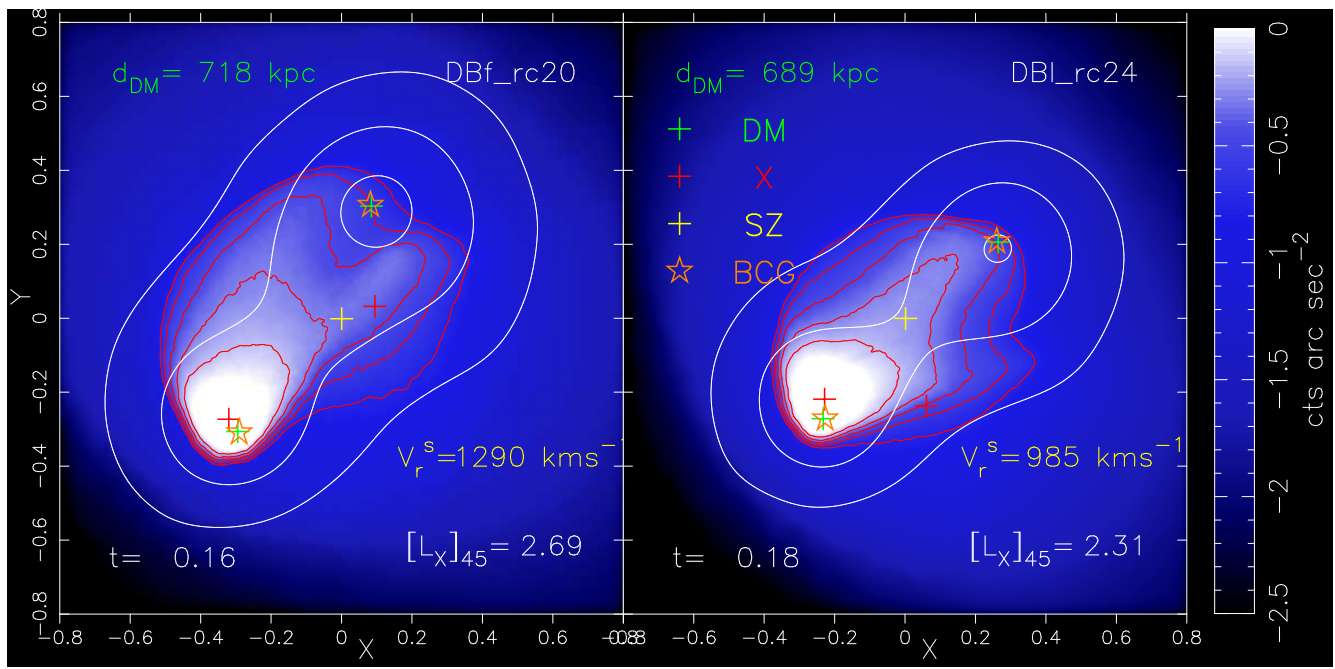}
\caption{  X-ray surface brightness images extracted from simulations 
of models DBf\_rc20 and DBl\_rc24. The merging parameters are the same as models
Bf\_rc20 and Bl\_rc24, respectively, but the runs were performed with 
 initial conditions  including  for the two clusters 
a stellar mass component describing a BCG 
(see Section \ref{subsec:icstar}).
In each panel the  open orange stars mark the  projected spatial location
 of the mass centroids of the star particles representing 
the BCGs. The value of $V^s_r$ is
relative mean radial velocity along the line of sight  between the 
  SE and NW  BCG components.
The meaning of the other symbols 
and the viewing direction is the same of the maps shown in 
Figure \ref{fig:planeA1}.
\label{fig:planeB4}
}
\end{figure*}

\subsection{Mergers with BCGs and self-interacting DM }
\label{sec:sidm}

One of the most interesting aspects of the merging cluster El Gordo is the 
measured  offset  between the projected positions of the 
X-ray emission and DM centroid of the SE cluster ($\sim 100 \kpc$).
As already said in the Introduction, the relative position of the X-ray peak 
with respect to the mass distribution is in contrast with what is expected
from dissipative arguments.
Moreover, nonzero offsets are also measured \citep{Kim21} between the BCG to DM 
centroid ($\sim 60\kpc$) and the BCG to X-ray peak ($\sim 70 \kpc$) centroid. 

According to the 
``central galaxy paradigm'' \citep{Van05}, BCGs are  expected to reside at the 
center of the DM halo, thus  tracing the bottom of the potential.
 In their simulation study, \citet{Martel14} concluded that the paradigm
is statistically satisfied, but that exceptions are common.
From a large survey ($\sim 10,000$) \citet{Zitrin12}  found a relatively small
mean offset ($\sim 15 \kpc$) between the BCG and DM cluster centroids. 
More recently, \citet{Deprop21}  showed that a significant fraction of BCGs are 
offset from the X-ray centroid, but are still aligned with the mass distribution.

In order to investigate the impact of a massive merging cluster 
like El Gordo on the central galaxy paradigm   
(i.e., that the  BCGs are located at the cluster centers), we re-simulated two 
of the fiducial models outlined in Section \ref{sec:opt}.
We chose fiducial models Bf\_rc20 and Bl\_rc24, whose simulation results 
produced the best fitting to the data overall and which have a mass of the 
primary that  brackets the observational range. 

Our new models are designated DBf\_rc20 and DBl\_rc24.\ The initial condition 
setup is the same as that of their respective model in Section \ref{sec:opt}, but 
each cluster now has a stellar mass component added to  its initial mass 
distribution.  A numerical realization of the star density profiles
that are supposed to represent the BCGs is described in Section 
\ref{subsec:icstar}.
The stellar masses of the primary and secondary cluster that 
we obtain from Equation (\ref{mstar.eq})  are
$M^{(1)}_{\star} = 2.2 \cdot 10^{12} \msun $ 
  and $M^{(2)}_{\star} = 1.6 \cdot 10^{12} \msun $,  respectively.
 Table \ref{bsdm.tab} lists the main merger parameters of the two models.

Figure \ref{fig:planeB4} shows the mock X-ray images extracted from the two runs 
at the simulation time identified as the present epoch. It is easily seen 
 that adding a BCG stellar component to the two clusters does not modify 
 in any way the X-ray morphology of the mergers. The two X-ray maps are  very 
similar to the corresponding maps  shown in the Bf\_rc20 and Bl\_rc24 panels of
Figure \ref{fig:planeB1} and Figure \ref{fig:planeB3}, respectively.

About the dynamics of the BCGs  two important results emerge from
Figure \ref{fig:planeB4}. The first concerns the measured velocity offsets
 $V_r^s$  along the line of sight between the two BCGs of each merger model.
As can be seen from Figure \ref{fig:profHV}, these values are, relatively 
 speaking,  very 
close to  the   velocities  $V_r$ extracted from the DM particle velocities 
 of the corresponding model without a star component.
This confirms that in a  collisionless DM  merging scenario the BCG star
 velocities can be assumed to be a fair proxy for the DM particle velocities.

The  other important result is that the mass centroids of the BCGs appear 
to be strictly aligned with their corresponding DM mass centroids.
This indicates that the off-center positions of the BCGs that we observe today 
 were not generated by the merging  process during the cluster collision.  
According to \citet{Martel14}  the spatial offset between the BCG mass 
centroid and that of its parent cluster is produced in major cluster mergers, 
as a result of the collisions. 
If at $z=0$ the cluster has reached an equilibrium configuration, the BCG will 
have settled at the bottom of the potential and its off-center location 
will be much reduced or even absent. We return 
to the likelihood of such a scenario for the merging cluster El Gordo in the conclusions.

An alternative scenario that could explain the observed spatial offsets 
between the different mass components involves SIDM that is 
self-interacting,
 these signatures being a characteristic  of SIDM in major mergers.
These expected properties have led many authors to investigate cluster 
collisions  in an SIDM 
framework \citep{Mark04,Kah14,Kim17,Rob17,Rob19,ZuH19,Fis22}.

We investigate here the possible outcomes of SIDM cluster mergers, using the 
implementation
of the SIDM modeling  described in Section \ref{subsec:icsidm}.
 We performed  SIDM merging simulations for each of the two merger models 
of Figure \ref{fig:planeB4}, and we generically 
refer to these models as XDBf\_rc20 and XDBl\_rc24,
respectively. Three  SIDM simulations were carried out for each model, 
with  the SIDM cross-section assuming  the following 
values: ${\sigma_{DM}}/{m_X} =1,~2,$ and $5 \sxu$, respectively.
 In Table \ref{bsdm.tab}, we list all models with their 
respective initial merger parameters.

\begin{figure*}[!ht]
\centering
\includegraphics[width=0.95\textwidth]{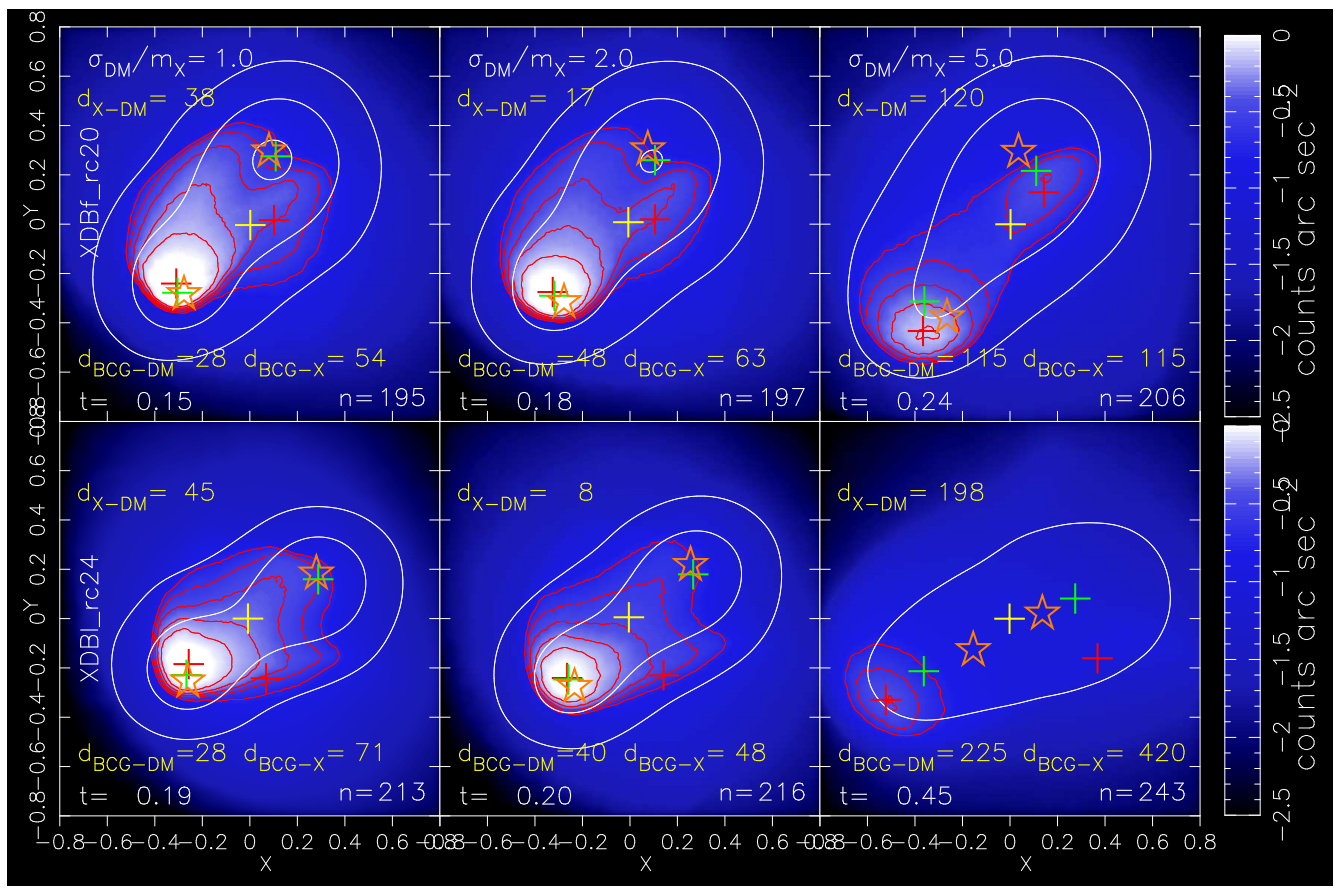}
\caption{ X-ray surface brightness maps extracted at the present 
epoch, $t,$ from two sets of SIDM  merging runs.
The simulations of the top (bottom)  panels have the same initial conditions 
of model DBf\_rc20 (DBl\_rc24) of Figure \ref{fig:planeB4}.
From left to right each row of panels corresponds to SIDM simulations performed 
for three  different values of the DM scattering cross-section: 
${\sigma_{DM}}/{m_X} =1,~2,$ and $5 \sxu$, respectively.
In each panel the distance $d_{\XDM}$ indicates the value in kpc of 
the projected  distance between the X-ray emission peak and the DM mass 
centroid, $d_{\BX}$ that between the mass centroid of the BCG galaxy and 
the X-ray emission peak, and finally $d_{\BDM}$ is the distance between 
the BCG and DM mass centroids. All of the centroids 
refer to the SE cluster. The meaning of the other symbols is the same
as in  Figure \ref{fig:planeB4}. 
 The thresholds and the spacing of the contour levels are the same
as those in Figure  \ref{fig:planeA1}.
In each panel the value of $n$  is the simulation step at which the map is 
extracted  and $t=0$ at the pericenter passage. The observer epoch is defined 
as in the other Figures.
\label{fig:planeSXa}
}
\end{figure*}

The mock X-ray maps extracted from the SIDM  runs are presented
in Figure \ref{fig:planeSXa}; the top panels are for
model XDBf\_rc20 and  the
bottom panels for model XDBl\_rc24. We also considered SIDM simulations with 
${\sigma_{DM}}/{m_X} =0.1 \sxu$, but the results are almost indistinguishable 
from the standard CDM mergers and are not shown here.

A first important conclusion to be drawn from the maps of Figure 
\ref{fig:planeSXa}  concerns the behavior of the X-ray gas morphology. 
In an SIDM scenario the interaction of DM leads to
an exchange of energy between the two clusters during the collision, which 
 implies the development of shallower DM potential wells
for the two DM halos \citep{Kap16,Kim17,ZuH19}.
This can be easily seen by contrasting the contour levels of the projected 
mass density: those displayed in the map of Figure \ref{fig:planeSXa} 
from  the ${\sigma_{DM}}/{m_X} =5 \sxu$  run of model XDBf\_rc20 appear much 
rounder 
than those exhibited in Figure \ref{fig:planeB1}  by their counterparts  from the 
standard CDM merger Bf\_rc20.

 The presence  of shallower DM potential wells leads in turn to a reduced 
resiliency  of the post-pericenter gas structures, 
which can now more easily escape from the potential wells of the original halos. 
As can be seen from Figure \ref{fig:planeSXa},  this effect is present in all of
the runs, but increases as ${\sigma_{DM}}/{m_X}$ gets higher and when
the mass of the primary gets lower. In fact, the bottom-right panel 
of Figure \ref{fig:planeSXa} shows that for ${\sigma_{DM}}/{m_X}=5  \sxu$
model XDBl\_rc24 has entirely lost its original X-ray morphology.
A complementary role in the process causing the disappearance of 
the post-pericenter gas structures is played by the  elapsed time  $t$ 
between the pericenter passage and the observer epoch. 
This time is much larger ($\sim 0.2 $ Gyr) than in the collisionless DM runs, 
and thus the X-ray structures generated during the collision are much weaker 
 when observed or even absent (see Section \ref{sec:return}).

This effect can be clearly seen in the panels of Figure \ref{fig:planeSXa}, where 
each 
panel reports the time $t$ since passing pericenter. 
For the XDBf\_rc20 simulations (top panels) the DM scattering cross-section 
increases from  ${\sigma_{DM}}/{m_X} =1\sxu$ up to ${\sigma_{DM}}/{m_X} =5\sxu$, 
while the time $t$ has a corresponding increase from  
$t=0.15$ Gyr up to $t=0.24$ Gyr. For  XDBl\_rc24 the effect is even more
pronounced, with $t$ now ranging from $t=0.19$ Gyr up to $t=0.45$ Gyr.
To better illustrate the difference from the standard CDM runs of 
Figure \ref{fig:planeB4}, we also report in each panel  the simulation step $n$.
From this one can recover the simulation time $t_n= n \Delta t_0$,   
at which the maps were extracted. 

We attribute this increasing dependence of the elapsed time, $t,$ on
the DM scattering cross-section, ${\sigma_{DM}}/{m_X,} $  to the mutual 
transfer of energy that occurs between the two DM halos during the collision.
In the off-axis mergers DBf\_rc20 and DBl\_rc24  the DM halos remain almost  
unperturbed, but in the SIDM runs the exchange of energy can be significant
when ${\sigma_{DM}}/{m_X} \sim 5 \sxu$.
As a  consequence of  these dissipative processes,  the post-pericenter 
cluster bulk velocities are now strongly reduced with respect the 
standard CDM model.  As shown below, this in turn will also impact on the 
BCG bulk velocities.

Another important feature that should be  present when observing 
merging clusters in an SIDM scenario are the expected offsets between the 
different mass components. Because of the dissipative behavior of the DM, 
during the merger the collisionless galaxy component   is expected to
decouple from its parent DM halo \citep{Kim17}. Moreover, the position of 
the X-ray peak 
in relation to that of the DM centroid is also expected to differ in a 
significant way from that  found in the corresponding standard CDM merger 
\citep{Tulin18}.

In the mock X-ray maps of Figure \ref{fig:planeSXa}, for each merging run we 
indicate in the corresponding panel the projected spatial positions of the 
different centroids.  As in the  X-ray maps of Figure \ref{fig:planeB4}, the red 
crosses indicate the positions of the X-ray peaks, while  the green crosses 
 are for the DM centroids.
The orange stars refer to the centroids of the BCGs.

From the merging simulations of Figure \ref{fig:planeSXa}, there are 
several noteworthy features that are clearly recognizable
by looking at the relative positions of the  different centroids.
As expected, the positions of the BCG centroids are now decoupled from  those
of their parent DM halos. Moreover, for the SE cluster the position of the 
X-ray peak is highly significant.   
 For the standard CDM mergers of Figure \ref{fig:planeB4} the DM peak is located farther 
away from the barycentre than the X-ray centroid, and the relative offset is negative. 
 For the  SIDM merging simulations of Figure 
\ref{fig:planeSXa} the relative separation between the two peaks initially 
decreases as ${\sigma_{DM}}/{m_X} $  increases, 
and becomes positive for  ${\sigma_{DM}}/{m_X} =5\sxu$. 
 For these runs  the X-ray centroid  is no longer 
lagging behind the mass centroid but is now clearly ahead of it.
\begin{table*}
\caption{IDs and initial merger parameters of the two SIDM merging simulations 
of Figure  \ref{fig:planeSXc}.$^{a}$}
\label{clsdm.tab}%
\centering
\scalebox{0.9}{   
\begin{tabular}{cccccccccc}
\hline
Model & $M^{(1)}_{\star} [\msun]$  & $M^{(2)}_{\star} [\msun] $ 
& $N^{(1)}_{\star} $ & $\varepsilon_{\star} [\kpc] $ & $\rcp [\kpc]$ &
$ \zeta$ & $ \sigma_{DM}/m_X [\sxu]$  & $(f_{g1},f_{g2})$ \\
\hline
XDBf\_{sa} &  $ 2.2 \times 10^{12}$ & $ 1.6 \times 10^{12}$ & 
        $ 16,785 $ &  9.5 & 406 & 1.7  &  5 &  (0.12,0.12) \\
XDBf\_{sb} &  $ 2.2 \times 10^{12}$ & $ 1.6 \times 10^{12} $ & $16,785$ &
  9.5 & 290 & 2.4  &  4 &  (0.12,0.14) \\
\end{tabular}
}
\begin{flushleft}
{\it Notes.} {$^a$ Columns from left to right:
 ID of the merging model, stellar mass of the BCG of the primary,  the same mass 
but for the secondary, number of star particles for the primary, 
  gravitational softening  length  of the star particles,
  gas core radius  of the primary, 
 dimensionless parameter $\zeta=r_s/r_c$,   
SIDM cross-section per unit mass, 
primary and secondary cluster gas mass fractions $f_g$
at  $r_{200}$. 
For the two SIDM merger models the  collision  parameters
 are those of model Bf in Table \ref{clparam.tab}: 
$\{ M^{(1)}_{200},~q,~V,~P \} =   
\{  1.6 \cdot 10^{15} \msun, ~2.32, ~2,500 \kms, ~ 600 \kpc \}  $
}.
\end{flushleft}
\end{table*}


For the same value of ${\sigma_{DM}}/{m_X}$,  the relative separations between 
the different centroids should be smaller in model XDBl\_rc24 than in model 
XDBf\_rc20,  the collision velocity  being  smaller ($V=2,000 \kms$)  in 
the former than in the latter ($V=2,500 \kms$).
Leaving aside the  ${\sigma_{DM}}/{m_X} =1\sxu$  runs, for which  this 
dependence is too weak to be detected, the two merging simulations with  
${\sigma_{DM}}/{m_X} =2\sxu$  (middle panels of Figure \ref{fig:planeSXa}) 
effectively reproduce this behavior. 

However, for the merging runs with 
${\sigma_{DM}}/{m_X} =5\sxu$ the separations between the centroids of 
model  XDBl\_rc24 are much larger that those of model  XDBf\_rc20.
We argue that this discrepancy is due to an observational effect, with
the elapsed time, $t,$  being much larger in model  XDBl\_rc24 ($t\sim 0.45$ Gyr) 
than in model XDBf\_rc20 ($t\sim 0.24$ Gyr).

 The same time difference is responsible of another noteworthy feature of 
model XDBl\_rc24 with ${\sigma_{DM}}/{m_X} =5\sxu$. 
From the bottom-right panel of Figure \ref{fig:planeSXa} it can be seen that 
this run exhibits negative BCG-DM offsets, that is, the DM peaks are farther 
from the barycentre than the BCGs.

As the clusters collide,  the DM particles will lose momentum through 
scatterings. The resultant drag force will lead to a deceleration of the DM halos 
and in turn to positive offsets between the DM and galaxy components
\citep{Mark04,Kim17,Rob17,Fis21,Fis22}.
However, this description applies only in proximity of the first pericenter
 passage. At later times the gravitational pull of a DM halo will act 
on the corresponding galaxy component as a restoring force, which will thus 
decelerate and reduce its bulk motion \citep{Harvey14}.

According to the relative strength of the  DM cross-section and the elapsed 
time t,  from initially positive the BCG-DM offsets   can then become 
negative, as seen in the bottom-right panel of Figure \ref{fig:planeSXa} 
for model XDBl\_rc24. 
These behaviors are in line with what is expected in a merging SIDM scenario, 
 due to the general nonlinear dependence on ${\sigma_{DM}}/{m_X} $  
of the relative distance between  the DM centroid and that of the other mass 
components.

A significant  conclusion that can be drawn from these findings is that 
model XDBf\_rc20 with ${\sigma_{DM}}/{m_X} =5\sxu$  is the only merging simulation
that is able produce  relative separations within the observational range 
($\sim 50-100 \kpc$). 
 The validity of such a conclusion being clearly dependent on the 
statistical significance of the size of the observed offsets
for the El Gordo cluster, we will come back  later to this issue.

As a consequence of this conclusion, the SIDM model adopted here  
to simulate the merging cluster El Gordo has no free parameters.\ The only 
admitted values for the  DM scattering cross-section are in the range 
${\sigma_{DM}}/{m_X} \sim 4-5\sxu$. 
The reason for this limitation in SIDM merging simulations of the cluster 
El Gordo is that the  relative separations between the centroids of 
the different mass components are locked to the distance 
$d_{DM}$ between the DM mass centroids, which fixes the present
epoch through the constraint  $d_{DM}\sim 700 \kpc$.

This is made clear by the middle panels of  Figure \ref{fig:planeSXa}, 
where for ${\sigma_{DM}}/{m_X} =2 \sxu$ both models exhibit a relative separation
between the different centroids of about $\sim 50 \kpc$, at approximately 
the same epoch $t$. However, for ${\sigma_{DM}}/{m_X} =5 \sxu$  only model 
 XDBf\_rc20 is able to reproduce the requested observational range for the 
relative separations, while model XDBl\_rc24 completely  fails to match the data.
This behavior as ${\sigma_{DM}}/{m_X} \rightarrow 5  \sxu$  is due to the 
relative impact of DM collisions, which is much higher in model XDBl\_rc24
than in XDBf\_rc20. As previously explained, for model 
XDBl\_rc24 this value in turn implies lower 
post-pericenter DM bulk velocities and a much higher elapsed time $t$.
We note that these conclusions also work in the other direction: 
for SIDM merging simulations of the El Gordo cluster, the best  agreement
with data is obtained  only for a primary mass of about  
$ M^{(1)}_{200} \sim1.6 \cdot 10^{15} \msun $.

Motivated by these results, we now turn our attention to the SIDM 
merging simulation of model XDBf\_rc20 with ${\sigma_{DM}}/{m_X} = 5  \sxu$,
which from the top-right panel of Figure \ref{fig:planeSXa} appears as the only
 viable candidate able to reproduce many observational properties of the
cluster El Gordo.  
From the  contour levels of the surface mass density (see the 
caption of Figure \ref{fig:planeA1}), we estimate for the SE cluster of this
model a peak value of the  projected  DM  mass surface density  of about 
 $\Sigma_{DM} \sim 0.3   \dms $.   The  cluster's scattering depth  is then 
$\tau=\Sigma_{DM} {\sigma_{DM}}/{m_X} \sim 1.5 $, which is marginally above the 
optically thick limit \citep{Mark04}. However, this is a peak value. By using an 
average,  one would obtain a lower value for $\tau$.

As already outlined, for this merging run
the  DM potential wells  of the two halos are strongly perturbed 
by the presence of  SIDM.
The top-right panels of Figure \ref{fig:planeSXa} demonstrate 
that with respect  the standard CDM merging simulation DBf\_rc20, 
the impact of SIDM on the post-pericenter X-ray structures  
is significant when ${\sigma_{DM}}/{m_X} = 5 \sxu$.

We are then forced to reconsider the original initial condition  parameters 
of the two baryonic halos of model Bf\_rc20, in order to reproduce 
in a ${\sigma_{DM}}/{m_X} = 5 \sxu$ SIDM merging simulation
the observed twin-tailed X-ray morphology of the El Gordo cluster. 
The initial collision parameters of this simulation are those of model Bf in 
Table \ref{clparam.tab}: $\{ M^{(1)}_{200},~q,~V,~P \} =   
\{  1.6 \cdot 10^{15} \msun, ~2.32, ~2,500 \kms, ~ 600 \kpc \} $.
\begin{figure*}[!ht]
\centering
\includegraphics[width=0.95\textwidth]{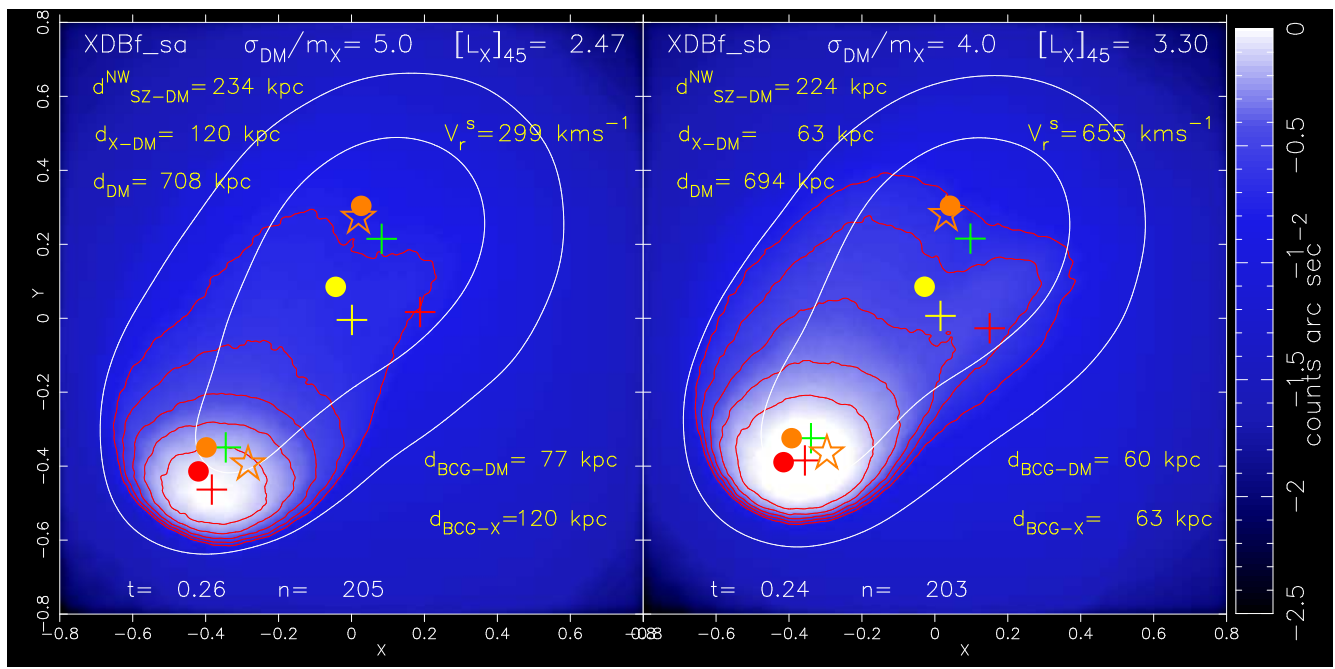}
\caption{X-ray images extracted at the observer epoch from two SIDM merging 
simulations. The initial conditions of the two runs are  given  in
Table \ref{clsdm.tab}. The meaning of the distances shown in each panel
is the same of the ones reported in Figure \ref{fig:planeSXa}, with 
the exception of $d^{NW}_{\SDM}$, which is the projected distance between the SZ 
peak and the DM mass centroid of the NW cluster.
As in Figure \ref{fig:planeB4}, $V^s_r$ is line-of-sight relative mean radial 
velocity between the two BCGs. The thresholds  and the spacing of the contour 
levels are the same as in Figure \ref{fig:planeA1}.
The filled circles indicate the peak locations from several observations, as
taken from Figure 6 of  \citet{Kim21}. Their spatial positions have been
normalized to the relative distance from the mass centroids (see the main text). 
The color coding of the circles is the same of the associated crosses, 
which indicate the projected positions of the corresponding centroids
as extracted from the simulations. 
\label{fig:planeSXc}
}
\end{figure*}

For this set of initial collision parameters we performed  SIDM merger 
simulations for runs with both ${\sigma_{DM}}/{m_X} = 5 \sxu$ and 
${\sigma_{DM}}/{m_X} = 4 \sxu$.
We found that in order to produce realistic X-ray morphologies with    
these simulations,   initial conditions with larger cluster gas fractions and
  gas core radii of the primary  are required than for standard CDM runs.

For the ${\sigma_{DM}}/{m_X} = 5 \sxu$ SIDM merging run 
it was found after  some testing that the best results are obtained 
for  $\rcp \sim 400 \kpc$  and cluster gas fractions 
$(f_{g1},f_{g2})=(0.12,0.12)$. We label this model as XDBf\_{sa}.
Similarly, for the ${\sigma_{DM}}/{m_X} = 4 \sxu$  simulation it was found that 
  $\rcp \sim 300 \kpc$  and  $(f_{g1},f_{g2})=(0.12,0.14)$ are the optimal 
settings
to match the observed X-ray morphology. We label this model as XDBf\_{sb}.
We report in Table \ref{clsdm.tab} the most relevant initial merger 
parameters of the two models; in both  cases the initial gas density profile 
of the secondary is the same as in  the standard CDM runs.
 Figure \ref{fig:planeSXc} shows the mock X-ray maps extracted from  these
two SIDM merger simulations.
\begin{figure*}
\centering
\includegraphics[width=17.2cm,height=8.2cm]{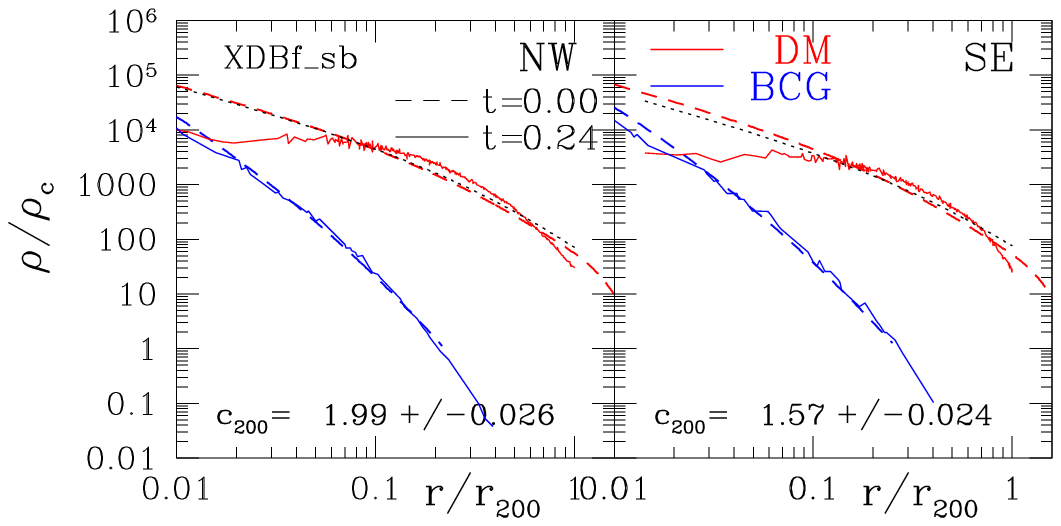}
\caption{Radial  density profiles  of the stellar and DM components
of  the merger model XDBf\_sb  of Figure \ref{fig:planeSXc}.
The left (right) panel is for the NW (SE) cluster. 
Solid lines refer to the present epoch, after the core passage,  and dashed lines 
correspond to the start of the simulation. The profiles of the different 
components are indicated by different colors,  red and blue are for 
 DM and  BCG, respectively. The origin of the profiles is centered on the 
 centroid of the corresponding mass component. 
In each panel is reported the value of $c_{200}$, together with its statistical
 error, as obtained by using  an NFW profile  to fit the cored DM density 
profile.  The adopted fitting procedure is the same as that described by 
  \citet{Du08}, the corresponding best-fit density profile is indicated by the 
black dots.
\label{fig:profDMh}
}
\end{figure*}

For the two merging clusters, a specific signature of  SIDM  is 
the expected offsets of their DM 
centroids from those of the other cluster mass components. For this reason,  
in each panel of Figure \ref{fig:planeSXc} we report for each simulation
the relative distance between the different centroids.

Specifically, for the SE cluster  we indicate the distance of the 
BCG to the DM centroid ($d_{\BDM}$), the BCG to the X-ray peak ($d_{\BX}$) 
and the DM to the X-ray peak ($d_{\XDM}$). For the NW cluster we report only 
the distance 
of the SZ to the DM centroid ($d^{NW}_{\SDM}$),  but not that of 
the DM to the X-ray peak because 
the gas structure of the primary was destroyed during the collision.

In order to assess how well the SIDM simulations can reproduce the 
observed offsets, we also show in Figure \ref{fig:planeSXc} the measured
positions of the different centroids.
Data points are extracted from the two panels of Figure 6 of \citet{Kim21}, the 
top (bottom) panel is for the NW (SE)  cluster. 
For the sake of comparison  we report here for the two clusters 
the relative location of several peaks, 
as measured with respect the positions 
 $(\Delta_x, \Delta_y)=(0.0^{\prime \prime} , 0.0^{\prime \prime} )$
 of the cluster mass centroids. 
 For the SE cluster the measured relative positions  are : 
${\Delta}^{\rm{SE-BCG}}=(-6.8^{\prime \prime} , 0^{\prime \prime} )$ and
${\Delta}^{\rm{SE-X}}=(-9.6^{\prime \prime} , -8.1^{\prime \prime} )$; 
 while for the NW cluster: ${\Delta}^{\rm{NW-SZ}}=(-16^{\prime \prime} ,
 -16.7^{\prime \prime} )$  and 
${\Delta}^{\rm{NW-BCG}}=(-7.15^{\prime \prime} , 11.4^{\prime \prime} )$. 

The NW cluster does not possess a BCG, but its galaxy number density peak
shows a large offset with respect the mass centroid position  \citep{Jee14}.
 With the notation ${\Delta}^{\rm{NW-BCG}}$ we then indicate the relative 
position of the galaxy number peak as shown in the top panel of
 Figure 6 of \citet{Kim21}. The measured relative offsets are indicated in 
Figure \ref{fig:planeSXc} with filled circles, their positions  
are with respect that of the DM centroids and 
in accordance with the adopted cosmology 
we set the spatial conversion of angular coordinates 
to $1^{\prime}\simeq470 \kpc$. 

 The absence  of observational errors  for the measured  offsets is clearly
a critical issue in order to set constraints on DM cross-sections as well as 
to rule out the null hypothesis of zero offsets.
According to \citet{Kim21}, the statistical significance 
of the X-ray peak offset  from the SE mass centroid is at 
  $\sim 2 \sigma$ level. Similarly, the galaxy number density peak is offset 
  from the NW (SE) mass centroid at $\sim 2 \sigma$ ($\sim 6.6 \sigma$)  level.
The statistical significance of the SZ peak  offset from the NW mass peak is
 at $\sim 3.7 \sigma$ level, although \citet{Kim21} caution that 
the interpretation of this offset is complicated by the   low angular 
resolution of the beam size ($\sim 1^{\prime}.4$). Overall, these thresholds 
allow us to  exclude with high significance zero size offsets.

For the two SIDM merging simulations the X-ray maps of Figure \ref{fig:planeSXc} 
show the spatial locations of the different centroids, as extracted 
from the simulations. Their relative positions constitute one of the  most 
interesting results of our study,  and we suggest that these findings 
 pose a severe challenge to the collisionless standard CDM model.

As already outlined,  the most striking feature 
is the position for the SE cluster of the X-ray peak,  which is 
now leading  the DM mass centroid.  Moreover, the relative 
distance $d_{\XDM}$ between the two centroids  varies 
from $d_{\XDM} \sim60 \kpc $  (${\sigma_{DM}}/{m_X} = 4 \sxu$)
up to $d_{\XDM} \sim120 \kpc $  (${\sigma_{DM}}/{m_X} = 5 \sxu$).
The agreement of this range of  values with the measured offset
$d^{SE}_{X-DM} \sim 100 \kpc $ can be considered significant, given the 
observational uncertainties (see below).

The offset  of the SZ peak is similarly affected in an SIDM 
merging scenario.  For the standard CDM merging simulations of 
Figure \ref{fig:planeB4}, 
the  distance of the SZ peak from the NW DM  centroid is  about 
$d^{NW}_{SZ-DM} \sim 330 \kpc $. This  value is similar to that 
found by   \citetalias{Zh15} for their model B and higher than 
the measured value  of about $\sim 150 \kpc$ \citep{Jee14}. 
However, for the SIDM simulations of 
Figure \ref{fig:planeSXc}  this distance is now reduced to
$d^{NW}_{SZ-DM} \sim 230 \kpc $,  in better agreement with the data.

Another important result that follows from an SIDM  scenario 
in major cluster mergers is the expected presence of BCG-DM offsets, 
with  BCG centroids that exhibit miscentred positions 
with respect to the centers of their  parent DM halos. 
Figure \ref{fig:planeSXc}  shows that 
according to value of ${\sigma_{DM}}/{m_X}$, 
the BCG-DM offsets for the SE cluster lie in the range 
$d_{\BDM} \sim 60-120 \kpc $, which agrees fairly well 
with the observed offset of about $\sim 60 \kpc$.
We note that for the NW cluster, the position of the simulated BCG strictly coincides
with that of the observed peak galaxy number density. 
\begin{figure*}
\centering
\includegraphics[width=17.2cm,height=8.2cm]{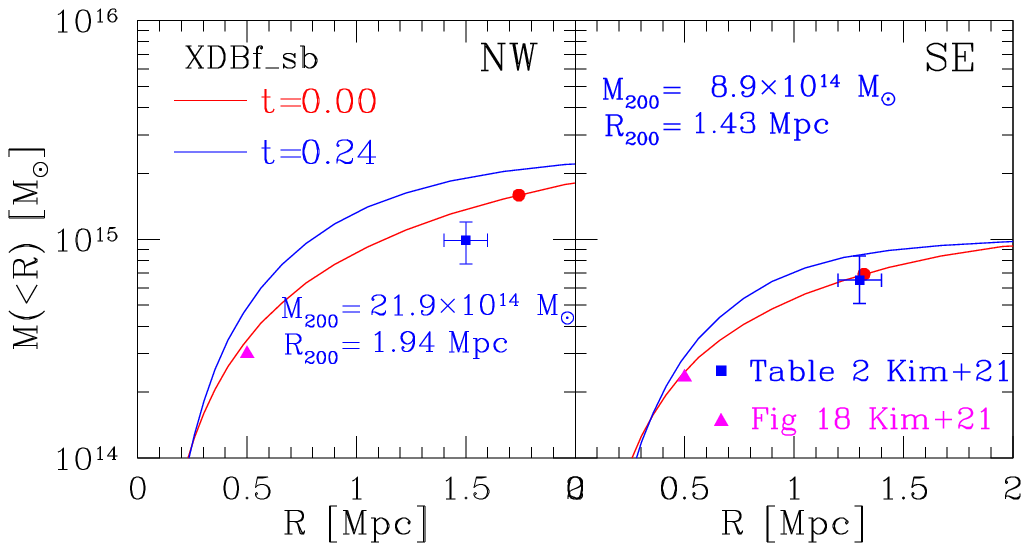}
\caption{Total mass as a function of radius, R, for the two clusters of the merger 
model XDBf\_sb. As in  Figure \ref{fig:profDMh}, the left (right) panel is for the 
NW (SE) cluster. 
Solid blue lines refer to the observer epoch,  and red lines at $t=0$. The quoted 
values of $M_{200}$ and  $r_{200}$ are determined according to Equation 
(\ref{mcl.eq}) from the cumulative mass profile .
In each panel the filled blue square refers to the lensing mass estimate of the 
corresponding cluster, as reported in Table 2 of \citet{Kim21}; 
 from the cumulative mass profile of their Figure 18  we also extrapolated the 
cluster mass within $500 \kpc$, as indicated by the  magenta triangle.
 For the same cluster the point denoted by a filled red circle 
corresponds   to the initial values of  $M_{200}$ and  $r_{200}$,
as given in Tables \ref{clparam.tab} and \ref{subcl.tab} for the  merger model 
Bf\_rc20.
\label{fig:profDMc}
}
\end{figure*}

In SIDM mergers,  positive BCG-DM offsets
are expected  after the first pericenter passage. 
This effect can be interpreted as a result of the deceleration of the DM 
component, this in turn is due to the  exchange of energy  between the DM halos 
that occurs during the cluster collision.  As mentioned 
in the  discussion of the results from the ${\sigma_{DM}}/{m_X} =5 \sxu$ SIDM 
simulations  
of Figure \ref{fig:planeSXa}, positive BCG-DM offsets are a transient phenomena.
As a BCG  starts to climb out of the cluster potential 
 after the pericenter passage, it will experience a gravitational pull 
\citep{Harvey14} from the DM mass component that  will reduce its 
forward momentum  as well as the size of the offset.

This restoring force  has the effect of reducing the BCG bulk velocities in 
the SIDM mergers of Figure \ref{fig:planeSXc},  similarly reducing the 
relative mean radial 
velocity  $V^s_r$ along the line of sight  between the  two SE and NW  
BCG components. As a result the BCG star velocities 
can  no longer be considered a fair proxy for the DM particle velocities.
This bias greatly helps reduce the discrepancy between the 
mean  relative velocities  $V_r$,  extracted from the DM particle velocities 
of the fiducial models of Section \ref{sec:opt} (see  Figure \ref{fig:profHV}), 
and those estimated from   galaxy-based spectroscopic measurements 
 \citep{Men12}.

In Figure \ref{fig:planeSXc} the measured velocity offsets $V_r^s$  between 
the two BCGs  are reported for the two  merger models. 
The values of $V_r^s$  range from $V_r^s\sim 300  \kms$ for 
 ${\sigma_{DM}}/{m_X} = 5 \sxu$   up to 
  $V_r^s\sim 650\kms$  when ${\sigma_{DM}}/{m_X} = 4 \sxu$.
These values can be contrasted with that given  by
 \citet{Men12}, who   reported  $V_r^s=598 \pm 96 \kms$ for the 
relative radial velocity of the SE cluster  with respect the NW component.
We therefore conclude that the SIDM  merger models of the El Gordo cluster 
presented here, predict mean  relative radial velocities between their stellar 
components  in much better agreement with data
than in the collisionless CDM cases.

Although the simulations of the two SIDM merger models are undoubtedly able 
to reproduce the observed offsets and the relative cluster peculiar 
velocity,  the two X-ray maps of Figure \ref{fig:planeSXc}  
 nonetheless exhibit  a twin-tailed X-ray morphology  much less defined than that displayed 
  by the merger model Bf\_rc20 in Figure \ref{fig:planeB1}. In fact, for model 
 XDBf\_sa  the tails behind the secondary are almost absent. 

Moreover, the X-ray emission in the outer regions behind the secondary is 
significantly less than in model  Bf\_rc20. This can be clearly  seen by the location of the third contour level ($2.9\cdot 10^{-1} \ctsn $ from the 
inside), which in the X-ray map of model Bf\_rc20 (top-right panel 
of Figure \ref{fig:planeB1}) has a minimum along the line joining the two DM 
centroids  at about $\sim 600 \kpc$ from the X-ray peak of the secondary. 
For the merger simulation of model XDBf\_sb this minimum is much less 
pronounced and at about $\sim 400 \kpc$ from the centroid of the X-ray 
emission.
\begin{figure*}[!ht]
\centering
\includegraphics[width=0.95\textwidth]{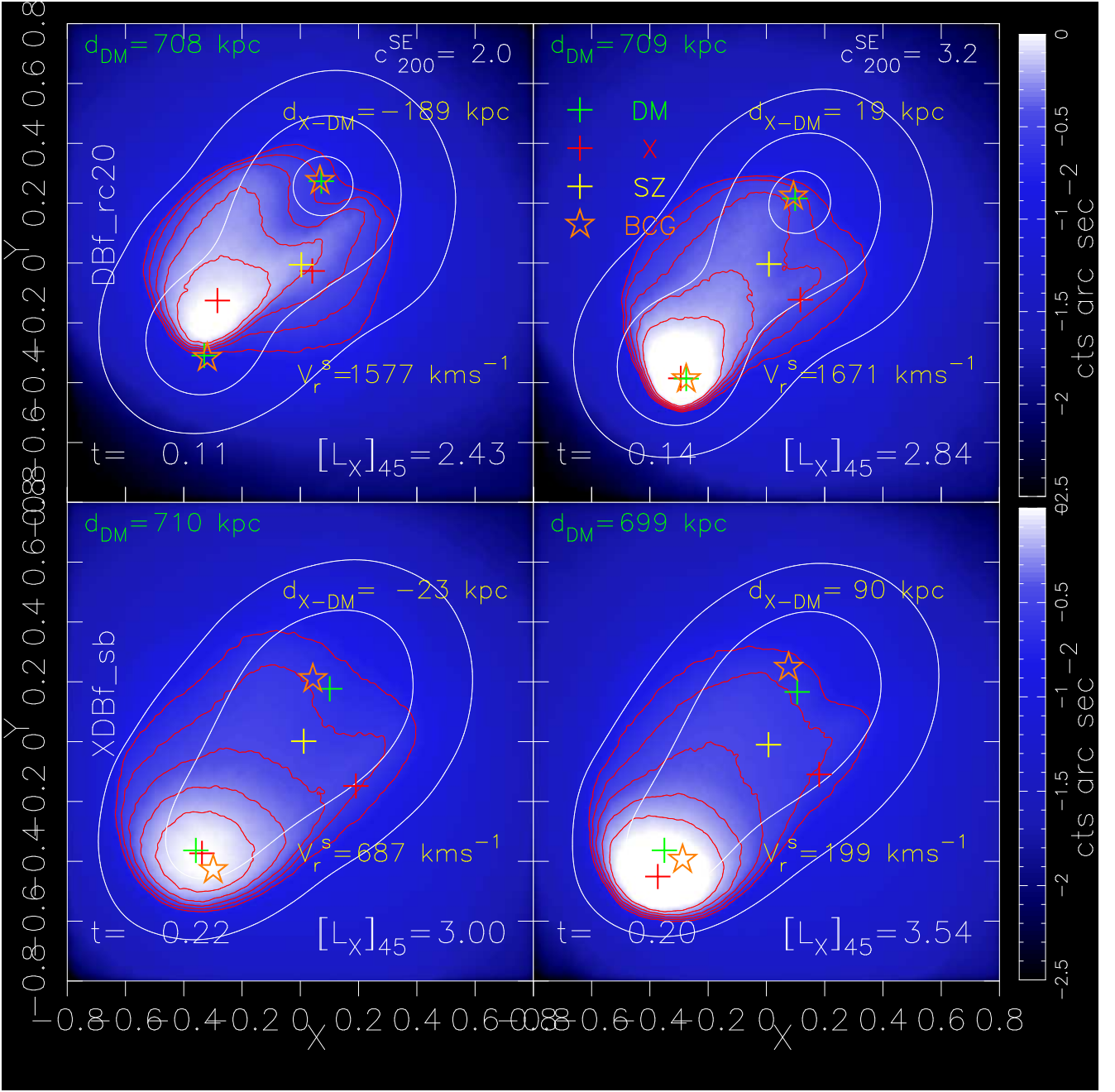}
\caption{X-ray images extracted  from two alternative simulations of 
the merger models DBf\_rc20 ( Figure \ref{fig:planeB4} ) and   XDBf\_sb  
( Figure \ref{fig:planeSXc} ).
 The only difference in the simulation setup is the 
 initial value of the  SE cluster halo concentration parameter, $c^{SE}_{200}$.
From Equation  (\ref{cfit.eq}) the concentration parameter of these models
 is fixed to  $c^{SE}_{200}=2.682$, while for this study we set here 
 $c^{SE}_{200}=2$  (left panels) and $c^{SE}_{200}=3.2 $ 
(right panels). Top panels are for model DBf\_rc20 and bottom panels 
 for model XDBf\_sb. 
A negative  $d_{\XDM}$ offset indicates that the 
X-ray peak is trailing the  DM centroid of the SE cluster.
\label{fig:planeB5}
}
\end{figure*}

These shortcomings of the two SIDM merger models  in reproducing  the 
observed X-ray morphology are present even after the adoption of 
initially higher  gas fractions and of larger gas scale radii
for the primary.
These initial condition parameters were purposely introduced to 
compensate for the much weaker resiliency of the gas structures 
 seen in the SIDM merging runs of Figure \ref{fig:planeSXa}.

We argue that this difficulty of SIDM merging simulations  in matching 
the observed X-ray morphology is strictly related to the complex 
dynamical interplay that occurs during the cluster collision between 
baryons and SIDM. We suggest that in order to extract from SIDM 
simulations X-ray maps more in agreement with observations, 
it is necessary to explore a much wider parameter space 
for the initial gas density profiles of the two clusters. 
The results presented here for these simulations 
should then be regarded as preliminary. We 
 postpone to  a future work a much more thorough study on the impact of SIDM 
on the X-ray morphology in cluster mergers.

{
Finally, one of the most significant prediction of DM collisionality is the 
flattening of the DM core density. It is therefore important to assess in SIDM 
mergers the consistency of the predicted mass distribution with 
observed mass profiles.
The scattering rate of DM particles is higher in 
high-density regions, and due to the collisions between DM particles 
the net effect is  a transfer of energy toward the inner regions.
This effect is evident from the  left panel of Figure \ref{fig:sidm_halo}, in 
which for different values of ${\sigma_{DM}}/{m_X} $ are shown 
at $t=1$ Gyr the density profiles of an isolated SIDM halo.
 As predicted,  the flattening of density  at inner radii  is higher as 
${\sigma_{DM}}/{m_X}$ increases.  

The development of flatter DM density profiles 
 is however expected to be significantly increased  in SIDM  cluster mergers.
 This is because the scattering probability (\ref{pijdm.eq}) depends on the 
relative velocity between DM particles.
 At  the pericenter  the relative velocity between the two clusters
 can be as high as  $v_{rel}\sim 4,000 \kms$, and  this in turn implies much 
 higher scattering rates than in the case of an isolated halo.

To explicitely verify this expectation, Figure \ref{fig:profDMh} shows 
for our model XDB\_sb the DM radial density profiles of each cluster.
For the sake of comparison  these are depicted at the beginning of the simulation
and after the core passage, at the observer epoch $t=0.24$ Gyr.
Additionally, at the same epochs  we also show  the density profiles of the BCG 
stellar component for the corresponding cluster.
The profiles of the NW and SE cluster are plotted separately in the left and 
right panel, respectively.

As expected, after the core passage the inner density profile of the two 
DM halos exhibits  a significant flattening   toward the cluster center.
Accordingly, for both of the DM density profiles  we now find  a core radius 
 of approximately  $ \sim 200 \kpc$.
It is interesting to make a comparison against observational estimates  
of the concentration parameters, as  obtained by fitting  the density profiles
 of each DM halo using an NFW model. We refer to 
  \citet{Du08} for a description of the adopted fitting procedure.
From the best fit of each halo we obtain for the concentration parameters 
and  their formal statistical errors $\concI=1.99\pm 0.026 $ and 
$\concII=1.57 \pm 0.024$ for the NW and SE cluster, respectively.

From their WL study on El Gordo cluster, in their Table 2 \citet{Kim21} report 
for the two clusters
$\concI(K)=2.54^{+0.97}_{-0.41}$ and $\concII(K)=3.2^{+1.17}_{-0.74}$.
For the NW cluster the observational estimate $\concI(K)$ of the 
concentration parameter is  marginally consistent at $1 \sigma$ level with 
the best fit value $\concI$, as extracted from the SIDM merger 
simulation XDBf\_sb.
On the contrary, for the SE cluster the estimate $\concII(K)\sim 3$ of 
\citet{Kim21} is significantly  higher than the best fit parameter 
$\concII\sim1.6$  predicted by our SIDM merger simulation.

The reason for such a large difference is in the adopted initial conditions 
of our merger simulations.
For the SE cluster the initial value $c_{200}=2.682$ of the concentration 
parameter is predicted by the \citet{Du08} relationship (\ref{cfit.eq}),  
and it was used in the merger simulations performed here to 
set up  the initial conditions of the SE DM halo.

For the SE cluster this then implies that  our initial setting of 
$c_{200}=2.682$ is much lower and marginally consistent, within the quoted 
uncertainties, with the  observational estimate $\concII(K)\sim 3$ of 
 the DM halo concentration parameter.
Therefore,  it is not surprising  that we find such a discrepancy  between the 
value of $\concII(K)$ and the best fit value  $\concII\sim1.6 $  extracted 
from  the SIDM merger simulation XDBf\_sb.

However, the presence of such a  discrepancy should not be considered
 particularly significant. 
As pointed out by \citet{Kim21}, estimates of the concentration 
parameters are poorly constrained due to their  low sensitivity 
to the lensing signal; moreover, the $M-c$ relation (\ref{cfit.eq}) has been 
derived by averages from a simulation ensemble of DM halos, so its 
application to a post-collision system such as El Gordo should be considered 
with caution.

Additionally, in each panel of  Figure \ref{fig:profDMh}  we also 
show the BCG density profile of the corresponding cluster.
The radial profile of each BCG is plotted at the present epoch (solid blue lines) 
and at $t=0$ (dashed blue line). We can see that for both of the halos 
the two profiles exhibit very
small differences, thus showing that the internal star distribution of the 
BCGs remains relatively unperturbed during the collision.

It is important to note that the profiles are centered on the mass centroids 
of the BCGs when they are calculated, so the location of 
their origin varies with time and a meaningful comparison between the BCG and DM
density profiles can only be performed  at $t=0$.
Using a sample of seven, massive, relaxed galaxy clusters
 \citet{New13} showed that in cluster central regions ($\simlt 10 \kpc$) 
the star mass component becomes dominant.

There would therefore  be  a disagreement between these findings and the 
behavior at $t=0$ of the radial density profiles of the two mass components. 
For each halo, Figure \ref{fig:profDMh}   shows that  at 
$r\sim10^{-2} r_{200}\sim 15 \kpc$
the star density $\rho_{\star}  $ is lower than $ \rho_{DM} $ by a factor 
of $\sim 4$. 
This discrepancy is partly due to a resolution effect, 
 the effective radius  of Section \ref{subsec:icstar}
being set to $r_e= 60 \kpc$. Moreover, 
 the profiles of Figure \ref{fig:profDMh}  are presented 
in the chosen radial range of $0.01 \simlt r/r_{200}\simlt 1$, 
  at the spatial resolution limit of $r=10^{-2.8} r_{200}\sim 3 \kpc$ the
 star density is  found to be  higher than the DM density:
 $\rho_{\star} \sim 8 \cdot 10^5 \rho_c \sim 4 \rho_{DM} $.

A more meaningful comparison can be made between
the cumulative total mass profiles of the two clusters as obtained from our
merger simulation XDBf\_sb, against those found by the lensing reconstruction
of the El Gordo cluster mass profiles.
For the NW and SE cluster  Figure 18 of \citet{Kim21} shows separately   
the total halo mass with a given radius, these profiles can be contrasted with 
the corresponding ones  extracted from  simulation XDBf\_sb and plotted in 
Figure \ref{fig:profDMc}. 
A visual comparison between the two Figures overall shows  that 
for each cluster the corresponding 
 mass profiles exhibit approximately the same radial behavior.
However, at large radii the mass profile of the NW cluster  
recovered from lensing data is not described well  by the corresponding profile
extracted  from the simulation, which  tends to be 
 systematically higher by about a factor of $1.5-2$.

To be more quantitative in each panel we quote the values of $M_{200}$ and 
 $r_{200}$, as determined 
according to Equation (\ref{mcl.eq}) from the cumulative mass profile of the 
cluster at the redshift $z=0.87$. These values can be compared with the 
corresponding mass estimates of \citet{Kim21}:  
 $ M^{NW}_{200}(K) =9.9^{+2.1}_{-2.2} \times 10^{14} \msun $ and 
 $ M^{SE}_{200}(K) =6.5^{+1.9}_{-1.4} \times 10^{14} \msun $ for the NW and SE 
cluster, respectively.

From the right panel of Figure \ref{fig:profDMc} it can be seen that 
the mass $ M^{NW}_{200} (t=0.24 ~{\rm Gyr}) \sim 2 \cdot 10^{15}  \msun $ 
of the NW cluster is a factor of $\sim 2$ higher than the corresponding 
lensing mass estimate $ M^{NW}_{200}(K) $. 
This is not surprising,  as the initial mass of the primary of model XDB\_sb 
was set to $ M^{(1)}_{200} =1.6 \cdot 10^{15} \msun$ because  its collision 
 parameters are the same  of model Bf\_rc20. 
According to the findings of Section \ref{sec:optoffa},  this is 
 the fiducial models that best matches the observations. 
The collision parameters of this model  (see Table \ref{clsdm.tab}) 
have  been  therefore specifically chosen 
to perform the  SIDM merger simulations  of Figure \ref{fig:planeSXc}.

For the SE  cluster we find 
$ M^{SE}_{200} (t=0.24 ~{\rm Gyr}) \sim 9 \cdot 10^{14}  \msun $, marginally
inconsistent at $1\sigma$ level with the estimate $ M^{SE}_{200}(K) $.
This mild tension is easily understood to originate from 
the assumed initial  masses  $ M_{200} $  for the two colliding clusters.
From Figure \ref{fig:profDMc} the present cluster masses at $r=r_{200}$ 
are found to be higher by a factor of $\sim 30\%$ compared to their
initial values. This is due to the flattening of the DM inner density profiles
 during the merger, and at the present epoch  this in turn leads   
to an average higher DM density at large cluster radii.
For the SE cluster initially $M^{SE}_{200}= 6.5 \cdot 10^{14} \msun $ and
  $r_{200} \sim 1.3 \mpc$; however due the presence of cored DM profiles
after the core passage, 
Equation  (\ref{mcl.eq}) is now solved  for 
$M^{SE}_{200}\sim 8.9 \cdot 10^{14} \msun $ and $r_{200} \sim 1.43 \mpc$.
 
These discrepancies between the mass profiles tend to become less severe
at inner radii.
We chose $R=500 \kpc$ as the minimum cluster radius at which the cluster mass
profiles of Figure 18 \citep{Kim21} start to appreciably diverge.
The masses of the NW and SE cluster at this radius are then estimated 
to be about $ M^{NW}_{\it extr} (R<500\kpc) \simeq 3 \times 10^{14} \msun $
and $ M^{SE}_{\it extr} (R<500\kpc)  \simeq 2.34 \times 10^{14} \msun $, 
respectively.

These masses  are indicated in the corresponding panels of 
 Figure \ref{fig:profDMc}  with magenta triangles.
For the NW cluster the difference between the cluster mass 
$ M^{NW}_{\it sim}  (R<500\kpc) $, as extrapolated from the mass profile of the 
simulation,    and $ M^{NW}_{\it extr}$ is still significant.
However, for the SE cluster the accord between 
$ M^{SE}_{\it sim} (R<500\kpc) $ and $ M^{SE}_{\it extr}$ is now 
much better.

We then conclude that present data do not allow us to firmly rule out 
the presence of cored DM profiles in the El Gordo cluster.
Moreover, we argue that these differences between the predicted and 
reconstructed mass profiles can be interpreted in terms of  the adopted initial 
conditions for model XDB\_sb.
For instance, we suggest that setting the initial mass of the secondary 
for the merger model XDB\_sb 
to approximately  $\sim 20\%$ less than  the initial adopted value  of 
$M^{(2)}_{200}\sim 6.5 \cdot 10^{14} \msun $ 
will provide, for the SE cluster, a mass profile in  better agreement 
with observational estimates.

However, it must be stressed that WL-based cluster mass measurements
\citep[see][and references cited therein ]{Um20} are subject to a  number of
uncertainties.  To break the mass-sheet degeneracy \citep{Um20} cluster masses 
must be estimated by assuming a specific mass profile, such as an NFW.
In a recent paper,  \citet{lee23} showed that in major  cluster mergers 
($\sim 10^{15} \msun$) departure from the assumed mass profile can be significant 
and the subsequent WL mass bias can be as high as $\sim 50\%$. They conclude that 
in massive mergers previous WL mass estimates can be significantly overestimated. 

Nonetheless, we suggest that in the case of El Gordo cluster such a large range 
of mass uncertainties should be taken with some caution.
This is motivated by recent,  independently based,  SL mass estimates. 
\citet{Kim21} found a WL total mass estimate of $\sim 2 \cdot 10^{15} \msun$ for 
the El Gordo cluster,
this is  roughly in the same range as the  corresponding mass estimates 
derived for El Gordo cluster  from recent  SL 
 analyses \citep{Cam23,Die23}.
}

\subsection{Testing the dependence of the X-ray to DM peak offset on the 
initial halo concentration and 
gas density profile of the SE cluster }
\label{sec:ictest}

As we have seen in Section \ref{sec:return}, a returning scenario fails
in reproducing the observed displacement $d^{SE}_{X-DM}$  
 of the X-ray peak from the SE mass centroid. On the contrary, the results of 
the previous Section demonstrate that a collisional DM scenario 
can naturally account for the observed offset. 
The size at the observer epoch of the observed offset $d^{SE}_{X-DM}$   
depends however on a number of additional factors that can complicate 
the interpretation of  $d^{SE}_{X-DM}$ in terms of an  SIDM scenario.

The first physical effect during the merger that can have a significant impact
on the motion of the cool core of the secondary is the depth of the potential well
 of the hosting DM halo. 
An initially shallower inner DM density profile 
will in turn lead to a less deep   potential in the cluster inner regions, 
thereby 
easing the motion of the SE cool-core during the cluster merger.

In order to assess the impact of a shallower DM potential on the final 
displacement of the SE cluster cool core, in this section we present 
results from merger simulations with initial condition setups  identical to those 
of previous runs, but with different values of the initial halo concentration 
parameter $\concII$ of the SE cluster.
To this end, we use as reference  the merger models DBf\_rc20 of
  Figure \ref{fig:planeB4}  and XDBf\_sb  of  Figure \ref{fig:planeSXc} .
 For both of these  models, for the mass of the secondary
one has from  Tables \ref{bsdm.tab}  and \ref{clsdm.tab}  
$M^{SE}_{200}= 6.9 \cdot 10^{14} \msun $. 
According to Equation (\ref{cfit.eq}), for these simulations  the initial  halo 
concentration parameter  is then set to $\concII=2.682$.

For each of  these two merger models,  we ran two  alternative simulations. 
The  initial condition setup of these simulations is identical to that of the 
parent simulation; the only difference is in the 
 initial value of the  halo concentration parameter for the SE cluster.
For the two simulations we set  this parameter to  
$\concII=2\equiv c_{\it low}$   and 
 to  $\concII=3.2\equiv c_{\it high} $, respectively.
This interval of values for the  concentration parameter 
 approximately corresponds to what is predicted by the standard deviation
 ($\sigma_c/c \sim 0.3$) of the  $M-c$ relation \citep{Bha14}, 
and it was specifically   chosen 
in order to bracket, for different shapes of the DM potential, 
the motion of the SE cool core during the merger.

Figure \ref{fig:planeB5} shows the X-ray maps, as extracted at the observer epoch
 from our  alternative merger simulations. We first discuss  results from 
the two alternative simulations (top panels) of  the parent merger model 
DBf\_rc20. 
A striking feature that emerges from the simulation results is that the final
size of the  $d^{SE}_{X-DM}$ offset is heavily influenced by the shape of the 
density profile of the hosting DM halo.
The top-left panel shows that for the DBf\_rc20 run with $\concII=2$ the X-ray 
peak is  trailing the DM centroid,  with the relative offset now negative 
($d^{SE}_{X-DM}\simeq -190 \kpc $). 

 This offset between the X-ray peak and the SE mass centroid was already 
present  in 
the merger models DBf\_rc20 of Figure \ref{fig:planeB4}, as expected   
because of the different behavior during the merger  between the dissipative 
ICM and the collisionless DM component. However, the size of $d^{SE}_{X-DM} $ 
is now much 
larger, thus showing that for the SE cluster the motion of the cool core during 
the merger  is very sensitive to the choice of relatively low values of 
$c_{200}$.
On the contrary, the top right panel of Figure \ref{fig:planeB5} shows that when 
$\concII=3.2 $ the size of $d^{SE}_{X-DM} $ is unaffected by the choice of 
$c_{200}$ and is very   similar to that of the standard run. 
\begin{figure}[!ht]
\includegraphics[width=8.2cm,height=8.2cm]{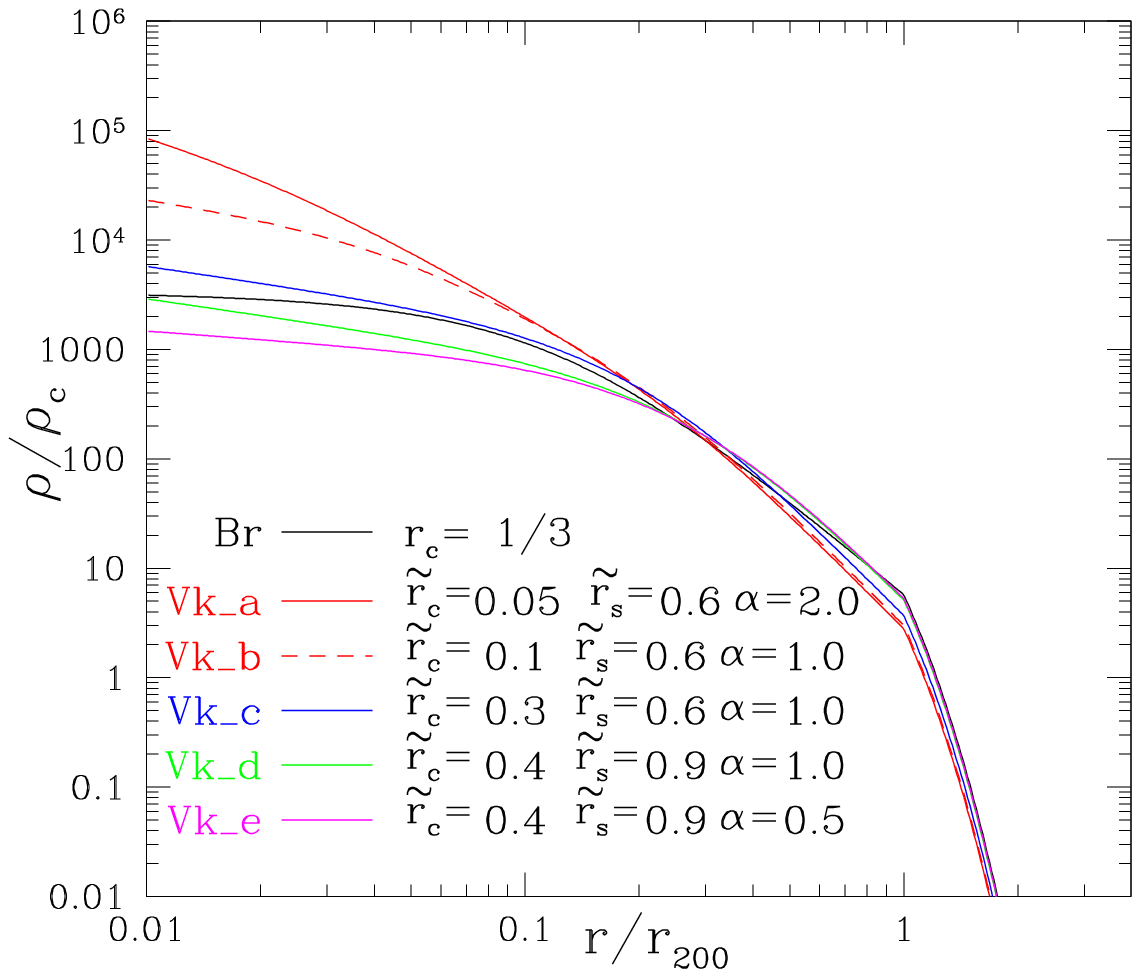}
\protect \caption{ Initial gas density profiles of the SE cluster for 
different models
of the radial gas distribution.  The Br profile refers to the Burkert 
 profile (\ref{rhogins.eq}), with the gas core radius  set to $r_c=r_s/3$ 
and  $r_s$ the 
NFW scale radius (\ref{rhodmins.eq}).
This is the initial gas density profile of the SE cluster 
 previously adopted  in all of the merger simulations. 
The label Vk generically refers to the \citet{Vikh06} gas density profile
described by the form  (\ref{rhogSE.eq}), different curves are for different
settings of the profile parameters 
$\{\protect\rtilde_c,~\protect\rtilde_s,~\alpha \}$.
The radii $\protect\rtilde_c$ and  $\protect\rtilde_s$ are in units of $a_s$, 
the  scale 
radius of the generalized 
sNFW profile (\ref{rhoDMSE.eq}). All of the profiles  are normalized to 
$f^{SE}_g=0.14$ at $ r = r_{200}$, as in model
XDBf\_sb of Table \ref{clsdm.tab}.
\label{fig:profSE}
}
\end{figure}

Model  XDBf\_sb is an SIDM merger model with $\sigma_{DM}/{m_X} = 4 \sxu$, 
the bottom-left and bottom-right  panels of Figure \ref{fig:planeB5} show the X-ray maps 
of the corresponding $\concII=2$   and  $\concII=3.2 $ simulations, respectively.
An important feature that is clearly seen from the bottom-left panel
 of Figure \ref{fig:planeB5}  is that for the  $\concII=2$  simulation of model 
 XDBf\_sb    the $d^{SE}_{X-DM} $ offset is now much smaller than 
in the corresponding  DBf\_rc20 run.

This behavior is the consequence of two competing effects: the easier motion of 
the cool core during the merger because of the shallower DM density profile 
(as in the BDf\_rc20), and 
the  deceleration of the DM halos (as in XDBf\_sb) due to the drag force  
induced  during the cluster collision by the scatterings of the 
 DM particles, with  the subsequent loss of momentum.

From the bottom-right panel of Figure \ref{fig:planeB5},  
 it can be seen that for the $\concII=3.2 $ simulation of model XDBf\_sb
the  $d^{SE}_{X-DM} $ offset is insensitive 
to the choice of $c_{200}$,  as in the corresponding  DBf\_rc20 run.
We argue that for values of $\concII \simgt 2.7  =c_{d_{X-DM}}$,  the size of the 
final offset between the X-ray and DM peak is weakly  sensitive to the 
form of the inner DM potential  of the SE cluster.


We also conclude that these findings are  not affected by observational 
constraints on the range of allowed   values for $\concII$.
As previously discussed, from their WL study on El Gordo cluster \citet{Kim21} 
obtained $\concII(K)=3.2^{+1.17}_{-0.74}$ for the posteriors of the 
concentration parameter of the SE cluster. 
Their Figure 7  (bottom panel) shows that their  estimate
$\concII(K)\sim3.2$  coincides approximately with $c_{\it high} $.
This shows that the $1 \sigma$  lower limit on $\concII(K)$  and the 
previously derived threshold $c_{d_{X-DM}} $  have approximately the same value, 
thus indicating that the size of the offset $d^{SE}_{X-DM} $ is not affected 
by  observational uncertainties for  $\concII$.
\begin{figure*}[!ht]
\centering
\includegraphics[width=0.95\textwidth]{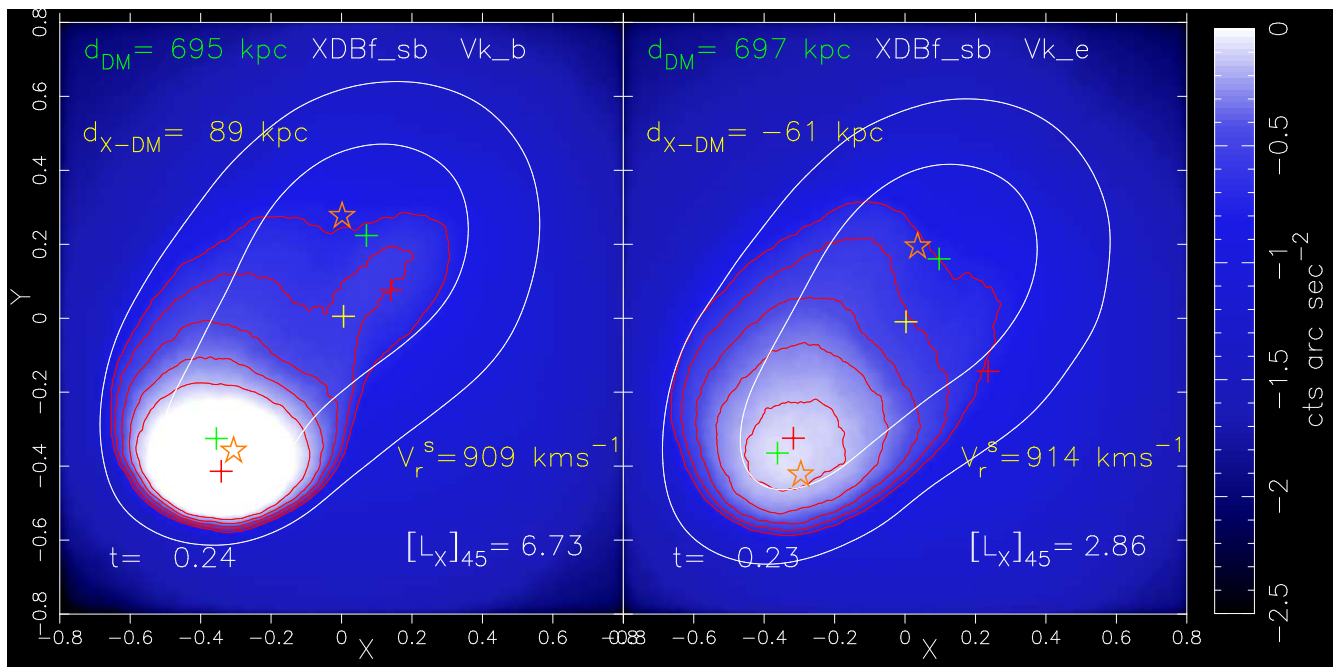}
\caption{X-ray maps extracted  from two alternative simulations of 
the merger model  XDBf\_sb.
 The only difference in the simulation setup is now in 
 the initial gas density profile for the SE cluster.
This is described according to models Vk\_b (left panel) and Vk\_e (right panel) 
of Figure \ref{fig:profSE}.
\label{fig:planeB7}
}
\end{figure*}

The  size of the  X-ray cool core  is another important factor 
 that contributes to its  final displacement  from the SE mass peak.
As outlined in the Introduction,  the simplest physical interpretation  of the 
El Gordo  merger \citep{Men12,Jee14} is  a scenario in which 
  the original  gas core of the primary has been destroyed by
the collision  with the compact core of the secondary, during its motion
 from NW to SE. Clearly, a very dense, compact, cool core will be less prone 
to the ram pressure experienced during its motion through the ICM of the primary.

Following \citetalias{Zh15}, for the initial gas density profile of the secondary
 we adopted the \cite{Bu95} density profile (\ref{rhogins.eq}),  and set
the gas core radius to  $r_c=r_s/3$. This is a reasonable choice that has been 
found to reproduced fairly well the observed X-ray luminosity $L_X$ (see 
Table \ref{clres.tab}), but there are other choices as well. 

We want then to study the impact on the final position of the SE cluster X-ray 
 peak of initially different gas density profiles.
To this end we used the merger model XDBf\_sb  as reference and  
we  performed simulations with the same initial condition setup but with different
 initial gas density profiles for the SE cluster.
In particular, we chose to model the gas density profile of the secondary
according to the modified beta profile of \citet{Vikh06}:

\begin{equation}
\begin{array}{lll}
\rho^2_g&\propto &\dfrac{(r/\rtilde_c)^{-\alpha}}
{\left[ 1+(r/\rtilde_c)^2 \right ]^{3\beta-\alpha/2}}
\dfrac{1}{\left [ 1+(r/\rtilde_s)^\gamma  \right ]^{\epsilon/\gamma}}\, .\\
 \end{array}
 \label{rhogSE.eq}
 \end{equation}

The shape of this profiles is controlled by six parameters :
 $\{ \alpha,~\beta,~\gamma,~\epsilon,~\rtilde_c,~\rtilde_s \}$, thus allowing 
a large amount of freedom in regulating its radial behavior.
Following \citet{Cha22}, we set here $\beta=2/3$, $\gamma=3$ and $\epsilon=3$.
The size of the cool core is then controlled by $\rtilde_c$ and $ \alpha$, while
the scale radius  $\rtilde_s$ regulates  the transition to 
$ \propto r^{-( 3 \beta + \epsilon/2)}$.

To ease the comparison with previous  findings \citep{Cha22}, it is convenient 
to express the radii $\rtilde_c$ and $\rtilde_s$ in units of  $a_s$,  the scale 
radius   of the  generalized ``super-NFW'' (sNFW) profile
\citep{Lill18}:  

\begin{equation}
\begin{array}{lll}
\rho_{DM}(r)&= &\dfrac{3M}{16 \pi a_s^3} \dfrac{1}{(r/a_s)(1+r/a_s)^{5/2}}\, .\\
 \end{array}
 \label{rhoDMSE.eq}
 \end{equation}The scale radius, $a_s,$ can be recovered from the  NFW  concentration 
parameter $c_{NFW}$ according to the linear relation   
 $c_{sNFW}=1.36+0.76 c_{NFW}$, where 
 $c_{sNFW}\equiv3 r_{200}/2a_s$ is the sNFW concentration parameter.  For the SE
 cluster of model  XDBf\_sb  one has $r^{SE}_{200} \sim  1.32 \mpc$, 
  $\concII = 2.68$  and $a_s \sim 0.58 \mpc$. 

 Figure \ref{fig:profSE}  presents for different values  of the 
profile parameters $\{\rtilde_c,~\rtilde_s,~\alpha \}$  a set of initial gas 
density profiles,  as  described by Equation (\ref{rhogSE.eq}). 
For the sake of comparison we also show the Burkert  profile (\ref{rhogins.eq}), 
previously adopted for the SE cluster in all of the merger models.
In the panel, the radii $\rtilde_c$ and  $\rtilde_s$ are expressed in 
units of $a_s$
and  all of the profiles  are consistently normalized to 
$f^{SE}_g=0.14$ at $ r=r_{200}$.

As can be seen, the size of inner cool core region and the cuspiness of the 
central gas density peak can be regulated by an appropriate choice of the 
profile parameters and range from that of a compact cool core (model Vk\_a: 
$\{\rtilde_c=0.05,~\rtilde_s=0.6,~\alpha=2 \}$ ) to that of an almost 
non-cool core (model Vk\_e: 
$\{\rtilde_c=0.4,~\rtilde_s=0.9,~\alpha=0.5 \}$ ).  
This  dependence of  the radial profile behavior  on the  set of  parameters 
$\{\rtilde_c,~\rtilde_s,~\alpha \}$  can be compared with similar plots 
shown in Figure 12 (left panel) of  \citet{Cha22}.

To explore the dependence of the observed  offset  $d^{SE}_{X-DM} $ 
on the initial gas density profile of the SE cluster,  as for the merger models 
of Figure \ref{fig:planeB5} we ran two  alternative simulations  
 of model  XDBf\_sb.  
The  initial condition setup of these simulations is identical to that of the 
parent simulation, the only difference being now  in the choice of
 initial gas density profile for the SE cluster.

In order to bracket a plausible range for the compactness of the cool core,
for the two alternative simulations  we chose  to model the gas density profile 
of the secondary according to models Vk\_b and Vk\_e of Figure \ref{fig:profSE}, 
respectively.  
The X-ray maps extracted from these  two alternative merger simulations are 
shown in Figure \ref{fig:planeB7}.

For model Vk\_b the left panel of Figure \ref{fig:planeB7} shows 
a  $d^{SE}_{X-DM} $ offset nearly identical  
 ($d^{SE}_{X-DM} \sim 90 \kpc$) to that of   the parent simulation XDBf\_sb  of 
 Figure \ref{fig:planeSXc}. This demonstrates that, above a certain threshold, 
the size of the final offset of the X-ray peak with respect to the DM centroid
 can be considered insensitive to  the cuspiness of the  SE cluster inner gas 
density.
On the other hand, it worth noting that for this model the X-ray luminosity 
($L_X \simeq  7\cdot  10^{45} \ergs  $) is now higher by a factor of $\sim 3$ 
with respect the corresponding observational range as reported in Table 
\ref{clres.tab}.

For model Vk\_e (right panel of Figure \ref{fig:planeB7}),
 the $d^{SE}_{X-DM} $ offset  is now negative  ($d^{SE}_{X-DM} \sim -60 \kpc$),   
with  the X-ray peak now  trailing the DM centroid.
We interpret this behavior as  a consequence of the much less  compact gas 
 density peak with respect that of model Vk\_b, this implying a larger ram 
pressure force,  and in turn  a larger deceleration,  
 experienced by the secondary's cool core  during the cluster 
collision   \citep{Ricker01,Poole06,Poole08,Mi09}.

To summarize, our results demonstrate that both the halo concentration parameter 
and the initial shape of the gas density profile of the SE cluster have an 
impact on the  offset $d^{SE}_{X-DM}$  of the X-ray peak relative to 
that of the  DM.
We found that  both of these dependences could act to reduce  
 the offset size, but are however unable to increase 
 it  above the observational estimate.

Our conclusion is therefore that the interpretation  
 of the observed offset between the X-ray peak and the DM centroid in an 
SIDM scenario remains valid. The alternative configurations of the 
initial condition setup analyzed here for the SE cluster are found to be
  ineffective in increasing the size of the offset.

Finally, it can be seen from Figure \ref{fig:planeB7} that very compact cool core 
gas density profiles can produce  $d^{SE}_{X-DM}$ offsets in good accord 
with the measured value, but they fail to account for the observed 
 X-ray luminosity.
This strongly suggests that the previous used  
Burkert profile (\ref{rhogins.eq}), with the gas core radius  set to $r_c=r_s/3$,
can be considered a good approximation to the  gas density profile of the SE 
cluster.

\section{Summary and conclusions}
\label{sec:discuss}
  We have constructed a large set of N-body/hydrodynamical 
binary cluster merger simulations aimed at studying  the  merging  cluster 
 El Gordo.  Specifically, the main goal of this paper is to assess whether  it is 
possible for merger simulations to reproduce the observed twin-tailed X-ray 
morphology while satisfying  other observational constraints.

To this end, we  considered a wide range of initial conditions, 
performing both off-axis and head-on merging runs. The initial 
conditions of the binary merging clusters were constrained to be 
consistent with recent lensing-based mass measurements \citep{Die20,Kim21}. We 
imposed the following range of values for the mass of the primary: 
$ 10^{15} \msun  \simlt M\simlt  1.6 \cdot 10^{15} \msun $.
A summary of our main results is as follows:

(i) The observed  twin-tailed X-ray morphology, as well as other observational 
constraints,  are well  matched  by the off-center fiducial merger models  
Bf\_rc20, Bg\_rc20, and Bl\_rc24  (Section \ref{sec:opt}), which have collision 
velocities and impact parameters in the range 
$2,000 \kms  \simlt V \simlt 2,500 \kms$  and 
$600 \kpc \simlt P \simlt 800 \kpc$, respectively.

These findings demonstrate  that in simulations with cluster masses lower 
than previously considered \citep{Donnert14,Molnar15,Zh15,Zh18},   
it is possible to consistently reproduce the observed twin-tailed X-ray 
morphology seen in the cluster El Gordo.
Finally, the observational constraints  discussed in 
Section \ref{sec:opt}  tend  to favor merger models with  a primary mass of about $  M\simeq  1.6 \cdot 10^{15} \msun $.
  
(ii) One of the most interesting  aspects of the galaxy cluster El Gordo is the 
odd location in the SE cluster of the X-ray peak, which in the outgoing scenario 
seems to be ahead of the DM lensing peak. To resolve this issue, a returning 
scenario was invoked \citep{Ng15} in which the merging cluster is 
observed in a post-apocenter phase and the X-ray peak is trailing  
the DM peak.

This scenario was extensively investigated (Section \ref{sec:return}). We constructed  a large set of head-on merging collisions  aimed
at reproducing the observed twin-tailed X-ray morphology of the El Gordo 
cluster. For the adopted initial collision  parameter space, the results shown in  
Figure \ref{fig:planeHO_2} demonstrate that a returning scenario should be 
considered  untenable, as the morphology  of all of the X-ray maps extracted 
from the simulations  
 appears inconsistent with the observed X-ray morphology.
 We argue in Section \ref{sec:return} that this behavior is strictly related to 
 the lifetime of the post-pericenter X-ray structures, which turned out to 
 be much less than the elapsed time needed for the  DM component of the 
 SE cluster to reach the apocenter and return.

 We conclude that the likelihood  of head-on collisions matching 
 the observational constraints for the merging cluster El Gordo is very low, 
 given the wide range of initial merging parameter space that has been considered.


(iii)  We re-simulated two of the fiducial models (Bf\_rc20 and Bl\_rc24) by adding a 
star particle distribution to the initial mass components of each of the two halos,  with the purpose of  representing  the BCG's  
mass contribution to the initial halo profiles.
The  resulting images for these two runs (models DBf\_rc20 and DBl\_rc24) 
are shown in Figure \ref{fig:planeB4}. The positions of the centroids 
of the different mass components clearly indicate  that,  in both of the 
mergers, the collision between the two clusters  has not produced 
any appreciable offset between the positions of the BCG centroids  relative 
to that  of the  DM halos.

In a collisionless CDM scenario, this leaves us with the problem of explaining 
 the presence and  magnitude of the BCG to DM 
offset seen in the El Gordo cluster. It must be emphasized that 
galaxy-DM and BCG-DM offsets are relatively rare but 
have been measured in other merging clusters
\citep[see, for example, Table 2  of][]{Kim17}.

According to \citet{Martel14}, such offsets 
are produced as a result of  violent cluster collisions
that strongly perturb  the  clusters and leave them in a 
non-equilibrium state, with  the BCGs being displaced from
the cluster centers. While in principle this scenario cannot be ruled out 
as an explanation of  the observed BCG to DM offset seen for the El Gordo SE cluster,
  we nonetheless argue that it appears difficult for the gas structure of the 
SE cluster to maintain its integrity after  a cluster collision
that must be  at the same time violent enough to move the BCG from its 
original position.

 Finally, it should be emphasized  that the observed BCG to DM offset 
constitutes  another serious shortcoming for
 the  returning scenario: as shown by the simulations presented in 
Figure \ref{fig:planeB4}, this offset  would  remain unexplained  in  
such a scenario. 

(iv)  Motivated by these findings, we studied the
results extracted from SIDM simulations of the El Gordo cluster, 
in which DM interactions are  described by a simple model 
 with a velocity-independent elastic DM cross-section (Section \ref{sec:sidm}).

The most important striking feature that emerges from the SIDM merging models
of the El Gordo cluster is that the observed offsets between the different 
mass components are well reproduced by  the simulations
 XDBf\_{sa} and XDBf\_{sb} (Table \ref{clsdm.tab}). These simulations 
 have a DM cross-section  in the range ${\sigma_{DM}}/{m_X} \sim 4-5\sxu$
and initial collision parameters of  model Bf\_rc20  
(Table \ref{subcl.tab}). 

Another significant aspect of these merger models is the value of 
 the mean relative  line-of-sight  radial velocity between the 
two BCGs, which in contrast with previous runs (see Figure \ref{fig:profHV}) 
is now on the 
order of several hundred $\kms$  and in much better 
agreement with data. As discussed in Section \ref{sec:sidm}, this is a
direct consequence of the 
 gravitational pull  experienced by the two BCGs 
because of the dissipative interactions now present between the two DM halos.

Although these models are not able to reproduce the observed 
X-ray morphology as well as in the standard CDM runs (Section \ref{sec:opt}),
 we nonetheless argue that these findings provide significant support for the 
possibility that the DM 
behavior in the El Gordo cluster exhibits  collisional properties.
Additionally, it must be stressed that a significant advantage of the SIDM
merging models is that they offer a natural explanation for all of the observed
offsets, while in the standard CDM scenario the observed separations between
the different centroids remain either unexplained (X-ray peak to DM) or 
fraught with difficulties (BCG to DM).

From these conclusions follow two main consequences that need to be discussed.
The first is that the two SIDM models of Table \ref{clsdm.tab} do not 
admit free parameters; the observed  offsets are well reproduced 
only for a DM scattering cross-section in the range 
${\sigma_{DM}}/{m_X} \sim 4-5\sxu$ and initial collision parameters 
given by model Bf\_rc20. This is a direct result of the post-pericenter 
evolution in an SIDM scenario, for which the observed separations between
the different centroids  are locked to the measured distance, 
$d_{DM}$, between the DM mass centroids, which must  be 
about $d_{DM}\sim 700 \kpc$ at the present epoch.

As a result, these observational constraints cannot be satisfied by simply 
rescaling the value of ${\sigma_{DM}}/{m_X} $,  if one assumes masses 
for the merging cluster lower than those in model Bf\_rc20: 
$( M^{(1)}_{200},~M^{(2)}_{200}) = ( 1.6, 0.69 ) \cdot 10^{15} \msun $.
This value of the primary's mass is higher than that obtained from recent
lensing-based estimates \citep[see Table 2 of][]{Kim21}, but still 
within the confidence limits. If more accurate measurements were to 
yield tighter limits, the SIDM merger models 
 presented here would clearly be in jeopardy.

The other striking feature of the SIDM merger models XDBf\_sa and  XDBf\_sb 
is that the best match to the data is obtained only for values of 
${\sigma_{DM}}/{m_X} $ in the very narrow range between 
$\sim 4\sxu$ and  $\sim 5\sxu$.  Such values are strongly excluded by the 
present upper limits on the SIDM cross-section, 
which on galaxy  cluster scales  have been derived independently  
by various authors \citep{Rob17,Rob19,Kim17,Wittman18,Harvey19,Shen22,Cross23}. 
These upper limits are at best of order unity 
(${\sigma_{DM}}/{m_X} \simlt 1 \sxu$), although \citet{Wittman18} claim 
that the constraint can be relaxed up to ${\sigma_{DM}}/{m_X} \simlt 2 \sxu$
by properly averaging galaxy-DM offsets in a sample of merging clusters.

For example, in the case of the Bullet Cluster, the absence, within the 
observational errors on the centroid positions, of an observed 
galaxy-DM separation \citep[$\simlt 20 \kpc$, ][]{Randall08} 
has been used in SIDM merging simulations \citep{Mark04,Randall08,Rob17} 
to constrain the allowed values of the DM cross-section to lie
below ${\sigma_{DM}}/{m_X} \simlt 1 \sxu$.
This evidently implies that an SIDM merging simulation of the Bullet Cluster, 
with the values derived here for  the DM cross-section 
(${\sigma_{DM}}/{m_X} \simlt 4-5 \sxu$),  would lead to simulation results 
clearly incompatible with the observational constraints.

 However, the difficulty of quantitatively assessing the 
statistical significance of the above findings must be emphasized. 
This is because of the absence of observational errors  for the 
various offsets derived from Figure 6 of \citet{Kim21}.

  \citet{Kim17} present  a set  of SIDM merging simulations of equal-mass 
   massive clusters,  with masses on the 
order of $\sim 10^{15} \msun$.
They compared the  expected galaxy-DM offsets from the SIDM simulations
against measured galaxy-DM offsets in observed mergers.
The authors   were able to derive 
 an upper limit  on the galaxy-DM offset  of  $\simlt 20 \kpc$ for 
${\sigma_{DM}}/{m_X} =1  \sxu$, in accordance with the  range of the spatial 
separations shown in the left panels of Figure \ref{fig:planeSXa}
for the ${\sigma_{DM}}/{m_X} =1  \sxu$ simulations.
As noticed by the authors, these bounds are an order of magnitude smaller than 
the measured offsets for these merging systems ($\sim 100-300 \kpc$), and 
thus in tension with their simulations. 
Their conclusions were, however, affected by the large observational uncertainties 
for the measured offsets in massive mergers of $\sim 10^{15} \msun$.

Nonetheless,  according to \citet{Kim21}, the presence of  offsets between the 
different mass components of the El Gordo cluster is statistically significant,  
  which in turn implies that  the null hypothesis of zero size offsets  is  
statistically disfavored.
On the other hand, is difficult  to assess the statistical 
significance of the offset sizes we found for the merger model XDBf\_sb  
(Figure \ref{fig:planeSXc}) with ${\sigma_{DM}}/{m_X} =4 \sxu$. This indeterminacy is caused by the absence  of positional errors 
for the various peak locations used to calculate the offsets.

In order to minimize the impact of statistical uncertainties, we now attempt to 
make a 
rough estimate of the positional errors for the offsets with the largest 
sizes. From the   relative location of the various peaks,  as extracted from 
Figure 6 of \citet{Kim21}, it is easily seen that the galaxy number density peak 
of the NW cluster is spatially offset from its parent mass centroid by 
about $\sim 120 \kpc$.
However, for a galaxy group, the determination of the galaxy number density peak 
location is heavily affected by shot noise. For a group with 
$\sim 100$ member galaxies, \citet{Kim17} extrapolate the expected  uncertainty 
to  $\sim 160 \kpc$.

The large offset ($d^{NW}_{SZ-DM} \sim 150 \kpc$)  between the SZ emission peak 
and the NW mass centroid is similarly influenced by the large  error
in the SZ centroid estimate 
($\sigma_{SZ} \sim 70 \kpc \sim 1^{\prime}.4/(S/N)$; \citetalias{Zh15}).
Moreover, for the SE cluster, the BCG-DM offset ($\sim 60 \kpc$)  is approximately 
within the $\sim 1 \sigma$ range of the mass centroid uncertainty: 
$\sigma_{DM}\sim 40 \kpc$. We crudely estimated $\sigma_{DM}$ from 
\citet{Kim21};  according to the authors, a $2 \sigma$ uncertainty
in the position of the mass  centroids corresponds 
to $\sim 10^{''}$  ($ \sim 80 \kpc$ at $z=0.87$).

The size of positional errors for the X-ray emission peak of the SE cluster
is expected to be less significant than for the other centroids, 
due to the squared dependence of the X-ray emission with the gas density.
These errors  are quantified  according to 
 the observational procedure adopted to process the raw X-ray images and to 
extract the positions of substructures in the X-ray emission, together with the 
 relative errors.
In the case of El Gordo, the effective positional error 
of the X-ray peak is relatively small  and  set by
the angular resolution ($\sim 0.5^{''}$) of the \textit{Chandra} X-ray image\footnote{J. Kim: private communication}.

The uncertainty in the size of the observed offset between the X-ray  and the 
SE mass peak is then dominated by the uncertainty $\sigma_{DM}$ in the position 
of the mass peak; this  allows us to place the offset  in the range 
$d^{SE}_{X-DM} \sim 100 \pm 40 \kpc $.
An important result that follows from this finding is that 
 SIDM merger simulations of the El Gordo cluster 
with ${\sigma_{DM}}/{m_X} \simlt 2 \sxu$  are marginally inconsistent with
the observed offset, $d^{SE}_{X-DM}$.
As can be seen from the maps of Figure \ref{fig:planeSXa}, for these models 
the offset of the X-ray peak is always limited 
to $d^{SE}_{X-DM} \simlt 40 \kpc $.
In the following we thus adopt the view  that 
the estimated offset sizes for the El Gordo cluster are in the closest agreement 
with SIDM merger simulations for values of the DM cross-section in the range 
   ${\sigma_{DM}}/{m_X} \sim 4-5\sxu$. 

To resolve the inconsistency  between this range of values for 
${\sigma_{DM}}/{m_X}$  and its current upper limits on cluster scales 
(${\sigma_{DM}}/{m_X} \simlt 1 \sxu$),  
 we suggest that the SIDM model proposed
so far should only be considered a first-order approximation to 
 the modeling of the physical processes  describing DM interaction.
Specifically, we argue that the capacity of DM to exhibit collisional properties
 during a cluster merger  is correlated  with some  other feature 
of the collision itself.

A simple logical choice to explain this behavior of DM  
 during the mergers is that the DM collisional properties  
are not always present, but are switched on in accordance with  the energy 
of the cluster collisions. This clearly implies that  an SIDM model based on 
scattering between DM particles should be regarded as a sort of low-order approximation to a much more complex physical process,  in which 
DM interactions between the two colliding DM halos come into play according to 
some energy threshold, $E_{\rm crit}$.
For  a merging cluster  with a collisional energy below this threshold, the 
DM of the two halos should stay in a sort of ground state  and 
will exhibit the usual collisionless properties during the cluster collision.

 It must be stressed that such an  energy behavior is completely different from the
mechanism previously proposed \citep{Kap16}, in which a power-law dependence of
$\sigma_{DM}/m_X$ on the collision  velocity has been postulated to 
satisfy its present upper limits over a wide range of scales.
Such a consideration also applies to other SIDM models \citep{Tulin18}
in which DM particles
scatter inelastically \citep{Vog19} or anisotropically   \citep{Rob17b}.

To provide a viable theoretical framework for the underlying physical processes
leading to such a step-like behavior for DM is beyond the scope of this work. Nonetheless, we can try to assess the validity of this hypothesis by looking 
at whether there is  some dependence of the observed offsets  on the energy of 
the merging system.

To validate the proposed picture, we took as reference the El Gordo 
cluster, for which the measured galaxy-DM and BCG-DM offsets lie
in the range $ 50 \sim 100 \kpc$. The best match to the data is obtained for models
 XDBf\_sa and  XDBf\_sb, which have the initial collision parameters 
$\{ M^{(1)}_{200},~q,~V,~d_{ini} \}$  of model Bf\_rc20:
$\{ 1.6 \cdot 10^{15} \msun, ~2.3, ~2,500 \kms, ~6.1 \mpc \}  $, 
where $d_{ini} = 2(r^1_{200}+r^2_{200}) \simeq 2(1.74+1.32)$. 
With these parameters we can  estimate using  Equation (\ref{ecoll.eq}) 
the energy of the collision:   
$ E_{\rm{EG}} \sim 1.4 \cdot 10^{64} {\rm{\,ergs}} $.

We then compared this finding to a merging system for which no appreciable
offsets have been detected between the DM and the stellar component: the Bullet Cluster. This is a well-known cluster merger  that 
has been the subject of many numerical studies \citep{Spr07,Mas08,La14,Lage15,Rob17}. 
We extracted the collision parameters $\{ M^{(1)}_{200},~q,~V,~d_{ini} \}$  from 
Table 1  (LF entry) of  \citet{Lage15}:
$\{ 1.9 \cdot 10^{15} \msun, ~7.3, ~2,800 \kms, ~2.8 \mpc \} $.
According to these collision parameters, 
the  total energy content (\ref{ecoll.eq}) of the Bullet Cluster is
$ E_{\rm{Bullet}} \sim 3 \cdot 10^{63} {\rm{\,ergs}} $.

According to our hypothesis, these simple estimates bracket the allowed 
values of $E_{\rm crit}$ to be approximately in the range 
$ E_{\rm{Bullet}} \simlt E_{\rm crit}   \simlt E_{\rm{EG}} $.
To further corroborate these findings, it would be interesting to verify 
if there are other major merger clusters that exhibit large galaxy-DM 
peak offsets ($\sim 100 \kpc$). If the merging clusters are as massive as 
El Gordo, this would clearly  lend support to the idea that there is a critical 
energy threshold,  $E_{\rm crit}$, around the value of $  E_{\rm{EG}} $.

From Table 2 of \citet{Kim17},  it appears that the Sausage Cluster CIZA 
J2242.8+5301 at $z=0.19$ is the only merging cluster that fits these 
constraints, apart from El Gordo.
This cluster is characterized by an elongated X-ray emission along the 
north--south merger axis, hence the name.  The morphology is suggestive of an 
almost head-on collision, with spectroscopic data \citep{Daw15} favoring  a 
collision plane close to that of the sky.
The total estimated mass \citep{Jee15} is about 
$\sim 2\cdot 10^{15} \msun$, with a mass ratio close to unity, and the  
DM centroids are separated by about $\sim 1 \mpc$.

For this cluster, Table 2 of \citet{Kim17} reports galaxy-DM offsets between 
$\sim 50\kpc$ and $\sim 300 \kpc$, albeit with large uncertainties. 
In the context of SIDM, it should  however be considered  particularly indicative 
that, for the northern group,  the  galaxy centroid appears to be leading the 
corresponding mass peak.
According to \citet{Jee15}, the detection of  this offset is highly 
significant, while for the southern cluster  the errors are relatively large.

We estimate the collision parameters     
of the  Sausage Cluster from  \citet{Jee15}: 
$\{ M^{(1)}_{200},~q,~V,~d_{ini} \}=
\{ 1.1 \cdot 10^{15} \msun, 
~1.12, ~1,300 \kms, ~4 \mpc \} $,
 where $d_{ini} = 2(r^1_{200}+r^2_{200})\sim 4 \mpc $  and $V$ is estimated 
from Equation (14) of \citet{Jee15} by setting $b=0$ and 
$t_{\rm impact}\sim 11 $ Gyr.  We can now find the energy of the merging,  which 
turns out to be about 
 $ E_{\rm{Sausage}} \sim 1.5 \cdot 10^{64} {\rm{\,ergs}} $.

This energy value is very close to that  of $ E_{\rm{EG}} $ and strongly
supports the hypothesis of the existence of an  energy threshold that regulates
DM behavior.\ This implies that, during the cluster collision,
DM can exhibit collisional properties  according to the amount of
collisional energy.

Numerical simulations of the Sausage Cluster  have previously been performed   \citep{Wee11,Donnert17,Mol17}, but without including the effect of 
SIDM. 
 From the  SIDM merging simulations of \citet{Kim17},  
the expected  galaxy-DM offset  for merging clusters of $\sim 10^{15} \msun$  
is on the 
order of  $\simlt 20 \kpc$ for 
${\sigma_{DM}}/{m_X} =1  \sxu$, which is smaller than 
the estimated offset for the Sausage Cluster by about a factor of 10.

We argue that these inconsistencies would be solved in the SIDM scenario proposed 
here, and that the study of the different offsets in major cluster mergers is 
going to provide significant clues as to the nature of DM.
In this context,  SIDM merging simulations of the Sausage Cluster, in connection 
with the results extracted here from the SIDM runs of the El Gordo cluster,  
 will most likely lead to meaningful insights into the modeling of DM 
self-interactions based on particle collisions.

%
\begin{acknowledgements}
The author  is grateful to  Congyao Zhang  for his many clarifying comments 
about his paper that have helped to produce this work. The author would also like
to thank Jinhyub Kim for helping with Figure \ref{fig:planeSXc},  John Miller is also  
acknowledged for carefully reading and improving the manuscript. 
 The computations of this paper were carried out  using the Ulisse cluster 
at SISSA and the Marconi cluster at CINECA (Italy), under a SISSA-CINECA agreement.
\end{acknowledgements}

\appendix
\section{Parallel implementation of the DM scattering algorithm 
and code validation } 
\label{appSIDM}
 We now describe the parallelization  of the Monte Carlo scheme 
described in Section \ref{subsec:icsidm}. 
We considered the whole set, $\{i\}{\subseteq \Omega} $, of the $N_{DM}$ 
  DM particles, which are 
distributed across the whole computational volume, $\Omega$. 
In a serial code scattering events will be sampled by applying 
 Equation (\ref{pijdm.eq}) to the whole set  $\{i\}{\subseteq \Omega} $ 
of DM particles. However, the scattering process is cumulative, so 
any scattering event requires knowledge from previous events. 
 If one implements  the algorithm in a serial code this condition is 
implicit, but with a parallel code this is not obvious.

The  Monte Carlo method is then implemented as follows. We  assumed that 
the parallel code  subdivides the computation into  $P$  tasks, 
and the simulation volume is correspondingly  split into
  $\Omega_p$   non-overlapping sub-volumes, with   $p=0,1,\ldots,P-1$ being 
the task index. The load balancing domain decomposition is done  by cutting 
segments 
of a globally ordered Hilbert space-filling curve, as in the code 
GADGET-2 \citep{Spr05}.

We next  evaluated the scattering probability 
  (\ref{pijdm.eq}) for the  subset  $\{i\}{\subseteq \Omega_p} $  of DM 
particles that are members of task $p$.  If a scattering event occurs,
then the tagged neighbor particle, $j,$ could be  a member of either $p$ or another task, $q$; in the latter  case, the particle  was imported 
by $p$ during the 
root-finding  neighbor search (\ref{rhodm.eq}).
 Moreover, even if $j$ is a member of the local task $p$, it could still be 
a neighbor member of a particle in another task $q$. 
The same applies to particle $i$. 
 Therefore,  a blind implementation of the  Monte Carlo algorithm  
in a parallel code  could lead  to the same particle  being  scattered 
 two or more times with the same initial velocity.

To avoid this problem, we then proceeded as follows. We first assumed 
that at any given time all   the scattering pairs $\{i,j\}$ of  
the $P$ tasks  have been identified. 
For  the generic pair $\{i,j\}_p$ in 
task $p$, there is then the possibility that the neighbor particle $j$ is a
member of another task $q$; otherwise, either $i$ or $j$ (or both) could be 
neighbors of another scattering pair in task $q$.
We now denote as $\{i,j\}^{in}_p$ all of the scattering pairs of $p$ that 
do not satisfy any of the above conditions (i.e., the scattering takes 
place entirely within task $p$). We then split the set of scattering 
pairs $\{i,j\}_{p}$ in task $p$ between the subset $\{i,j\}^{in}_p$  
and all the other pairs, which we denote as $\{i,j\}^{oth}_p$.

To avoid multiple scatterings, it is now necessary that each task $p$ knows if
 a member of a scattering pair $\{i,j\}^{oth}_p$ has undergone  a scattering
 event in another task, this in order to
  accordingly update its velocity.
To accomplish this  for each pair  $\{i,j\}^{oth}_p$  we associate a 64 bit 
integer key defined as   $ID^{(i_K,j_K)}_p=i_K(i)+M j_K(j)$, 
where $i_K(i)$ denotes an integer key 
of the local DM particle $i$.
 The integer key $i_K(i)$ is  unique for each DM particle  and it is assigned
 initially; it is a DM particle property that moves with the particle 
 when  this migrates  between processors.
The integer $M$ ($2^{31}$) must be significantly larger   than $N_{DM} $; 
therefore, the integers $ID^{(i_K,j_K)}_p$ are unique for each scattering pair in the $P$ tasks.

 To consistently incorporate multiple scatterings in the parallel code, 
we now perform a global sorting of the $P$ sets of integers 
$ID^{(i_K,j_K)}_p$, with $p=0,1,\ldots,P-1$.
 After the sorting each task $p$ knows the global list of sorted keys, 
 and therefore all of the other scatterings and subsequently in which task 
they  are taking place.

Finally, the post-scattering velocities (\ref{vdmscatt.eq}) of each 
scattering pair in task $p$ are evaluated as follows. 
After the global sorting each task will  process its local set
of scattering pairs by looping iteratively over its local list of sorted 
keys,  this in increasing key  order because of the sorting. Pairs that 
 exhibit a  scattering dependence  with 
 $ID^{(i_K,j_K)}_p > ID^{(j_K,m_K)}_q$  are put in a waiting list until the 
updated velocities of  particle $j$ have been received from task $q$.
We note that this implementation allows for a single particle to 
undergo multiple scatterings.
 
To improve code performances, we make use of asynchronous communications 
 to optimize data transfer between processors. We exploit the 
advantage of using  this communication mode by splitting   
the list of local pairs $\{i,j\}^{in}_p$  and by   scattering 
 a subset of these pairs
whenever the task is waiting updated particle velocities from other tasks.

To test the SIDM part of the code we evolve an isolated  DM NFW halo, 
initially at equilibrium, for different values of the 
SIDM cross-section.  We considered the following values:
 ${\sigma_{DM}}/{m_X} =0,~1,$ and $10 \sxu$.
For the DM halo, we set $M_{200}=1.3 \cdot 10^{15} \msun$  at $z=0.87$, 
so $R_{200}=1.63$ \mpc ~and for the concentration parameter
$c_{200}=2.55$ from Equation  (\ref{cfit.eq}). 
The initial condition setup of the DM halo is described in Section
\ref{subsec:icdm}.

For the different runs  we  show here at a given time  $t$ 
 the spherically averaged DM radial density profile $\rho_{DM}(r,t)$  
and the scattering rate per particle 
$\Gamma(r,t)$.  The  theoretical   expression of  $\Gamma(r,t)$ at the 
time $t$ is evaluated as \citep{Rob15} 

 \begin{equation}
 \Gamma_{th}(r)=\frac{\sigma_{DM}}{m_X}  < v_{pair}(r) > \rho_{DM}(r)
    \label{gammadm.eq}
 ,\end{equation}
where $\rho_{DM}(r)$ is the local DM density and $< v_{pair}(r) > $ the mean 
pairwise velocity.  Assuming a Maxwell-Boltzmann distribution 
 this  quantity  can be written as \citep{Rob15} 
 $< v_{pair}(r) >= (4 /\sqrt{\pi}) \sigma_{1D}(r)$, where 
$\sigma_{1D}(r)$ is the 1D velocity dispersion 
of the NFW halo  \citep{Lok01}.

To contrast  the predicted rate,  $\Gamma_{th}(r),$ with the corresponding
rate extracted from the simulations,   $\Gamma_{sim}(r),$  we measured 
 $\Gamma_{sim}(r)$  as follows. We  first accumulated the number of 
scattering events as a function of radius in radial bins  together with the 
number of particles  in the same bin. At epoch $t$, the binned scattering rate, 
 $\Gamma_{sim}(r),$  was then obtained by dividing for each bin the number of
scattering events   by  the number of particles.  
It must be stressed that during the simulations the rates $\Gamma_{sim}(r)$ are 
obtained
by evaluating the local scattering probability (\ref{pijdm.eq}), but 
without updating the particle velocities (\ref{vdmscatt.eq}) of  
the pair. This is to avoid any modification to the DM density profile, which
otherwise  will make Equation (\ref{gammadm.eq}) not applicable.

\begin{figure*}[!ht]
\includegraphics[width=0.95\textwidth]{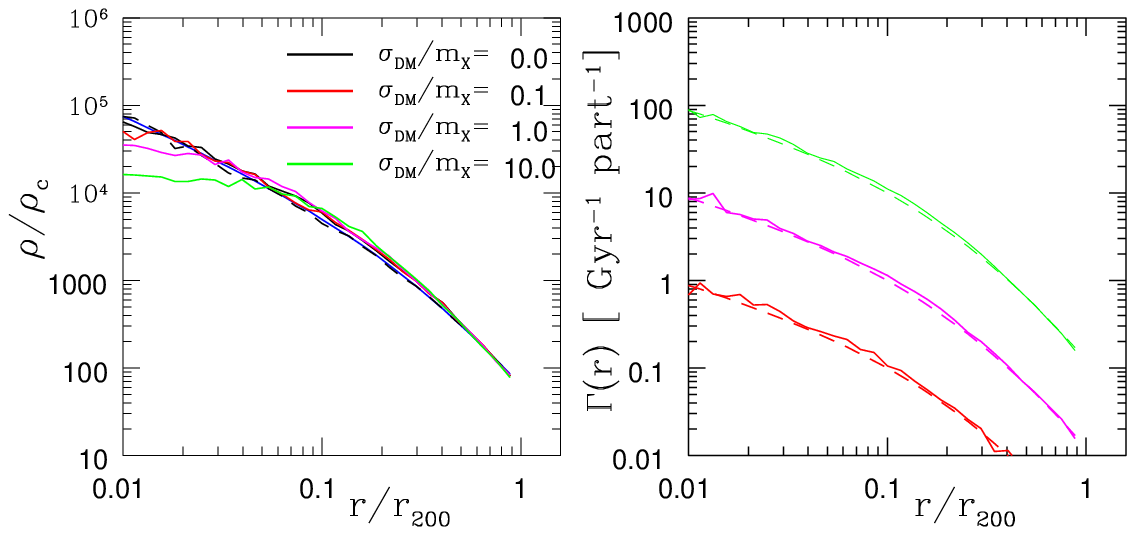}
\caption{ Radial density (left panel) and scattering rate (right panel) 
profiles at $t=1$ Gyr for three  SIDM runs with different 
DM scattering cross-sections: ${\sigma_{DM}}/{m_X} =0,~1,$ and $10 \sxu$.
The tests were carried out for a NFW DM halo with 
 $M_{200}=1.3 \cdot 10^{15} \msun$   and  $R_{200}=1.63$ \mpc, 
initially at equilibrium.
In the left panel we also show the density profile of a run without 
SIDM, ${\sigma_{DM}}/{m_X} =0$, at $t=0$ ( dash black line) and
at $t=1$ Gyr ( solid black line).  The solid blue line is the 
NFW theoretical profile.
\label{fig:sidm_halo}}
\end{figure*}

The results we obtain after $t=1$ Gyr are shown  in Figure 
\ref{fig:sidm_halo}, in which for the different cases are plotted the DM 
density profiles (left panel) and  the corresponding  scattering rates 
(right panel). For completeness, in the left panel we also show 
 the density profile (solid black line) of a halo without SIDM 
(${\sigma_{DM}}/{m_X} =0$). As expected, there is the development
of  cored density profiles for which at inner radii  the flattening of 
density  
is higher as ${\sigma_{DM}}/{m_X}$ increases. For the considered 
SIDM cases, the scattering rates  are in full agreement with the theoretical 
predictions.

The results of similar tests have been shown 
by various authors \citep{Vog12,Rob15,Rob17,ZuH19,Fis21}, we compare our 
density profile results with those of  a similar  test carried out 
by \citet{ZuH19}.  The authors adopt the same  SIDM implementation 
\citep{Vog12} employed here  and, moreover, assume the same  approximations 
for the  DM cross-section ${\sigma_{DM}}$. 
In their Section 3.1 the authors investigate the time evolution of an
isolated  halo  comprising gas and DM. They assume a cluster gas fraction 
of $0.12$,  we then consider dynamically negligible the impact 
of the gas when comparing at inner radii the  different DM density 
profiles.

For the halo DM density profile \citet{ZuH19}  chose an 
Hernquist profile with  $M^{Z}_{200}=1.25 \cdot 10^{15} \msun$  
and $a^Z_H=600  \kpc $ for the cluster mass and profile scale 
radius, respectively.
Our halo mass approximately coincides  with $M^{Z}_{200}$, so 
our rescaled profiles can be directly compared with the corresponding ones 
of  \citet{ZuH19}.

 Following  \citet{Spr05b}  we relate the scale radius $r_s$ and concentration 
parameter $c$ of the NFW profile to the scale length $a_H$  of an Hernquist 
density profile with the same mass within $r_{200}$:
 
 \begin{equation}
 a_H=r_s  \sqrt{ 2[ \ln(1+c)-c/(1+c)]}.
  \label{ahern.eq}
 \end{equation}

For the halo parameters we used $r_s =r_{200}/c\simeq 0.64  \mpc$
and  $a_H \simeq 660 \kpc$, which is only $\sim 10\%$ higher than
  $a^Z_H $.
Thus, at inner radii our profiles can be compared 
with the DM halo profiles displayed in top-left panel of  Figure 1  of 
 \citet{ZuH19}. From the left panel  of Figure \ref{fig:sidm_halo} 
at the initial epoch ($t=0$)
we estimate at the radius $ r \sim 10^{-2} r_{200} \simeq 16 \kpc$ 
 a central DM density of $\rho_{DM} /\rho_c \simeq 8 \cdot 10^4$.
For the chosen redshift ($z=0.87$) this value 
corresponds to $ \rho_{DM} \simeq 3 \cdot 10^7 \msun kpc^{-3}$, in 
accordance with the corresponding value estimated from Figure 1 of 
 \citet{ZuH19}.

We now estimate at $t=1$ Gyr  the central value of the halo DM density 
for the SIDM run with ${\sigma_{DM}}/{m_X} =10 \sxu$. 
For this test from  Figure \ref{fig:sidm_halo} 
we obtain  at the radius $ r \sim 10^{-2} r_{200}$   
 a central DM density of $\rho_{DM} /\rho_c \simeq 1.8 \cdot 10^4$, 
which is equivalent to $ \rho_{DM} \simeq 6.6 \cdot 10^6 \msun kpc^{-3}$.
 For the same SIDM run  this value is in close agreement  with the 
corresponding DM density,   estimated 
at the same radius and at the same epoch from  Figure 1 of \citet{ZuH19}.
The results of our tests then validate  our implementation of DM  
self-interactions, and demonstrate that the resulting code 
can be safely used to perform  the SIDM simulations presented here.

%

\end{document}